\documentclass[draft]{agujournal2019}
\usepackage{url} 
\usepackage{lineno}
\usepackage[inline]{trackchanges} 

\usepackage{soul}
\usepackage{units}

 \usepackage[makeroom]{cancel}
 \usepackage{amssymb, latexsym, amsmath}
 \usepackage{graphicx}
 \usepackage{epstopdf}
 \usepackage{color}
\usepackage{placeins}
\usepackage{float}
\usepackage{lscape}
\usepackage{xcolor}

\usepackage{ifpdf}
\usepackage{epsfig}
\usepackage[latin1]{inputenc}
\usepackage{amsfonts}
\usepackage{amsbsy}
\usepackage{amstext}
\usepackage{amsopn}
\usepackage{amsgen}

\usepackage{tikz}
\usepackage{adjustbox}  
   
\soulregister\citeA7
\soulregister\cite7
\soulregister\ref7
\soulregister\eqref7
\soulregister\it7
\soulregister\textit7
\soulregister\footnote7
\soulregister\caption7
\soulregister\rm7
\soulregister\unit7
\soulregister\label7

\newcommand{\s}{\nobreak\hspace{.11em}\nobreak}

 \newcommand{\be}{\begin{equation}}
 \newcommand{\ee}{\end{equation}}
 \newcommand{\ba}{\begin{eqnarray}}
 \newcommand{\ea}{\end{eqnarray}}
 \newcommand{\bs}{\begin{subequations}}
 \newcommand{\es}{\end{subequations}}

\draftfalse

\journalname{JGR: Planets}

\begin{document}

\include{macros}

%
%



\title{Is there a semi-molten layer at the base of the lunar mantle?}

%
%




\authors{Michaela Walterov\'{a}\affil{1}, Marie B\v{e}hounkov\'{a}\affil{2}, Michael Efroimsky\affil{3}}


\affiliation{1}{Institute of Planetary Research, German Aerospace Center (DLR), Berlin 12489 Germany}
\affiliation{2}{Department of Geophysics, Faculty of Mathematics and Physics, Charles University, Prague 12116 Czech Republic}
\affiliation{3}{US Naval Observatory, Washington DC 20392 USA}




\correspondingauthor{Michaela Walterov\'{a}}{kanovami@gmail.com}




\begin{keypoints}


\item A lunar mantle governed by the Andrade model fits selenodetic
constraints only with a very weak frequency dependence of tidal dissipation
\item We seek the parameters of two more complex models that may explain the anomalous frequency dependence of tidal Q measured by LLR
\item Both a dissipative basal layer and elastically-accommodated
grain-boundary sliding in the deep mantle can result in the same tidal response

\end{keypoints}

%
%

%
%


\begin{abstract}
Parameterised by the Love number $k_2$ and the tidal quality factor $Q$, and inferred from lunar laser ranging (LLR), tidal dissipation in the Moon follows an unexpected frequency dependence often interpreted as evidence for a highly dissipative, melt-bearing layer encompassing the core-mantle boundary. Within this, more or less standard interpretation, the basal layer's viscosity is required to be of order $10^{15}$ to $\unit[10^{16}]{Pa\; s}$ and its outer radius is predicted to extend to the zone of deep moonquakes. While the reconciliation of those predictions with the mechanical properties of rocks might be challenging, alternative lunar interior models without the basal layer are said to be unable to fit the frequency dependence of tidal $Q$.
    
The purpose of our paper is to illustrate under what conditions the frequency-dependence of lunar tidal $Q$ can be interpreted without the need for deep-seated partial melt. Devising a simplified lunar model, in which the mantle is described by the Sundberg-Cooper rheology, we predict the relaxation strength and characteristic timescale of elastically-accommodated grain boundary sliding in the mantle that would give rise to the desired frequency dependence. Along with developing this alternative model, we test the traditional model with a basal partial melt; and we show that the two models cannot be distinguished from each other by the available selenodetic measurements. Additional insight into the nature of lunar tidal dissipation can be gained either by measurements of higher-degree Love numbers and quality factors or by farside lunar seismology.
\end{abstract}

\section*{Plain Language Summary}
As the Moon raises ocean tides on the Earth, the Earth itself gives rise to periodic deformation of the Moon. Precise measurements of lunar shape and motion can reveal those deformations and even relate them to our natural satellite's interior structure. In this work, we discuss two interpretations of those measurements. According to the first one, the lunar interior is hot and a small part of it might have melted, forming a thick layer of weak material buried more than 1000 km deep under the lunar surface. According to the second one, there is no such layer, and the measured deformation can be explained by the behaviour of solid rocks at relatively low temperatures. We show that the two possibilities cannot be distinguished from each other by the existing data.

%
%

%


%
%
%
%

 \section{Motivation\label{motivation}}

 Fitting of the lunar laser ranging (LLR) data to the quality-factor power scaling law $\s Q\sim\chi^{p}\s$ rendered a small {\it{negative}} value of the exponential: $\s p\s=\s-\s 0.19~$
 \cite{williams2001}. Further attempts by the JPL team to reprocess the data led to $\s p\s =\s -\s 0.07~$. $\,$According to \citeA{WilliamsBoggs2009},\\
 ~\\
 ``{\it{
 $\,Q\,$ for rock is expected to have a weak dependence on tidal period, but it is expected to decrease with period rather than increase.
 }}"
 \vspace{3mm}

 The most recent estimates of the tidal contribution to the lunar physical librations \cite{williams2015} still predict a mild increase of $Q$ with period: from $Q=38\pm4$ at one month to $Q=41\pm9$ at one year, yielding $p=-0.03\pm0.09$. \citeA{Efroimsky2012a,Efroimsky2012b} suggested that since the frequency-dependence of $k_2/Q$ always has a kink shape, like in Figure \ref{Fig1}, the negative slope found by the LLR measurements could be consistent with the peak of the kink residing between the monthly and annual frequencies. This interpretation entails, for a homogeneous Maxwell or Andrade lunar model, very low values of the mean viscosity, indicating the presence of partial melt.

 Our goal now is to devise an interpretation based on the Sundberg-Cooper model. Within that model, the kink contains not one but two peaks, and we are considering the possibility that the negative slope of our interest is due to the monthly and annual frequencies bracketing either this peak or the local inter-peak minimum.

\section{Introduction\label{intro}}

 \subsection{Overview of Previous Works}

 The knowledge of the interior structure of the Moon is essential for understanding its thermal, geochemical, and orbital evolution as well as the coupled evolution of the Earth-Moon system. The proximity of our natural satellite to the Earth has also made it a frequent target of geophysical exploration. A large amount of data was collected by lunar seismic stations, deployed by the Apollo missions, which were functional for several years between 1972 and 1977 \cite<for a review, see, e.g.,>[]{garcia2019,khan2013,nunn2020}. Other constraints are being placed by selenodetic measurements or by geochemical and petrological considerations. However, the deepest interior of the Moon still remains somewhat mysterious. Although different models based on the inversion of seismic travel times generally agree on the lunar mantle structure down to $\sim\unit[1200]{km}$, below these depths they start to diverge greatly \cite{garcia2019}.

 After the acquisition of the first data by the lunar seismic network, it was pointed out by \citeA{nakamura1973,Nakamura} that direct shear waves from the farside of the Moon are not being detected by some of the near-side seismometers. Moreover, deep moonquakes, a class of tidally-triggered seismic events originating at around $\unit[1000]{km}$ depth, were almost absent on the farside. This puzzling phenomenon was interpreted by \citeA{nakamura1973} as an indication of a shear-wave shadow zone caused by a highly attenuating region around the core. \citeA{Nakamura2005} further reported efforts to find farside moonquakes among the deep moonquake nests that had not been located previously. Having identified about 30 likely farside nests, his updated analysis still demonstrated that either the region of the Moon's deep interior within about $40$ degrees from the antipodes (the centre of the farside) is nearly aseismic or a portion of the lunar lower mantle severely attenuates or deflects seismic waves. Lunar seismic data were also reprocessed by \citeA{Weber} and \citeA{garcia2011}. However, while \citeA{Weber} also found evidence for deep mantle layering and a strongly attenuating zone at the mantle base, \citeA{garcia2011} did not find evidence for such a feature in their analysis. The discussion about the seismic evidence for a strongly attenuating zone is thus still ongoing \cite{garcia2019}.\\

Several authors argued for the existence of a low-velocity zone (LVZ) at the base of the mantle also on other than seismological grounds. They linked it to partial melting in the deep lunar interior, which might be triggered either by tidal dissipation \cite{harada2014}, or by the presence of incompatible, radiogenic elements buried after an ancient mantle overturn \cite{khan2014}. The idea of an overturn has been suggested by numerical modelling of magma ocean solidification with the emplacement of ilmenite-bearing cumulates above core-mantle boundary. Moreover, it is potentially supported by observations of near-surface gravity anomalies that point at an early lunar expansion triggered by radiogenic heating of the deep interior \cite{zhang2013}.

 Evidence for a low-rigidity/low-viscosity zone has also been sought in the lunar libration signal obtained by LLR \cite<e.g.,>[]{williams2001,williams2015}, and
 in selenodetic measurements (including orbiter tracking) that are sensitive to the lunar gravity field and tidal deformation \cite<e.g.,>[]{konopliv2013,lemoine2013}. One of the most surprising findings resulting from fitting the LLR data was the low value and unexpected frequency dependence of the tidal quality factor $Q$, as mentioned in Section \ref{motivation} above. The inferred frequency dependence can be explained by a low effective viscosity of the Moon \cite{Efroimsky2012a,Efroimsky2012b}, or by the presence of a secondary peak in the dissipation spectrum \cite<e.g.,>[]{williams2015}, possibly caused by the putative basal layer \cite{harada2014}. The thickness, rheological properties, composition, and thermal state of that layer have been explored in a large number of studies \cite{harada2014,harada2016,khan2014,matsumoto2015,raevskiy2015,williams2015,matsuyama2016,tan2021,kronrod2022,xiao2022,briaud2023,briaud2023nat} and are reviewed in greater detail in the Supporting Information (SI) to this text. A summary is also provided in Table \ref{tab:layer_references}. The typical value of the basal layer's viscosity is $\sim\unit[10^{16}]{Pa\;s}$, the outer radius is mostly below $\unit[600]{km}$, and the predictions for the rigidity range from about $\unit[16]{GPa}$ \cite{khan2014,xiao2022} to solid-like values \cite{raevskiy2015,matsuyama2016,kronrod2022}. Earlier results from LLR indicated that the lunar core-mantle boundary (CMB) might still be out of equilibrium, which would imply long relaxation times and high lower-mantle viscosities, in contradiction to the presence of partial melt. However, this hypothesis is not supported by more recent evaluations of LLR data \cite{viswanathan2019}, that indicate that the CMB is in hydrostatic equilibrium.\\

\begin{table}[]
    \centering
    \begin{tabular}{l c c c c p{2.5cm}}
        \hline
        Reference & Viscosity & Rigidity & Radius & Thickness & Rheology \\
         & [$\unit{Pa\;s}$] & [$\unit{GPa}$] & [$\unit{km}$] & [$\unit{km}$] & \\
        \hline
        \citeA{harada2014} & $2\times10^{16}$ & $35$ & $500$ & $170$ & Maxwell \\[0.3em]
        \citeA{khan2014} & --- & $\sim16$ & $340-490$ & $150-200$ & power law$^{\rm{a}}$ \\[0.3em]
        \citeA{raevskiy2015} & --- & $30-55$ & $530-550$ & $\sim180$ & elastic \\[0.3em]
        \citeA{williams2015} & $\sim5\times10^{16}$ & $35$ & $\geq535$ & $\geq205$ & Andrade/Burgers + Debye peak$^{\rm{b}}$ \\[0.3em] 
        \citeA{matsumoto2015} & $\left(2.5^{+1.5}_{-0.9}\right)\times10^{16}$ & $\sim30$ & $\sim570$ & $>170$ & Maxwell \\[0.3em] 
        \citeA{matsuyama2016} & --- & $43^{+26}_{-9}$ & $300-700$ & $197^{+66}_{-186}$ & power law$^{\rm{a}}$ \\[0.3em] 
        \citeA{harada2016} & $3\times10^{16}$ & $35$ & $540-560$ & $210-230$ & Maxwell \\[0.3em] 
        \citeA{tan2021} & $3\times10^{16}$ & $35$ & $560-580$ & $230-250$ & Maxwell \\[0.3em]
        \citeA{kronrod2022} & $\sim10^{16}$ (?) & $30-60$ &  & $100-350$ & Maxwell \\[0.3em] 
        \citeA{xiao2022} & $5\times10^{16}$ & $\sim16$ & $\sim600$ & $282\pm5.4$ & Andrade$^{\rm{b}}$\\[0.3em] 
        \citeA{briaud2023} & $\left(4.5\pm0.8\right)\times10^{16}$ & $25$ & $500\pm1$ & $80-170$ & Maxwell + viscous core \\[0.3em]
        \citeA{briaud2023nat} & $10^{16.99\pm1.22}$ & $25$ & $545\pm35$ & $230\pm65$ & Maxwell + viscous core \\[0.3em]
         & $10^{17.98\pm1.06}$ & $25$ & $560\pm34$ & $198\pm49$ & Maxwell + viscous outer core + inner core \\
        \hline
    \end{tabular}
    \footnotesize
    $^{\rm{a}}$ By ``power law", we mean the anelastic correction of $k_2$ suggested by \citeA{zharkov2005}.\\
    $^{\rm{b}}$ Multiple rheological models were considered. We only list the preferred ones.
    \caption{An overview of the recent predictions for the basal layer's properties.}
    \label{tab:layer_references}
\end{table}

 Despite the relative consistency of the evidence for and the theoretical expectation of a highly dissipative basal layer, alternative models of a ``melt-free'' Moon have also been proposed \cite{nimmo2012,karato2013}. In particular, \citeA{nimmo2012} showed that the employment of a realistic, microphysically substantiated model of the tidal response can explain the low tidal $Q$ and the observed $k_2$ of the Moon without requiring the existence of a weak basal layer. Nevertheless, the lunar models considered by those authors were not able to fit the frequency dependence of the tidal $Q$. Another argument for high values of lower-mantle viscosities comes from the observations of deep moonquakes. \citeA{kawamura2017} reevaluated an ensemble of moonquakes occurring at depths between $750$ and $\unit[1200]{km}$ and found a brittle-ductile transition temperature of approximately ${1240}\,$--$\,\unit[1275]{K}$, implying a cold lunar interior with temperatures below the solidus of dry peridotite.

 As indicated in the previous paragraph, a feature of the selenodetic measurements that is difficult to explain without the existence of a highly dissipative basal layer is the aforementioned frequency dependence of the lunar $Q$, repeatedly derived from LLR measurements in the series of works by \citeA{williams2001,WilliamsBoggs2009,williams2014}, and \citeA{williams2015}. Even an independent implementation of the LLR software by \citeA{pavlov2016} predicts the same value of $Q$ for the monthly period as for the annual period, which is still not consistent with the expected frequency dependence of tidal dissipation in melt-free silicates.

 In the absence of other than LLR-based data on the lunar $Q$, the most plausible explanation for the unexpected frequency dependence might still be an observational uncertainty, rather than an effect contained in a tidal model. Nevertheless, in this work, we shall explore two possible implications of the frequency dependence under the explicit assumption that the fitted values are a result of a natural phenomenon and not of a model's limitations or an observation error.
 
 \subsection{Lunar $k_2$ and $Q$} \label{subsec:obs}

 We will use the potential tidal Love number derived from the GRAIL mission tracking data. Two independent analyses performed by the JPL group \cite[the GL0660B solution]{konopliv2013} and the GSFC group \cite[the GRGM660PRIM solution]{lemoine2013} yielded two possible values of the parameter: $k_2=0.02405\pm0.000176$ and $k_2=0.02427\pm0.00026$, respectively. The unweighted mean of the two alternative values is $k_2=0.02416\pm0.000222$ for a reference radius of $\unit[1738]{km}$, and $k_2=0.02422\pm0.000222$ for the actual mean radius of $\unit[1737.151]{km}$ \cite{williams2014}. For comparison, the recent analysis of the data from the Chang'e 5T1 mission gives $k_2=0.02430\pm0.0001$ \cite{yan2020}. We note that the value obtained from satellite tracking data corresponds, in particular, to the real part of the complex Love number introduced later in Subsection \ref{sub:complex_kl}. The GRAIL data are dominated by one-month tidal effects, and the resulting $k_2$ is thus interpreted as indicative of the deformation at the monthly frequency (A. Konopliv, private communication).

 The tidal quality factor $Q$ was obtained by fitting tidal contribution to lunar physical libration measured by LLR \cite{williams2001,williams2014,williams2015}. Interpreting the measurements of physical libration presents a highly complex problem, depending on cross interactions of tides raised by the Earth and the Sun, precise modeling of the lunar orbit and of the instantaneous positions of the Earth-based stations and the Moon-based retroreflectors, and on adequate incorporation of the lunar core-mantle friction \cite{williams2001}. In practice, the tidal time delay at a monthly period and the dissipation-related corrections to the periodic latitudinal and longitudinal variations in the Moon's orientation are output and related analytically to linear combinations of $k_2/Q$ at a number of loading frequencies. Since many of the loading frequencies are close to each other, the periodic corrections enable approximate estimation of the leading dissipation terms. Specifically, the strongest correction (compared to its uncertainty) is related to the annual longitudinal libration. Assuming a fixed $k_2$ at the monthly frequency, equal to the above-mentioned unweighted average, and using a complex rheological model best fitting the dissipation-related corrections to libration angles, \citeA{williams2015} derived the following frequency-dependent values of tidal quality factor: $Q=38\pm4$ at the period of 1 month, $Q=41\pm9$ at 1 year, and lower bounds of $Q\geq74$ at 3 years and $Q\geq58$ at 6 years.
 
 \citeA{williams2015} also attempted to find the frequency-dependence of $k_2$; however, the effect could not be detected by existing measurements. We note that in contrast to the unexpected frequency dependence of $Q$ found with the JPL-based software \cite{williams2001,williams2014,williams2015}, an independent implementation of the fitting tool with different preset solutions for part of the geophysical phenomena \cite{pavlov2016} predicted $Q=45$ at both the monthly and the annual frequencies. Moreover, \citeA{williams2015lpsc} reported $k_2/Q$ derived from the GRAIL data (sensitive to the monthly tidal variations) that indicate $Q_{\rm{monthly}}=41\pm4$.

 As an additional, though a relatively weak constraint on the lunar interior structure, we consider the degree-$3$ potential tidal Love number $k_3$
 and the degree-2 deformational Love number $h_2$ corresponding to radial deformation. The $k_3$ number has been derived from GRAIL mission tracking data and, as with $k_2$ above, we adopt the unweighted average of the two existing independent solutions \cite{lemoine2013,konopliv2013}: $k_3=0.0081\pm0.0018$. The $h_2$ number has been measured by LLR and by laser altimetry \cite{mazarico2014,pavlov2016,viswanathan2018,thor2021}, the most recent value, presented by \citeA{thor2021}, being $h_2=0.0387\pm0.0025$.
 
 We finally note the reason why the constraints on the lunar deep interior from the measurements of $k_3$ are weak. A degree-$l$ component of the internal tidal potential is proportional to $r^l$, where $r$ is the distance between the centres of mass of the tidally perturbed body and the perturber. For this reason, with increasing degree $l$, the shallower depths contribute more and more to the Love numbers $k_l$. The sensitivity of the higher-degree Love numbers to the deep interior is, therefore, limited as compared to degree $2$.

\subsection{Outline of This Work}

 After an overview of the models and interpretations proposed in recent literature (with a focus on the last ten years of the discussion), we are ready to continue with the central part of this project. Our plan is to provide an interpretation of the unexpected frequency dependence of tidal $Q$ which does not require partial melting \cite<in a way similar to>[]{nimmo2012} and compare it with a model containing a highly dissipative basal layer \cite{harada2014,matsumoto2015}. Section \ref{sec:rheo} introduces and gives a justification for the rheological model employed. Namely, it discusses the Sundberg-Cooper extension of the Andrade model and the dissipation related to elastically accommodated grain-boundary sliding (GBS). The following Section \ref{sec:love} links the non-elastic rheology to Love numbers and tidal quality factors. In Section \ref{sec:moon}, we first illustrate the expected position of a secondary peak in the dissipation spectrum of a homogeneous Moon, and then attempt to find the parameters of multi-layered lunar models that would produce the values of the monthly tidal $Q$ and annual $k_2/Q$ reported by \citeA{williams2015}. At the same time, we fit the empirical values of lunar $k_2$, $k_3$, and $h_2$ given in Subsection \ref{subsec:obs} and the total mass and moment of inertia of the Moon. Section \ref{sec:discussion} discusses implications of our models, and the results are briefly summarised in Section \ref{sec:concl}.

 \section{General Facts on Rheologies} \label{sec:rheo}

 \subsection{Constitutive Equation}

 Rheological properties of a material are encoded in a constitutive equation interconnecting the present-time deviatoric strain tensor $\,u_{\gamma\nu}(t)\,$ with the values that have been assumed by the deviatoric stress $\,{\sigma}_{\gamma\nu}(t\,')\,$ over the time period $\,t\,'\,\leq\,t\,$. Under linear deformation, the equation has the form of convolution, in the time domain:
 \begin{linenomath*}
 \begin{eqnarray}
 2\,u_{\gamma\nu}(t)\,=\,\hat{J}(t)~\sigma_{\gamma\nu}\,=\,\int^{t}_{-\infty}\stackrel{\;\centerdot}{J}(t-t\,')~
 {\sigma}_{\gamma\nu}(t\,')\,d t\,'~~,~~~
 \label{I12_4}
 \label{E1}
 \end{eqnarray}
 \end{linenomath*}
 and the form of product, in the frequency domain:
 \begin{linenomath*}
 \begin{eqnarray}
 2\;\bar{u}_{\gamma\nu}(\chi)\,=\;\bar{J}(\chi)\;\bar{\sigma}_{\gamma\nu}(\chi)\;\;.
 \label{LLJJKK}
 \label{E2}
 \end{eqnarray}
 \end{linenomath*}
 Here $\,\bar{u}_{\gamma\nu}(\chi)\,$ and $\,\bar{\sigma}_{\gamma\nu}(\chi)\,$ are the Fourier images of strain and stress, while the complex compliance $\,\bar{J}(\chi)\,$
 is a Fourier image of the kernel $\,\dot{J}(t-t\,')\,$ of the integral operator (\ref{I12_4}), see, e.g., \citeA{Efroimsky2012a,Efroimsky2012b} for details.
 
 \subsection{The Maxwell and Andrade Models\label{solids}\label{section3.2}}

 At low frequencies, the deformation of most minerals is viscoelastic and obeys the Maxwell model:
 \begin{subequations}
 \begin{linenomath*}
 \begin{eqnarray}
 \stackrel{\centerdot}{\mathbb{U}}\,=\,\frac{1}{2\,\mu}\;\stackrel{\centerdot}{\mathbb{S}}\,+\,\frac{1}{2\,\eta}\;{\mathbb{S}}~~~
 \end{eqnarray}
 \end{linenomath*}
 or, equivalently:
 \begin{linenomath*}
 \begin{eqnarray}
 \stackrel{\centerdot}{\mathbb{S}}\,+\;\frac{1\;}{\tau_{\rm{_M}}}\,{\mathbb{S}}
 ~=~2\,\mu\,\stackrel{\centerdot}{\mathbb{U}}~~~,
 \end{eqnarray}
 \end{linenomath*}
 \label{these}
 \end{subequations}
 ${\mathbb{U}}\,$ and $\,{\mathbb{S}}\,$ being the deviatoric strain and stress; $\,\eta\,$ and $\,\mu\,$ denoting the viscosity and rigidity. (Below, we shall address the question as to whether $\mu$ is the unrelaxed or relaxed rigidity.)
 The {\emph{Maxwell time}} is introduced as
 \begin{linenomath*}
 \begin{eqnarray}
 \tau_{\rm{_M}}\;\equiv\;\frac{\,\eta\,}{\,\mu\,}\,~.
 \label{Maxwell}
 \end{eqnarray}
 \end{linenomath*}
 For this rheological model, the kernel of the convolution operator (\ref{E1}) is a time derivative of the compliance function
 \begin{linenomath*}
 \begin{eqnarray}
 ^{\textstyle{^{(M)}}}J(t\,-\,t\,')\,=\,\left[\,J_{\rm{e}}\,+\,\left(t\;-\;t\,'\right)\;\frac{1}{\eta}\,\right]\;\Theta(t\,-\,t\,')~~~,
 \label{Max}
 \end{eqnarray}
 \end{linenomath*}
 where $\s\Theta(t\,-\,t\,')\s$ is the Heaviside step function, while the elastic compliance $\s J_{\rm{e}}\s$ is the inverse of the shear rigidity $\mu\,$:
 \begin{linenomath*}
 \begin{eqnarray}
 J_{\rm{e}}~\equiv~\frac{\,1\,}{\,\mu\,}\;\,.
 \end{eqnarray}
 \end{linenomath*}
 In the frequency domain, equation (\ref{these}) can be cast into form (\ref{LLJJKK}), with the complex
 compliance given by
 \begin{linenomath*}
 \begin{eqnarray}
 ^{\textstyle{^{(M)}}}{\bar{\mathit{J\,}}}(\chi)~=~J_e\,-\,\frac{i}{\eta\chi}~=~J_{\rm{e}}\,\left(\,1~-~\frac{i}{\chi\,\tau_{\rm{_M}}}\right)\;\,,
 \label{don}
 \label{LL42}
 \label{E8}
 \end{eqnarray}
 \end{linenomath*}
 and the terms $\,J_{\rm{e}}\,$ and $\,-\,{i}/(\eta\chi)\,$ being the elastic and viscous parts of deformation, correspondingly. So a Maxwell material is elastic at high frequencies, viscous at low.

 More general is the combined Maxwell-Andrade rheology, often referred to simply as the Andrade rheology. It comprises inputs from elasticity, viscosity, and anelastic processes:
 \begin{linenomath*}
 \begin{equation}
  ^{\textstyle{^{(A)}}}J(t-t^{\,\prime}) =\,\left[\s  J_{\rm{e}}\s +\s\beta\s (t-t^{\,\prime})^\alpha + \frac{t-t^{\,\prime}}{\eta}\s\right]\,\Theta(t-t^{\,\prime})\,\;,
 \end{equation}
 \end{linenomath*}
the corresponding complex compliance being
 \begin{subequations}
 \begin{linenomath*}
 \begin{eqnarray}
 ^{\textstyle{^{(A)}}}{\bar{\mathit{J\,}}}(\chi)&=&J_{\rm{e}}\,+\,\beta\,(i\chi)^{-\alpha}\;\Gamma\,(1+\alpha)\,-\,\frac{i}{\eta\chi}
  \label{112_1}
  \label{E3a}\\
 \nonumber\\
 &=& J_{\rm{e}}\,+\,\beta\,(i\chi)^{-\alpha}\;\Gamma\,(1+\alpha)\,-\,i\,J\,(\chi\,\tau_{\rm{_M}})^{-1}
 ~~,
 \label{112_2}
 \label{E3b}
 \end{eqnarray}
 \end{linenomath*}
 \end{subequations}
 where $\,\Gamma\,$ is the Gamma function,  while $\,\alpha\,$ and $\,\beta\,$ denote the dimensionless and dimensional Andrade parameters.

 Expressions (\ref{112_1} - \ref{112_2}) suffer an inconvenient feature, the fractional dimensions of the parameter $\s\beta\s$. It was therefore suggested in \citeA{Efroimsky2012a,Efroimsky2012b} to shape the compliance into a more suitable form
 \begin{linenomath*}
 \begin{equation}
  ^{\textstyle{^{(A)}}}J(t-t^{\,\prime}) =  \left[\s J_{\rm{e}}\, +\,J_{\rm{e}}\s \left(\frac{t\s-\s t^{\,\prime}}{\tau_{\rm{_A}}}\right)^\alpha + J_{\rm{e}}\,\frac{t\s-\s t^{\,\prime}}{\tau_{\rm{_M}}}\right]\;\Theta(t\,-\,t\,')\,\;,
\label{equation}
\end{equation}
\end{linenomath*}
\begin{linenomath*}
  \begin{eqnarray}
 ^{\textstyle{^{(A)}}}{\bar{\mathit{J\,}}}(\chi)~=~J_{\rm{e}}\,\left[\,1\,+\,(i\,\chi\,\tau_{\rm{_A}})^{-\alpha}\;\Gamma\,(1+\alpha)~-~i~(\chi\,\tau_{\rm{_M}})^{-1}\right]\,\;,~~~
 \label{112_3}
 \label{E3c}
 \end{eqnarray}
 \end{linenomath*}
 with the parameter $\,\tau_{_A}\,$ christened as $\s ${\it{the Andrade time}}$ \s $ and linked to $\beta$ through
 \begin{linenomath*}
 \begin{eqnarray}
 \beta\,=\,J_{\rm{e}}~\tau_{\rm{_A}}^{-\alpha}~~.
 \label{beta}
 \label{E4}
 \end{eqnarray}
 \end{linenomath*}
 Compliance (\ref{E3c}) is identical to (\ref{112_1}) and  (\ref{112_2}), but is spared of the parameter $\beta$ of fractional dimensions.

 \subsection{Why the Maxwell and Andrade Models Require Refinement\label{refinement}}

 In the literature, it is common to postulate that both the rigidity and compliance assume their $\,${\it{unrelaxed}}$\,$ values denoted with $\s\mu_{\rm{U}}\s$ and $\s J_{\rm{U}}\s$.

 This convention is reasonable for sufficiently high frequencies:
 \begin{linenomath*}
 \begin{equation}
  \chi~\mbox{~is~\,high}\qquad\Longrightarrow\qquad\mu~=~\mu_{\rm{U}}\qquad\mbox{and}\qquad J_{\rm{e}}~=~J_{\rm{U}}\quad.
 \end{equation}
 \end{linenomath*}
 The convention, however, becomes unjustified for low frequencies. In that situation, the material has, at each loading cycle, enough
 time to relax, wherefore both the rigidity modulus and its inverse assume values different from the unrelaxed ones. In the zero-frequency limit, they must acquire the relaxed values:
 \begin{linenomath*}
 \begin{equation}
 \chi~\rightarrow~0\qquad\Longrightarrow\qquad\mu~\rightarrow~\mu_{\rm{R}}\qquad\mbox{and}\qquad J_{\rm{e}}~\rightarrow~J_{\rm{R}}~~.
 \end{equation}
 \end{linenomath*}
 This fact must be taken care of, both within the Maxwell and Andrade models.

 \subsection{Generalisation of the Maxwell and Andrade Models,\\ according to Sundberg and Cooper (2010)} \label{Sund} 

The simplest expression for the time relaxation of the elastic part of the compliance is
 \bs
 \begin{linenomath*}
 \begin{eqnarray}
 J_{\rm{e}}(t)&=&J_{\rm{U}}\,+\,(J_{\rm{R}}\,-\,J_{\rm{U}})\,\left[1\,-\,e^{-t/\tau}\,\right]
 \label{sc1}\\
 \nonumber\\
 &=& J_{\rm{U}}\s\left[\,1\,+\,\Delta  \,\left( 1 -   e^{-\;{t}/{\tau} }  \right)
\right]\;\,,
\label{sc2}
\end{eqnarray}
\end{linenomath*}
\label{sc}
 \es
  where the so-called relaxation strength is introduced as
  \begin{linenomath*}
  \begin{equation}
     \Delta \equiv \frac{\textstyle J_{\rm{R}}}{\textstyle J_{\rm{U}}}-1~~,
 \end{equation}
 \end{linenomath*}
 while $\s\tau\s$ is the characteristic relaxation time. When relaxation of $\s J_{\rm{e}}\s$ is due to elastically accommodated grain-boundary sliding, this time can be calculated as
 \begin{linenomath*}
 \begin{eqnarray}
 \tau\;=\;\tau_{\rm{gbs}}\,=\;\frac{\eta_{\rm{gb}}\,d}{\mu_{\rm{U}}\;\delta}\,\;,
 \label{eq:tau_gbs}
 \end{eqnarray}
 \end{linenomath*}
 where $\,\eta_{\rm{gb}}\,$ is the grain-boundary viscosity, $\,d\,$ is the grain size, while $\,\delta\,$ is the structural width of the grain boundary. Details of energy-dissipation regimes associated with grain-boundary sliding are given, e.g., in \citeA{jackson2002,jackson2010,jackson2014}.

 In the frequency domain, this compliance is written as
 \begin{linenomath*}
  \begin{eqnarray}
 \bar{J}_{\rm{e}}(\chi)\;=\;J_{\rm{U}}\s\left[\s 1\;+\;\frac{\Delta}{1\:+\;\chi^2\;\tau^2}\;+\;i\;\frac{\chi\;\tau\;\Delta}{1\;+\;\chi^2\;\tau^2}\,\right]\;\,,
 \label{14}
 \end{eqnarray}
 \end{linenomath*}
 its imaginary part demonstrating a Debye peak. Our goal is to trace how this Debye peak translates into the frequency-dependence of the inverse tidal quality factor $1/Q$ and of $k_2/Q$ of a near-spherical celestial body.\\

 Substitution of formula (\ref{14}) into the overall expression (\ref{E3c}) for the Andrade complex compliance will produce the  \citeA{Sundberg} rheology:
 \begin{subequations}
 \begin{linenomath*}
 \begin{eqnarray}
 {\bar{\mathit{J\,}}}(\chi)
 &=&J_{\rm{U}}\left[1+\frac{\Delta}{1+\chi^2\tau^2}\,-\,i\;\frac{\chi\;\tau\;\Delta}{1+\chi^2\tau^2}
  +(i\chi\tau_{\rm{_A}})^{-\alpha}\s\Gamma(1+\alpha)-i(\chi\tau_{\rm{_M}})^{-1}\right]\;\qquad
  \label{15a}\\
 \nonumber\\
 \nonumber\\
  &=&
  \nonumber
 J_{\rm{U}}\s \bigg [ 1 \,+\,\frac{\Delta}{1\s+\s\chi^2\,\tau^2} \,+\,\Gamma(1+\alpha)\;\zeta^{-\alpha}\,(\chi\tau_{\rm{_M}})^{-\alpha}\,\cos \left(\frac{\alpha\pi}{2} \right) \bigg]
   ~\\   \label{15b}\\  \nonumber &-&
 i\,J_{\rm{U}}\s \bigg [ \frac{\chi\;\tau\;\Delta}{1\s+\s\chi^2\,\tau^2}\;+\;\Gamma(1+\alpha)\;\zeta^{-\alpha}\, (\chi\tau_{\rm{_M}})^{-\alpha}\,\sin \left(\frac{\alpha\pi}{2} \right) \,+\, (\chi \tau_{\rm{_M}})^{-1} \bigg]\,~,~~~
 \end{eqnarray}
 \end{linenomath*}
 \label{15}
 \end{subequations}
 where we introduced the dimensionless Andrade time
 \begin{linenomath*}
 \begin{eqnarray}
 \zeta\,=\;\frac{\tau_{\rm{_A}}}{\tau_{\rm{_M}}}~~.
 \label{dimensionlessAndrade}
 \end{eqnarray}
 \end{linenomath*}
 Be mindful that in expression (\ref{equation}) it is only the first term, $J_{\rm{e}}$, that is changed to function (\ref{sc2}).
 Accordingly, in equation (\ref{E3c}), it is only the first term, $J_{\rm{e}}$, that is substituted with function (\ref{14}).
 In the other terms, both the Maxwell and Andrade times are still introduced through the unrelaxed value $J_{\rm{e}}= J_{\rm{U}}\,$:
 \begin{linenomath*}
 \begin{eqnarray}
 \tau_{\rm{_M}}\,\equiv\;\eta\,J_{\rm{U}}\;\;,\;\;\;
 \tau_{\rm{_A}}\,\equiv\;\left( \frac{J_{\rm{U}}}{\beta}  \right)^{1/\alpha}
 ~~.
 \label{def}
 \end{eqnarray}
 \end{linenomath*}

Had we combined the elastic relaxation rule (\ref{14}) with the Maxwell model (\ref{LL42}) instead of Andrade, we would have arrived at the Burgers model~--- which would be equation (\ref{15}) with the Andrade terms omitted, i.e. with $\tau_{\rm{_A}}\longrightarrow\infty$. Simply speaking, in the absence of transient processes, Andrade becomes Maxwell, while Sundberg-Cooper becomes Burgers.

The presently standard term ``Sundberg-Cooper rheology'' was coined by \citeA{RenaudHenning2018} who studied tidal heating in mantles obeying this rheological law. This rheological law was later employed for Mars \cite{Mars} and for Pluto and Charon \cite{BAGHERI2022114871}.

 Along with the dimensionless Andrade time $\zeta$, 
 below we shall employ the relative relaxation time
 \begin{linenomath*}
 \begin{equation}
     t_{\rm{rel}} = \frac{\tau}{\tau_{\rm{_M}}}
     \label{along}
 \end{equation}
 \end{linenomath*}
 relating the relaxation timescale for the compliance $J_{\rm{e}}$ to the Maxwell time.

 \subsection{Further Options}

 The characteristic relaxation time $\tau$ can be replaced with a distribution $D(\tau)$ of times spanning an interval from a lower bound $\tau_{\rm{L}}$ to an upper bound $\tau_{\rm{H}}$. So the relaxation of the elastic part of the compliance will be not
 \begin{linenomath*}
 \begin{eqnarray}
 J_{\rm{e}}(t)\s=\s J_{\rm{U}}\s\left[\,1\,+\,\Delta  \,\left( 1 -   e^{-\;{t}/{\tau} }  \right)\right]\;\,
 \end{eqnarray}
 \end{linenomath*}
 but
 \begin{linenomath*}
  \begin{equation}
 \label{Eq:ExtBurgJ}
 J_e(t) = J_U \left[ 1 + \Delta \int_{\tau_L}^{\tau_H} D(\tau) \left[1 - \exp \left(- \;\frac{t}{\tau} \right)\right] d\tau
 \right]\,\;.
 \end{equation}
 \end{linenomath*}
 If the relaxation is due to elastically-accommodated GBS, this distribution would be a consequence of variable grain-boundary viscosity, grain sizes and shapes, and non-uniform orientation of grain boundaries with respect to the applied stress \cite<see also>[]{lee2010}.

 Insertion of expression (\ref{Eq:ExtBurgJ}) in the Maxwell model (\ref{Max}) or in the Andrade model (\ref{equation})
 produces the {\it extended Burgers model} or the {\it extended Sundberg-Cooper model}, correspondingly. For details, see \citeA{TidalReview} and references therein.
 
 \section{Complex Love Numbers and Quality Functions} \label{sec:love}

 The perturbing potential wherewith the Earth is acting on the Moon can be decomposed in series over Fourier modes $\omega_{lmpq}$ parameterised with four integers $lmpq$. If the tidal response of the Moon is linear, both the produced deformation and the resulting additional tidal potential of the Moon are expandable over the same Fourier modes, as proved in \citeA[Appendix C]{EfroimskyMakarov2014}. The proof is based on the fact that a linear integral operator (convolution) in the time domain corresponds to a product of Fourier images in the frequency domain.

 While the Fourier modes can be of either sign, the physical forcing frequencies in the body are
 \begin{linenomath*}
  \ba
  \chi_{lmpq}\,=\;|\s\omega_{lmpq}\s|\;\,.
  \label{frequency}
  \ea
  \end{linenomath*}
  An extended discussion of this fact can be found in Section 4.3 of \citeA{EfroimskyMakarov2013}.
  
  Wherever this causes no confusion, we omit the subscript to simplify the notation:
  \begin{linenomath*}
  \ba
  \omega\equiv\omega_{lmpq}~~,\quad
  \chi\s\equiv\s\chi_{\textstyle{_{lmpq}}}\;\,.
  \ea
  \end{linenomath*}

  \subsection{The Complex Love Number}\label{sub:complex_kl}

 Writing the degree-$l\,$ complex Love number as
 \begin{linenomath*}
  \ba
 \bar{k}_{l}(\omega)\;=\;
 \Re
 \left[\bar{k}_{l}(\omega)\right]\;+\;i\;
 \Im
 \left[\bar{k}_{l}(\omega)\right]\;=\;|\bar{k}_{l}(\omega)|\;
 e^{\textstyle{^{-i\epsilon_{l}(\omega)}}}~~,
 \label{L36}
 \ea
 \end{linenomath*}
 we conventionally denote the phase as $\,-\,\epsilon_l\,$, with a ``minus" sign. This convention imparts $\,\epsilon_l\,$ with the meaning of phase lag. We also introduce the so-called {\it{dynamical Love number}}
 \begin{linenomath*}
 \ba
 {k}_{l}(\omega)~=~|\bar{k}_{l}(\omega)|~~.
 \ea
 \end{linenomath*}
 A key role in the tidal theory is played by the {\it{quality functions}}
 \bs
 \begin{linenomath*}
 \ba
 {K}_{l}(\omega)
 \;\equiv\;-\;
 \Im
 \left[\,\bar{k}_{\it{l}}(\omega)\,
 \right]\,=\,
 \bar{k}_{\it{l}}(\omega)\;\sin\epsilon_{l}(\omega)
 \ea
 \end{linenomath*}
   entering the series expansions for tidal forces, torques, dissipation rate \cite{EfroimskyMakarov2014}, and orbital evolution \cite{BoueEfroimsky2019}
 
 Since
     ~Sign$\,\epsilon_l(\omega)$ = ~Sign$\,\omega\,$ \cite{EfroimskyMakarov2013}, they can be written as
\begin{linenomath*}
 \ba
 {K}_{l}(\omega)
 \;\equiv\;-\;
 \Im
 \left[\,\bar{k}_{\it{l}}(\omega)\,
 \right]\,=\,
 \frac{{k}_{l}(\omega)}{Q_{l}(\omega)}\;\,\mbox{Sign}{\,\omega}
 ~~,
 \label{ggffrr}
 \ea
 \end{linenomath*}
 \es
 where the tidal quality factor is introduced via
 \begin{linenomath*}
 \ba
 Q_l^{-1}(\omega)\,=\,|\s\sin\epsilon_{l}(\omega) \s|\;\,.
 \ea
 \end{linenomath*}

  The dependency $\sin\epsilon_l(\omega)$ being odd, the function $Q_l(\omega)$ is even. Also, even is the function $k_l(\omega)$. Therefore, for any sign of $\omega$ and $\epsilon_l$, it is always possible to treat both  $Q_l(\omega)$ and $k_l(\omega)$ as functions of the forcing frequency $\chi\equiv|\omega|\;$:
  \begin{linenomath*}
  \ba
  Q_l(\omega)\,=\,Q_l(\chi)~~,\quad k_l(\omega)\,=\,k_l(\chi)~~~.
  \ea
  \end{linenomath*}

 Often attributed to \citeA{biot}, though known yet to Sir George \citeA{darwin}, the so-called
  $\s${\it{correspondence principle}}$\s$, or the $\s${\it{elastic-viscoelastic analogy}}, is a valuable key to numerous problems of viscoelasticity. It enables one to derive solutions to these problems from the known solutions to analogous static problems. In application to bodily tides, this principle says that
  the complex Love number of a uniform spherical viscoelastic body, $\,\bar{k}_{l}(\chi)\,$, is linked to the complex compliance $\,\bar{J}(\chi)\,$ by the same algebraic expression through which the static Love number $\,{k}_{l}\,$ of that body is linked to the relaxed compliance $\,{J}_{\rm{R}}\,$:
  \begin{linenomath*}
  \ba
 \bar{k}_{l}(\chi)
 ~=~\frac{3}{2\,(l\,-\,1)}\;\,\frac{\textstyle 1}{\textstyle 1\;+\;{\cal{B}}_{l}/\bar{J}(\chi)}
 ~~~,~\quad~
  \label{k2bar}
 \ea
 \end{linenomath*}
 where
 \begin{linenomath*}
 \ba
 {\cal{B}}_{l}\,\equiv~\frac{\textstyle{(2\,{\it{l}}^{\,2}\,+\,4\,{\it{l}}\,+\,3)}}{\textstyle{{\it{l}}\,\mbox{g}\,
 \rho\,R}}~=\;\frac{\textstyle{3\;(2\,{\it{l}}^{\,2}\,+\,4\,{\it{l}}\,+\,3)}}{\textstyle{4\;{\it{l}}\,\pi\,
 G\,\rho^2\,R^2}}~~~,\quad
   \label{B}
   \ea
   \end{linenomath*}
   $\rho$, $R$, and $g$ being the density, radius, and surface gravity of the body, and $G$ being  Newton's gravitational constant.

 As an aside, we would mention that while $\,-\Im\left[k_l(\omega)\right]$ emerges in the tidal torque, the real part of the complex Love number, $\s\Re\left[k_l(\omega)\right] = k_l(\omega)\,\cos \epsilon_l(\omega)\s$, shows up in the expansion for the tidal potential. Not considered further in the present study, the general expression for this product and its version for the Maxwell and other rheologies can be found in \citeA[Appendix A6]{Efroimsky2015}.

 \subsection{$k_l(\chi)/Q_l(\chi)$ and $1/Q_l(\chi)$ for an Arbitrary Rheology}

 Expression (\ref{k2bar}) entails: 
 \begin{linenomath*}
 \ba
 {K}_{{l}}(\chi)\;=\;
 {k}_{{l}}(\chi)\;\sin\epsilon_{l}(\chi)~=~-~\frac{3}{2(l-1)}~\frac{{\cal{B}}_{\textstyle{_l}}\;\Im\left[\bar{J}(\chi)\right]
 }{\left(\Re\left[\bar{J}(\chi)\right]+{\cal{B}}_{\textstyle{_l}}\right)^2+\left(
 \Im
 \left[\bar{J}(\chi)\right]\right)^2}~~~,
 \label{E10a}
 \label{22}
 \ea
 \end{linenomath*}
 the coefficients $\,{\cal{B}}_l\,$ rendered by equation (\ref{B}). We see that for a homogeneous incompressible sphere, the information needed to calculate the quality function comprises the radius, the density, and the rheological law $\s\bar{J}(\chi)\,$.

 The inverse tidal quality factor of degree $\s l\s$ is given by \cite{Efroimsky2015}
 \begin{linenomath*}
 \ba
 Q_l(\chi)^{-1}\equiv\;|\s\sin\epsilon_l(\chi)\s| \;\;,
 \ea
 \end{linenomath*}
 \begin{linenomath*}
 \ba
 \label{23}
 \sin\epsilon_l(\chi)\,=\,-\;\frac{
 {\cal{B}}_{\textstyle{_l}}\;\;
 \Im
 \left[\bar{J}(\chi)\right]
 }{
 \sqrt{ \left(\,
  \Re
 \left[\bar{J}(\chi)\right]\,\right)^2+\,\left(\,
  \Im
 \left[\bar{J}(\chi)\right]\,\right)^2\,}~
 \sqrt{\left(\,
  \Re
 \left[\bar{J}(\chi)\right]+{\cal{B}}_{\textstyle{_l}}\right)^2+\,\left(
  \Im
 \left[\bar{J}(\chi)\right]\right)^2}\,
 }\quad.\quad
 \ea
 \end{linenomath*}
 
   All new is well-forgotten old. As we were writing this paper, it became known to us that for the Maxwell rheology, the frequency-dependence of $\sin\epsilon_2$ was studied yet by \citeA{Gerstenkorn} in a work that went virtually unnoticed. Because of different notation and Gerstenkorn's terse style, it is not apparent if his values for the peak's magnitude and location are the same as ours. However, the overall shape of the dependence  $\sin\epsilon_2(\chi)$ obtained by \citeA[Fig. 2]{Gerstenkorn} seems right.

\subsection{Notational Point: $Q$ and $Q_2$}

In publications where both seismic and tidal dissipation are considered, it is necessary to distinguish between the seismic and tidal quality factors. In that situation, the letter $Q$ without a subscript is preserved for the seismic factor.

In the literature on tides, it is common to employ $Q$ as a shorter notation for the quadrupole tidal factor $Q_2$. We shall follow the latter convention:
\begin{linenomath*}
\ba
Q\,\equiv\,Q_2\;\,,
\ea
\end{linenomath*}
and shall use the two notations intermittently.

 \subsection{The frequency-dependencies of $\;k_l/Q_l\,$ and $\,1/Q_l\,$\\
 for the Maxwell and Andrade models}

\begin{figure}[htbp]
 \vspace{2.5mm}
 \centering
 \includegraphics[angle=0,width=0.68\textwidth]{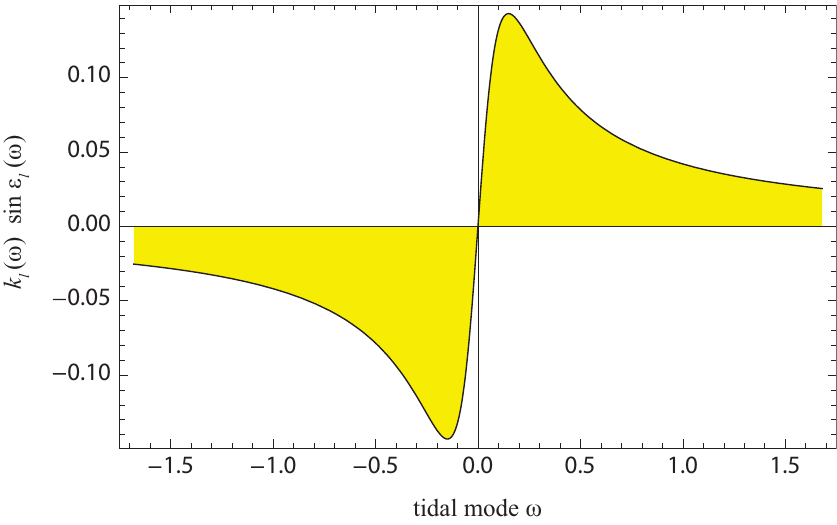}
 \caption{\small{~A typical shape of the quality function $\,K_{\textstyle{_l}}(\omega)\s=\s k_l(\omega)\,\sin\epsilon_l(\omega)\,$, ~where $\,\omega\,$ is a shortened notation for the tidal Fourier mode
 $\,\omega_{\textstyle{_{lmpq}}}\,$. \cite<From>{2014Icar..241...26N}.
 \label{Fig1}}}
 \end{figure}

 For a homogeneous sphere composed of a Maxwell or Andrade material, the quality function $K_{\textstyle{_l}}(\omega)$
 has a kink form, as in Figure \ref{Fig1}. The function $\s\sin\epsilon_{\textstyle{_l}}(\omega)\s$
 is shaped similarly.

 Insertion of expression (\ref{E8}) into equation (\ref{22}) shows that for a spherical Maxwell body the extrema of the kink  $\,K_l(\omega)\,$ are located at
 \begin{linenomath*}
  \ba
 {\omega_{\rm{peak}}}_{\textstyle{_l}}\,=\;\pm\;
 \;\frac{\tau_{\rm{_M}}^{-1}}{1\,+\,{\cal{B}}_l\,\mu}\;
 \label{wh}
 \ea
 \end{linenomath*}
 the corresponding extrema assuming the values
 \begin{linenomath*}
 \ba
 K_l^{\rm{(peak)}}\s=\;\pm\;\frac{3}{4\s(l-1)}\;\frac{ {\cal B}_l\,\mu }{ 1 + {\cal B}_l\,\mu }\,\;,
 \label{see}
 \ea
 \end{linenomath*}
 wherefrom $|K_l|\s<\s\frac{\textstyle 3}{\textstyle 4\s(l-1)}\,$.
 
 Inside the interval between peaks, the quality functions are near-linear in $\omega\,$: 
 \begin{linenomath*}
 \ba
 \label{444}
 |\s\omega\s|\;<\;|\s{\omega_{\rm{peak}}}_l\s|\quad\Longrightarrow\quad K_l(\omega)\;
 \;\simeq\;\frac{3}{2(l-1)}\;\frac{{\cal B}_l\,\mu}{1\,+\,{\cal B}_l\,\mu}
 \;\frac{\omega}{|\s{\omega_{\rm{peak}}}_l\s|}\,\;.
 \ea
 \end{linenomath*}
 Outside the inter-peak interval, they fall off as about $\omega^{-1}\,$:
 \begin{linenomath*}
 \ba
 |\s\omega\s|\;>\;|\s{\omega_{\rm{peak}}}_l\s|\quad\Longrightarrow\quad
 K_l(\omega)
 \;\simeq\;\frac{3}{2(l-1)}\;\frac{{\cal B}_l\,\mu}{1\,+\,{\cal B}_l\,\mu}\;\frac{|\s{\omega_{\rm{peak}}}_l\s|}{\omega}\;\;.
 \label{do}
 \label{555}
 \ea
 \end{linenomath*}
 
 While the peak magnitudes (\ref{see}) are ignorant of the viscosity $\eta$, the spread between the peaks scales as the inverse $\s\eta\s$, as evident from expression (\ref{wh}). The lower the mean viscosity, the higher the peak frequency $|{\omega_{\rm{peak}}}_l|$.

 It can be demonstrated using equation (\ref{23}) that for a homogeneous Maxwell body
 the extrema of  $\s \sin\epsilon_l(\omega)\s$ are located at
 \begin{linenomath*}
 \ba
 \omega_{\rm{peak~of}~\sin\epsilon_l}\,=\;\pm\;\frac{\tau_{\rm{_M}}^{-1}}{\sqrt{1+{\cal B}_l\mu}}
 \,\;.
 \label{peakofQ}
 \ea
 \end{linenomath*}
 For the Moon, this peak is located within a decade from its counterpart for $\s K_l\s$ given by formula (\ref{wh}).

 In many practical situations, only the quadrupole ($l=2$) terms matter. The corresponding peaks are located at 
 \begin{linenomath*}
  \ba
 {\omega_{\rm{peak}}}_2\,=\;
 \pm\;\;\frac{\tau_{\rm{_M}}^{-1}}{1\,+\,{\cal{B}}_2\,\mu}\,\approx\;
 \pm\;\frac{1}{{\cal{B}}_{2}\;\eta}\;=\;\pm\;\frac{\textstyle{8\,\pi\,
 G\,\rho^2\,R^2}}{\textstyle{57\;\eta}}\quad.  
 \label{peak}
 \ea
 \end{linenomath*}
 The approximation in this expression relies on the inequality $\s{\cal B}_l\s\mu\gg 1\s$, fulfillment whereof depends on the size of the body. For a Maxwell Moon with $\,\mu= 6.4\times 10^{10}$~Pa \,and $\,G(\rho R)^2\approx 2.24\times 10^9\,$~Pa, we have  $\s{\cal B}_2\s\mu\s\approx 64.5$, so the approximation works.

 While for the Maxwell and Andrade models each of the functions $K_l(\omega)$ and $\sin\epsilon_l(\omega)$ possesses only one peak for a positive argument, the situation changes for bodies of a more complex rheology. For example, the existence of an additional peak is ensured by the insertion of the Sundberg-Cooper compliance (\ref{15}) into expressions (\ref{22}) or (\ref{23}).
 
  \section{Application to the Moon} \label{sec:moon}

 \subsection{The ``Wrong'' Slope Interpreted with the Maxwell Model}

 As we explained in Section \ref{motivation}, fitting of the LLR-obtained quadrupole tidal quality factor $\s Q=Q_2\s$ to the power law $Q\sim\chi^p$ resulted in a small negative value of the exponential $p$ \cite{williams2015}. An earlier attempt to explain this phenomenon implied an identification of this slightly negative slope with the incline located to the left of the maximum of the quality function $(k_2/Q_2)(\chi)$, see Figure \ref{Fig1}. Within this interpretation, $\s\chi_{\rm{peak}}\equiv|\omega_{\rm{peak}}|\s$ should be residing somewhere between the monthly and annual frequencies explored in \citeA{williams2015}. As was explained in \citeA{Efroimsky2012a}, this sets the mean viscosity of the Moon as low as
 \begin{linenomath*}
 \ba
 \eta\,\approx~3\,\times\,10^{15}~\mbox{Pa~s}~~,
 \ea
 \end{linenomath*}
 
 The extrema of $(1/Q_2)(\chi)$ are close to those of $(k_2/Q_2)(\chi)$, as can be observed from equations (19) and (45) of \citeA{Efroimsky2015}. Therefore, had we used instead of the maximum of $k_2/Q_2$ given by (\ref{peak}) the maximum of $1/Q_2$ given by (\ref{peakofQ}), the ensuing value would have been only an order higher:
 \begin{linenomath*}
 \ba
 \eta\,\approx~4\,\times\,10^{16}~\mbox{Pa~s}~~.
 \ea
 \end{linenomath*}
 Such values imply a high concentration of the partial melt in the mantle -- quite in accordance with the seismological models by \citeA{Nakamura} and \citeA{Weber}.

 However, employment of a rheology more realistic than Maxwell may entail not so low a viscosity~--- in which case the existence of a semi-molten layer may be questioned.
 
 \subsection{Frequency Dependence of Tidal Dissipation in the Sundberg-Cooper Model}

 The Debye peak emerging in the imaginary part of $\,\bar{J}_{\rm{e}}\,$ (equation (\ref{14})) will, obviously, show itself also in the shape of the imaginary part of the overall $\,\bar{J}\,$, the bottom line of equation (\ref{15b}). Consequently, substitution of expression (\ref{15}) in equations (\ref{22}) and (\ref{23}) will entail the emergence of a Debye warp on the kinks for $\,k_l/Q_l\,$ and $\,1/Q_l\,$. Where will the additional peak be located for realistic values of the relaxation timescale $\,\tau\,$? What values for the mean viscosity will it entail?

\begin{figure}[htbp]
     \centering
     \includegraphics[width=\textwidth]{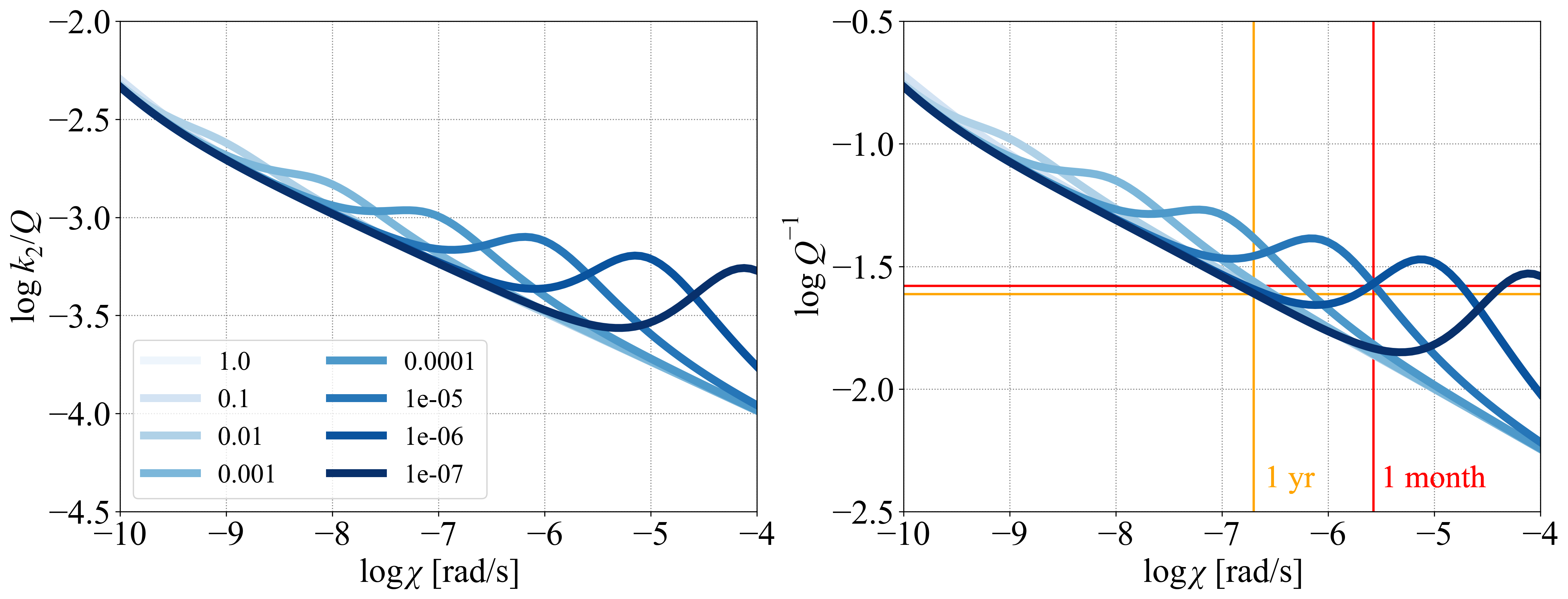}
     \caption{The negative imaginary part of the Love number (left) and the inverse quality factor (right) for different ratios between the timescale $\tau$ and the Maxwell time $\tau_{\rm{_M}}$ (indicated by the shades of blue). The yellow and red vertical lines show the $Q_2$ values given by \citeA{williams2015} for the annual and the monthly component, respectively. In this case, we consider a homogeneous lunar interior model governed by the Sundberg-Cooper rheology. The mantle viscosity was set to $\unit[10^{22}]{Pa\;s}$ and the mantle rigidity to $\unit[80]{GPa}$.}
     \label{fig:sc_where}
 \end{figure}

 In the end of Section \ref{Sund}, we introduced the relative relaxation time as $\s t_{\rm{rel}} \equiv {\tau}/{\tau_{\rm{_M}}}\s$.
 Figure \ref{fig:sc_where} illustrates specifically the effect of $t_{\rm{rel}}$ in the Sundberg-Cooper model on the position of the additional Debye peak for a homogeneous lunar interior with an arbitrarily chosen high mean viscosity $\eta_{\rm{Moon}}=\unit[10^{22}]{Pa\; s}$. The emergence of another local maximum in the $k_2/Q_2$ and $1/Q_2$ functions may naturally explain the increase in dissipation (or decrease in the quality factor $Q$) with frequency, even within a homogeneous and highly viscous model.

\begin{figure}[htbp]
    \centering
    \includegraphics[width=\textwidth]{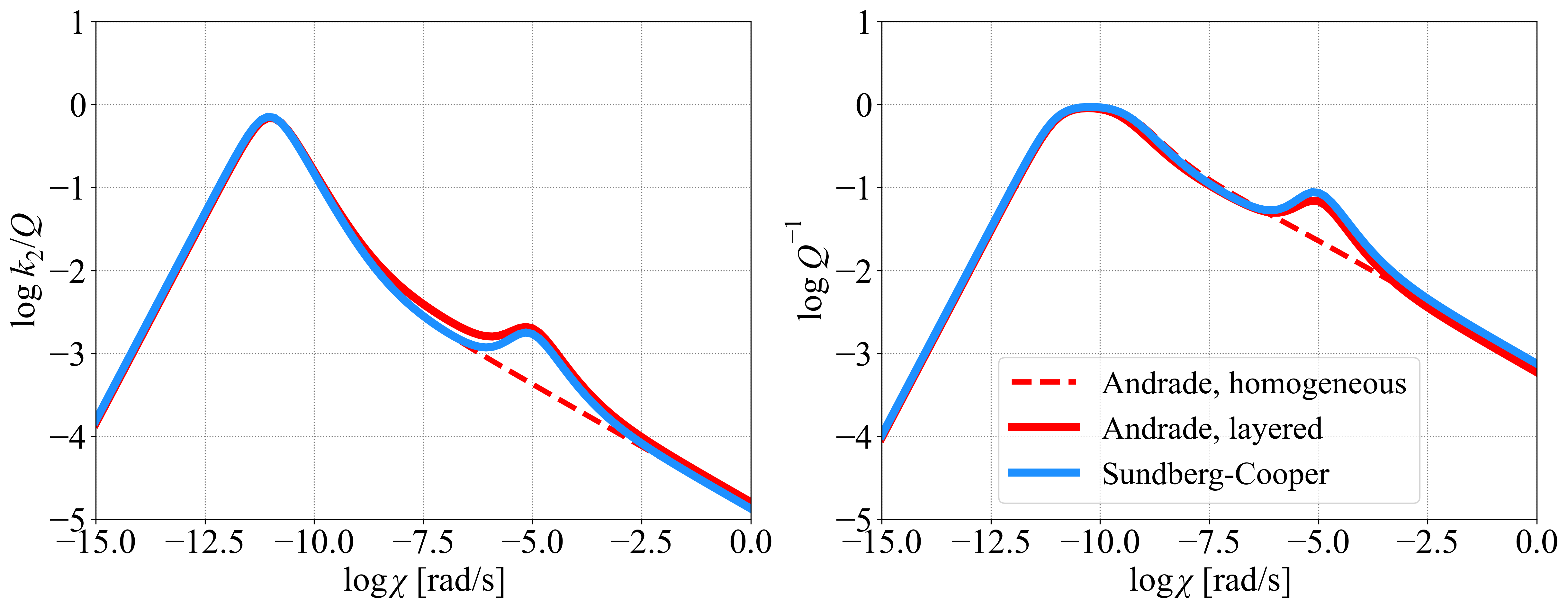}
    \caption{The tidal quality function (left) and inverse quality factor (right) for three model cases: a homogeneous Andrade model (dashed red line), a homogeneous Sundberg-Cooper model (blue line), and a three-layered model (solid red line)  comprising a core, an Andrade mantle and a Maxwell semi-molten layer at the base of the mantle.
        }    
    \label{fig:sc_vs_andrade}
\end{figure}

\begin{figure}[htbp]
    \centering
    \includegraphics[width=\textwidth]{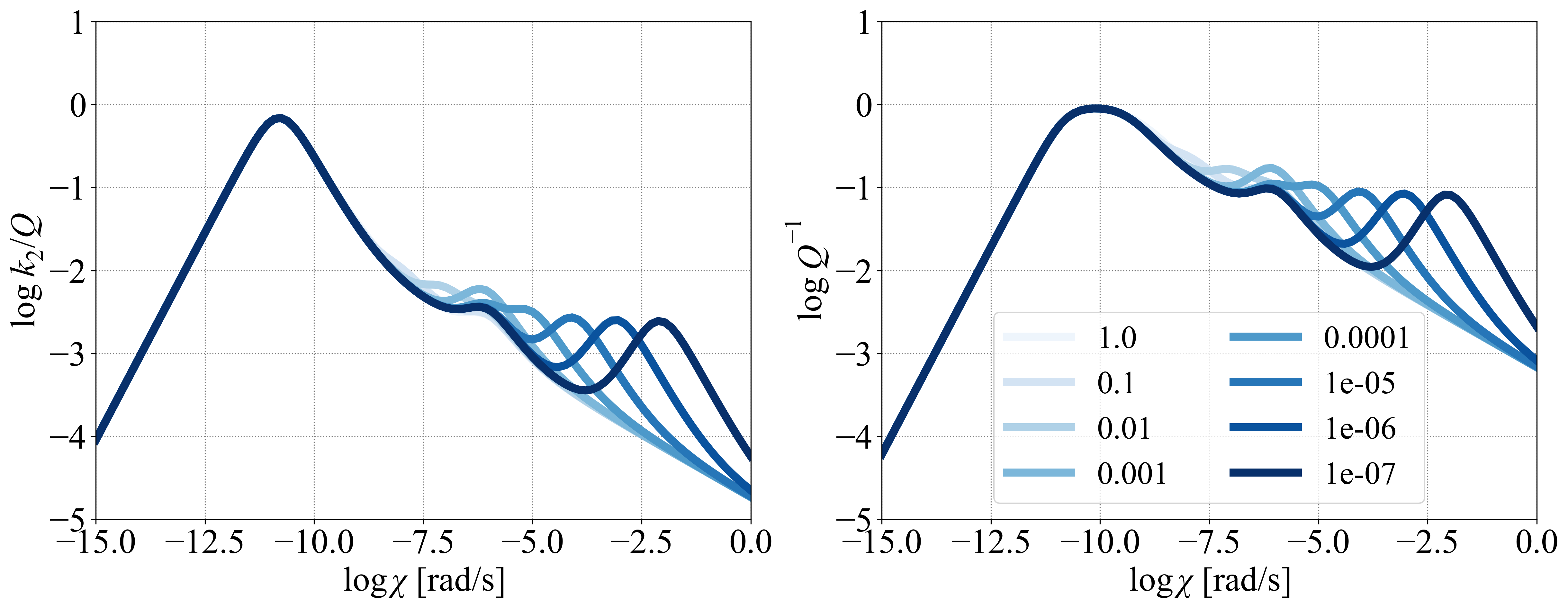}
    \caption{The tidal quality function (left) and inverse quality factor (right) of a three-layered lunar model comprising a core, a Sundberg-Cooper mantle, and a Maxwell semi-molten basal layer. Different shades of blue correspond to different ratios between the timescale $\tau$ and the Maxwell time $\tau_{\rm{_M}}$. For illustrative purposes, the semi-molten basal layer is made unrealistically thick ($\unit[500]{km}$).}
    \label{fig:sc_layered}
\end{figure}

  Furthermore, as was recently shown by \citeA{gevorgyan2021}, the tidal response of a homogeneous Sundberg-Cooper planet mimics the response of a body consisting of two Andrade layers with different relaxation times. This kind of aliasing may, in principle, be also demonstrated by the Moon. Figure \ref{fig:sc_vs_andrade} depicts the tidal quality function $k_2/Q_2$ and the inverse quality factor $1/Q_2$ as functions of frequency, for a homogeneous Sundberg-Cooper moon and for a differentiated lunar interior with a rheologically weak layer at the base of the mantle. In the second case, the basal layer is described by the Maxwell model and the overlying mantle by the Andrade model. Both cases follow the same frequency dependence, implying that the existence of a weak basal layer cannot be confirmed unequivocally by the tidal data. In a layered model containing a core, a Sundberg-Cooper mantle, and a Maxwell basal semi-molten layer, the tidal response would be characterised by three peaks (Figure \ref{fig:sc_layered}).\\

 \subsection{Constructing a Multi-layered Model}

Section \ref{sec:love} introduced the complex Love number $\bar{k}_l(\chi)$ for an arbitrary linear anelastic or viscoelastic rheology assuming a homogeneous incompressible sphere. While such a model can reasonably approximate the response of the Moon with a homogeneous mantle and a small core, its application to a body with a highly dissipative basal layer would not be accurate \cite{bolmont2020}. A planetary interior with a highly dissipative layer can still be approximated by a homogeneous model with an additional absorption peak or band in the underlying rheological law. However, we would need to know the mapping between the parameters of the dissipative layer and the parameters of the additional peak \cite{gevorgyan2021,gevorgyan2023}.

Therefore, in the following sections, we will replace the homogeneous model with three models consisting of three or four layers and we will calculate the corresponding complex Love numbers numerically, using a matrix method based on the normal mode theory \cite<e.g.,>[]{takeuchi1972,wu1982,sabadini2004}. For the sake of simplicity, we consider all layers in the numerical model (linearly) viscoelastic and we mimic the response of liquid layers by the Maxwell model with $J_{\rm{e}}$ in equation (\ref{E8}) approaching $0$. This method has also been tested against another implementation of the same model, in which the liquid layers were inputted through different boundary conditions; the results obtained within the two approaches are virtually the same. Using the output complex Love numbers for various rheological parameters, we then proceed by fitting the empirical values.

If not stated differently for illustrative purposes, the three alternative models will always comprise an elastic crust of constant density ($\rho_{\rm{cr}}=\unit[2550]{kg\;m^{-3}}$) and thickness ($D_{\rm{cr}}=\unit[40]{km}$), consistent with the gravity and topography data \cite{wieczorek2013}, and a liquid core with a low viscosity ($\eta_{\rm{c}}=\unit[1]{Pa\;s}$). Although the existence of an inner core is possible and even indicated by the stacked seismograms presented by \citeA{Weber}, its response to tidal loading would be decoupled from the rest of the mantle, and it would contribute to the resulting tidal deformation only negligibly. Therefore, the inner core is not included in our modelling. We note that the recent study of \citeA{briaud2023nat} shows that an inner core might be required even by tidal and mineralogical data. However, their model uses a different rheological model of the mantle and also predicts much higher outer core viscosity than considered in our work.

Subsection \ref{sub:uncertainty} makes use of a three-layered model (Model 1) consisting of the liquid core, a homogeneous mantle described by the Andrade rheology, and the elastic crust. The density and radius of the liquid core, as well as the density, rigidity, viscosity, and the Andrade parameters of the mantle, are treated as free parameters and fitted to the data.

The second model (Model 2), considered in Subsection \ref{subsec:sc_inversion}, is essentially similar to the previous one except that its mantle is governed by the Sundberg-Cooper rheological model. In addition to the previous set of parameters, we now also seek the values of the relaxation strength $\Delta$ and the relative relaxation time $t_{\rm{rel}}$. 

Finally, the model with a basal dissipative layer (Model 3), which is discussed in Subsection \ref{subsec:melt_inversion}, contains a core, an elastic crust, and a two-layered mantle. Each layer of the mantle is assumed to be homogeneous. The basal layer is described by the Maxwell model with fitted rigidity $\mu_{\rm{LVZ}}$, viscosity $\eta_{\rm{LVZ}}$, and density $\rho_{\rm{LVZ}}$; additionally, we fit its outer radius $R_{\rm{LVZ}}$. For the overlying bulk mantle, we consider the Andrade model with fitted viscosity $\eta_{\rm{m}}$, rigidity $\mu_{\rm{m}}$, density $\rho_{\rm{m}}$, and the Andrade parameters $\alpha$, $\zeta$. The reason for using the simple Maxwell model instead of the Andrade model in the basal layer is the following: in order to fit the measured tidal quality factor $Q$ at the monthly and the annual frequency, the peak dissipation from the basal layer should be located either between these frequencies or above the monthly frequency. At the same time, in the vicinity of the peak dissipation, the Andrade and Maxwell rheologies are almost indistinguishable from each other.
(Comparing the last two terms on the final line of equation (\ref{15}), we observe that the viscous term exceeds the Andrade term when $\tau_{\rm{_M}}\chi \ll \left(\tau_{\rm{_A}}/\tau_{\rm{_M}}\right)^{\alpha/(1-\alpha)}\s$. In realistic situations, $\tau_{\rm{_M}}\chi_{\rm{peak}}$ satisfies this condition safely. So, near the peak the Andrade term is virtually irrelevant, and the regime is almost Maxwell.)
Hence, we chose the simpler of the two rheological models. This decision will also facilitate the comparison of our results for the basal layer's characteristics with the predictions by \citeA{harada2014,harada2016}, and \citeA{matsumoto2015}, who likewise modeled the basal layer with the Maxwell rheology. In contrast to our study, they applied the same model to the mantle as well.

\subsection{Explored parameter ranges}

The three alternative models considered consist of a small number of homogeneous interior layers. In this work, we are not predicting the mineralogy of the mantle~---~and the composition of the basal layer, if present, is only briefly discussed in Subsection \ref{subsec:disc_layer}. Our use of a homogeneous mantle layer (or two homogeneous mantle layers) reflects our lack of information on the exact chemical and mineralogical composition, the grain size, the thermal structure, and the presence of water. Instead, we characterise the mantle with a single, ``effective", rigidity and viscosity, which can be later mapped to a detailed interior structure \cite<see also>[who discussed the effect of approximating a radially stratified mantle with a homogeneous one for Venus and terrestrial exoplanets]{dumoulin2017,bolmont2020}. Furthermore, we neglect any lateral heterogeneities in the lunar interior. We also assume that the lunar mantle is incompressible and can be reasonably described by a linear viscoelastic model~---~which is valid at low stresses. Given the magnitude of tidal stresses in the Moon, this assumption might have to be lifted in future works, though \cite{karato2013}.

For the effective mantle viscosity, we consider values ranging from $\unit[10^{15}]{Pa\; s}$ up to $\unit[10^{30}]{Pa\;s}$. The effective viscosity of the basal layer in Model 3 is varied between $\unit[1]{Pa\; s}$ and $\unit[10^{30}]{Pa\; s}$. Lunar mantle rigidity is linked to the speed of S-waves in the medium, which has been constrained by lunar seismic experiments. Assuming that the effective tidal rigidity is not too different from the seismologically-determined values, we only vary the effective mantle rigidity in a tight range from $60$ to $\unit[90]{GPa}$, consistent with the seismic wave velocities reported in the VPREMOON model of \citeA{garcia2011}. For the basal layer in Model 3, we require that $\mu_{\rm{LVZ}}$ be always smaller than $\mu_{\rm{m}}$ and greater than $\unit[0]{GPa}$. While the viscosities are varied on the logarithmic scale, the rigidities are only varied on the linear scale.

The core size and core density in our study are mainly constrained by the mean lunar density and the moment of inertia. We adopt a range of values consistent with previous works, following Table 1 of \citeA{garcia2019}. For the core size, we assume $R_{\rm{c}}\in[0, 450] \unit{km}$ and for the core density $\rho_{\rm{c}}\in[4000, 7000] \unit{kg\; m^{-3}}$. The mantle density is varied in the range from $3000$ to $\unit[4000]{kg\; m^{-3}}$.

An essential ingredient of the complex rheological models used in this study are the parameters $\alpha$, $\zeta$, $\Delta$, and $\tau$ (or $t_{\rm{rel}}$). These parameters are only weakly constrained by laboratory measurements or geodetic and seismological observations. Therefore, we explore a wide range of their values. The Andrade parameter $\alpha$, which characterises the time dependence of transient creep in a medium \cite{andrade1910}, typically lies in the interval $0.2-0.4$, although values outside this range have also been observed \cite{kennedy1953,jackson2010,castillo2011,Efroimsky2012a}. Geodetic measurements performed on the Earth favour a narrower interval of $0.14 - 0.2$, and the currently accepted model of tides in the solid Earth, presented in the IERS Conventions on Earth Rotation, employs the value of $\alpha = 0.15$ \cite[eqn 6.12 and a paragraph thereafter]{petit2010}. Here, we consider an interval of $0-0.5$ for the simplest model with a homogeneous Andrade mantle (Model 1) and a more realistic interval of $0.1-0.5$ for the other two models.

The mean value of the dimensionless Andrade time $\zeta$ was found to be close to unity in polycrystalline olivine under laboratory conditions \cite{castillo2011}. However, the individual fits to laboratory data obtained with olivine, periclase, and olivine-pyroxene mixtures also allow values few orders of magnitude smaller or greater \cite<e.g.,>{tan2001,jackson2002,barnhoorn2016,qu2021}. To account for our lack of knowledge, we consider $\log\zeta\in[-5,5]$. The relaxation time of elastically-accommodated GBS, required by Model 2 and given by equation (\ref{eq:tau_gbs}), is linked to the relative thickness of grain boundaries with respect to the grain size, the material's rigidity, and the grain-boundary viscosity. Both the relative grain-boundary thickness and the grain-boundary viscosity are largely unknown. The relative relaxation time, $t_{\rm{rel}}$, can be expressed as

\begin{linenomath*}
\begin{equation}
    t_{\rm{rel}} = \frac{\eta_{\rm{gb}}}{\eta_{\rm{m}}} \frac{d}{\delta}\; ,
\end{equation}\\
\end{linenomath*}
\noindent
where $\eta_{\rm{gb}}$ is expected to be much smaller than $\eta_{\rm{m}}$. \citeA{jackson2014} derives grain-boundary viscosities between $10^{5}$ and $\unit[10^{8}]{Pa\;s}$ for pure olivine at different temperatures and mentions values around $\unit[1-100]{Pa\;s}$ for a grain boundary filled with basaltic melt \cite{murase1973}. Grain boundary thicknesses typically correspond to a few atomic layers and studies of polycrystalline olivine report values around $\unit[1]{nm}$ \cite{marquardt2018}. Grain sizes can span from $\sim\unit[1]{\mu m}$ to $\sim\unit[1]{cm}$. Having these ranges in mind, we see that the relative relaxation time can only be constrained as $t_{\rm{rel}}\ll1$, as is also mentioned in both experimental and theoretical studies \cite<e.g.>{morris2009,lee2011,jackson2014}. Here, we adopt a similar range as was used by \citeA{morris2009} and set $t_{\rm{rel}}\in[10^{-10},1]$.

Finally, the relaxation strength of the elastically-accommodated GBS is reported by \citeA{Sundberg} to be in the range between $\approx0.2$ and $1.91$, following different assumptions on the grain shapes and different modelling approaches. To allow for a slightly wider range of values, we let the parameter $\Delta$ vary on a logarithmic scale between $10^{-2}$ and $10$.

In the inversions presented below, we are fitting the three alternative models of the lunar interior to the total mass of the Moon, the moment of inertia factor (MoIF), and the tidal parameters, namely $k_2$ and tidal $Q$ at the monthly frequency, $k_2/Q$ at the annual frequency, and $k_3$, $h_2$ at the monthly frequency. For the samples consistent with the geodetic constraints, we also estimate the seismic $Q$ of the mantle and compare it with seismological literature \cite{nakamura1982,gillet2017,garcia2019}, although this additional constraint is not used to reject models. A list of the model parameters of the three models discussed in the following sections is presented in Table \ref{tab:model_params}. The empirical values considered are then given in Table~\ref{tab:observables}.

\begin{table}[htbp]
    \centering
        
    \begin{tabular}{l l l l}
    \hline
    Parameter & Type & Value & Unit \\
    \hline
    \multicolumn{4}{c}{Common parameters} \\
    \hline
    Crustal thickness $D_{\rm{cr}}$ & const. & $40$ & $\unit{km}$ \\
    Crustal density $\rho_{\rm{cr}}$ & const. & $2,550$ & $\unit{kg\;m^{-3}}$ \\
    Core size $R_{\rm{c}}$ & fitted & $0-450$ & $\unit{km}$\\
    Core viscosity $\eta_{\rm{c}}$ & const. & $1$ & $\unit{Pa\;s}$ \\
    Core density $\rho_{\rm{c}}$ & fitted & $4,000-7,000$ & $\unit{kg\;m^{-3}}$ \\
    Mantle viscosity $\eta_{\rm{m}}$  & fitted & $10^{15}-10^{30}$ & $\unit{Pa\;s}$ \\
    Mantle rigidity $\mu_{\rm{m}}$    & fitted & $60-90$  & $\unit{GPa}$\\
    Mantle density $\rho_{\rm{m}}$ & fitted & $3,000-4,000$ & $\unit{kg\;m^{-3}}$ \\
    Andrade parameter $\zeta$ & fitted & $10^{-5}-10^{5}$ & ---\\
    \hline
    \multicolumn{4}{c}{\textbf{Model 1} (Andrade mantle)} \\
    \hline
    Andrade parameter $\alpha$ & fitted & $0-0.5$ & ---\\
    \hline
    \multicolumn{4}{c}{\textbf{Model 2} (Sundberg-Cooper mantle)} \\
    \hline
    Andrade parameter $\alpha$ & fitted & $0.1-0.5$ & ---\\
    Relaxation strength $\Delta$ & fitted & $10^{-2}-10^{1}$ & ---\\
    Relative relaxation time $t_{\rm{rel}}$ & fitted & $10^{-10}-10^{0}$ & ---\\
    \hline
    \multicolumn{4}{c}{\textbf{Model 3} (Andrade mantle + basal layer)} \\
    \hline
    Andrade parameter $\alpha$ & fitted & $0.1-0.5$ & ---\\
    Upper radius of the basal layer $R_{\rm{LVZ}}$ & fitted & $R_{\rm{c}}-700$ & $\unit{km}$\\
    Viscosity of the basal layer $\eta_{\rm{LVZ}}$ & fitted & $10^{0}-10^{30}$ & $\unit{Pa\;s}$ \\
    Rigidity of the basal layer $\mu_{\rm{LVZ}}$ & fitted & $0-\mu_{\rm{m}}$ & $\unit{Pa}$ \\
    Density of the basal layer $\rho_{\rm{LVZ}}$ & fitted & $\rho_{\rm{m}}-\rho_{\rm{c}}$ & $\unit{kg\;m^{-3}}$ \\
    \hline
    \end{tabular}
    \caption{Parameters of the three models considered in this work.}
    \label{tab:model_params}    
\end{table}

\begin{table}[htbp]
    \centering
    \begin{tabular}{l l l}
    \hline
    Parameter & Value & Reference \\
    \hline
    MoIF & $0.392728\pm0.000012$ & \citeA{williams2014}\\
    $M$ & $\unit[(7.34630\pm0.00088)\times10^{22}]{kg}$ & \citeA{williams2014}\\
    $k_2$, monthly & $0.02422\pm0.00022$ & \citeA{williams2014}\\
    $Q$, monthly & $38\pm4$ & \citeA{williams2015}\\
    $k_2/Q$, annual & $(6.2\pm1.4)\times10^{-4}$ & \citeA{williams2015}\\
    $k_3$, monthly$^{\rm{a}}$ & $0.0081\pm0.0018$ & \citeA{konopliv2013,lemoine2013}\\
    $h_2$, monthly & $0.0387\pm0.0025$ & \citeA{thor2021}\\
    \hline
    \end{tabular}\\[1em]
    \footnotesize
    $^{\rm{a}}$ Listed is the unweighted mean of the values given in references.

    \caption{Observational constraints used in this work.}
    \label{tab:observables}
    
\end{table}

\subsection{Applicability of the Andrade Model} \label{sub:uncertainty}

Before discussing the two complex interior models able to fit the anomalous frequency dependence of lunar tidal dissipation, we first attempt to use the set of parameters given in Table \ref{tab:observables} to constrain a simpler model, which only contains a liquid core and a viscoelastic mantle governed by the Andrade rheology (equation (\ref{E3c})). Such a model, accounting neither for a basal dissipative layer nor for elastically-accommodated GBS, might still be able to fit the data. Thanks to the large uncertainty on the lunar quality factor \cite<more than $10\%$ at the monthly frequency and $20\%$ at the annual frequency,>{williams2015}, we may not need to introduce any additional complexities to interpret the tidal response of the Moon. The error bars of the tidal quality factors are so wide that they allow, at least in principle, for a situation where $Q_{2,\,\rm{annual}}$ is smaller than $Q_{2,\,\rm{monthly}}\s$.

To find the parameters of this preliminary model, we performed a Bayesian inversion using the MCMC approach and assuming Gaussian distributions of observational uncertainties \cite<e.g.,>[]{mosegaard1995}. In particular, we employed the \textit{emcee} library for \textit{Python} \cite{foreman2013}, which is based on the sampling methods proposed by \citeA{goodman2010}. The algorithm was instructed to look for the mantle viscosity $\eta_{\rm{m}}$, the mantle rigidity $\mu_{\rm{m}}$, the core and mantle densities, and the Andrade parameters fitting the empirical values of $k_{2, \rm{monthly}}$, $k_{3, \rm{monthly}}$, $h_{2, \rm{monthly}}$, $Q_{2, \rm{monthly}}$, and $(k_2/Q_2)_{\rm{annual}}$, MoIF, and total mass $M$. We generated $232,000$ random samples until the model converged. Specifically, the convergence was tested against the autocorrelation time of each variable in the ensemble, the total length of all chains being required to exceed $100$ times the longest autocorrelation time. In order to filter out the influence of initial conditions, we neglected the first $4,640$ samples (our burn-in period was, therefore, twice the autocorrelation time).\\

\begin{figure}[htbp]
\includegraphics[width=0.8\textwidth]{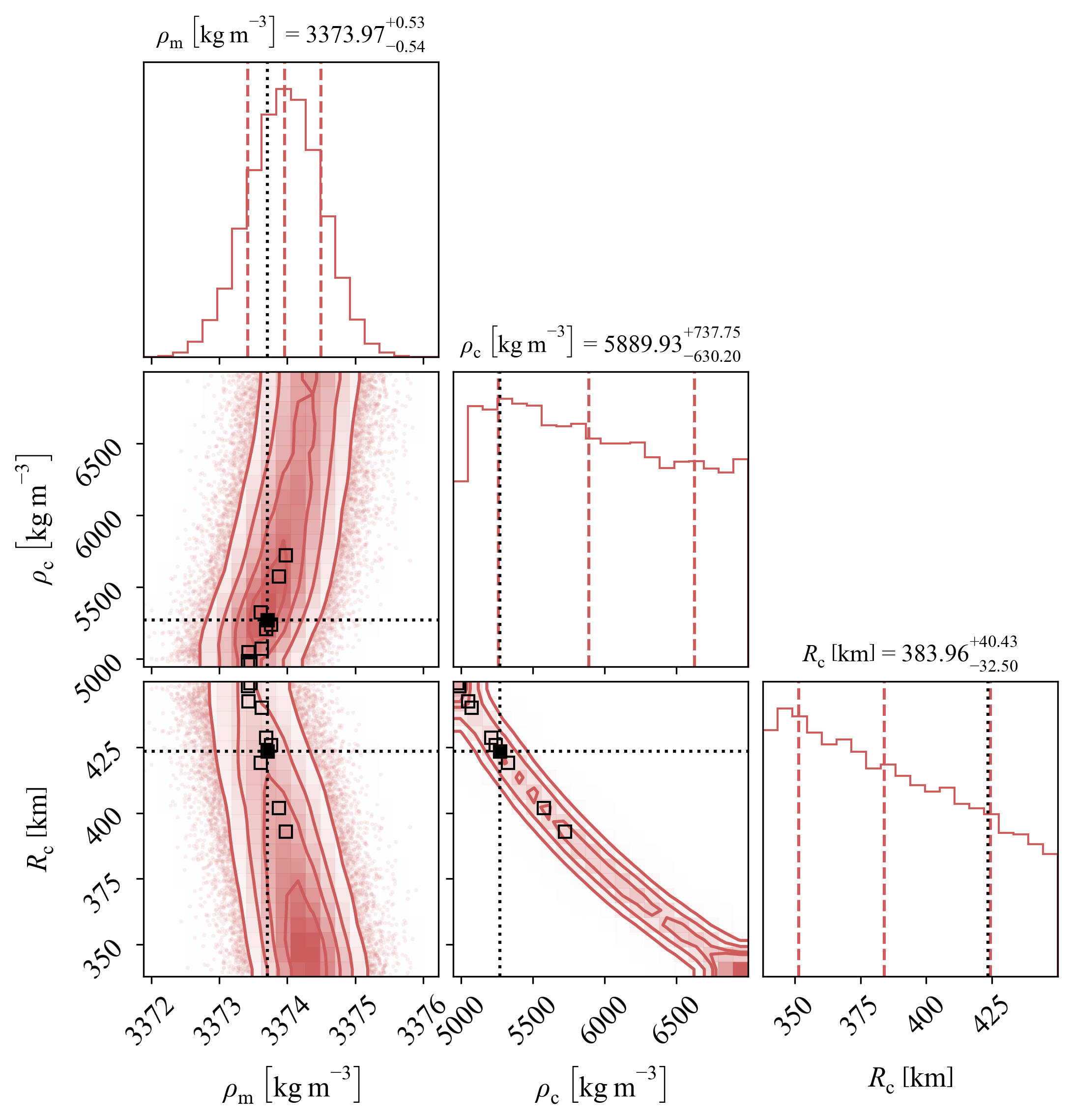}    
    \caption{The posterior probabilities of the mantle density $\rho_{\rm{m}}$, the core density $\rho_{\rm{c}}$, and the outer core radius $R_{\rm{c}}$ of Model 1, satisfying the full set of observational constraints (Table \ref{tab:observables}). The full black square and the dotted black lines indicate the parameters of the best-fitting sample; the empty squares with a black edge are the ten best-fitting samples. The vertical dashed lines plotted over the marginal posterior distributions stand for the $16$th, $50$th, and $84$th percentiles, respectively.}
    \label{fig:Andrade-GEO}
\end{figure}

The interior structure of the lunar model, i.e., the core radius and the densities of the core and the mantle (Figure \ref{fig:Andrade-GEO}), are primarily determined by the mean density and the MoIF of the Moon, with a small contribution from the tidal parameters. Since the mean density and MoIF are known with high precision, we can readily obtain a precise estimate of the mean mantle density. The combination of the simplified interior model and the model constraints used in this study results in a mean mantle density of $\unit[3,373.97^{+0.53}_{-0.54}]{kg\; m^{-3}}$. The estimation of the properties of the core is obscured by the trade-off between the core radius and core density: smaller cores are required to be denser and bigger cores need to be less dense to match the total mass. Figure \ref{fig:Andrade-GEO} shows that the predicted core radii range from $\unit[325]{km}$ up to the maximum considered value of $\unit[450]{km}$ and that the smaller core sizes are slightly preferred. Core densities fall into the range from $5,000$ to $\unit[7,000]{kg\; m^{-3}}$.

The full black square on Figure \ref{fig:Andrade-GEO} (as well as on other similar figures in this Section) indicates the parameters of the best-fitting sample. For Model 1, this sample has $\chi^2=2.09$ and corresponds to an interior model with a relatively large core ($R_{\rm{c}}=\unit[423]{km}$) and a relatively low core density ($\rho_{\rm{c}}=\unit[5,270]{kg\;m^{-3}}$). The empty black squares in Figure \ref{fig:Andrade-GEO} symbolise the ten best-fitting combinations of parameters. An overview of the best-fitting samples is also provided in Table S1 of the SI. All of them demonstrate core sizes close to or greater than $\unit[400]{km}$ and correspondingly reduced core densities.

\begin{figure}[htbp]
    \centering
    \includegraphics[width=1\textwidth]{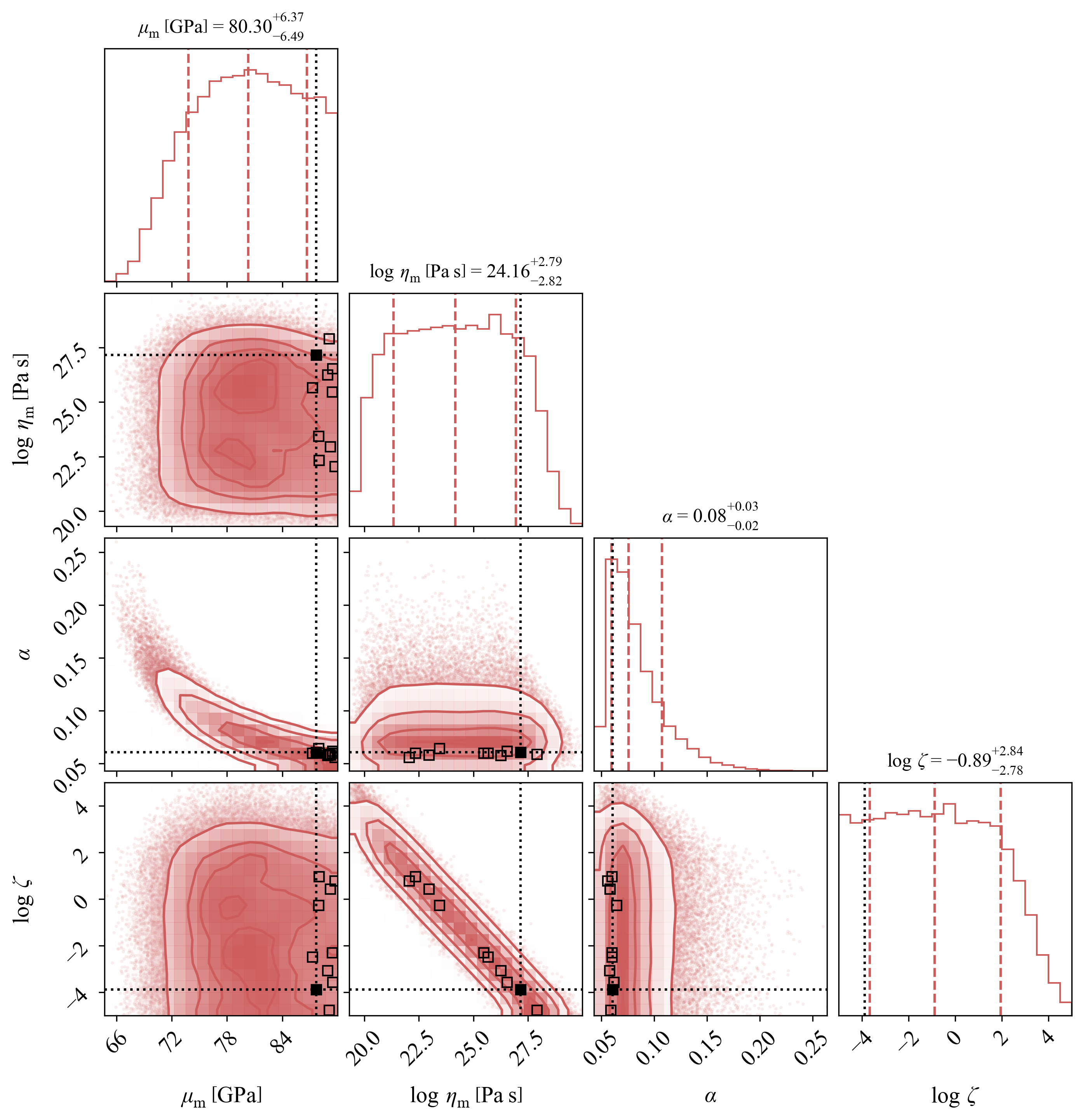}
    \caption{Same as Figure \ref{fig:Andrade-GEO}, but for the effective mantle rigidity $\mu_{\rm{m}}$, the mantle viscosity $\eta_{\rm{m}}$, and the Andrade parameters $\alpha$ and $\zeta$.}
    \label{fig:Andrade-RHEO}
\end{figure}

The posterior probabilities of the fitted rheological parameters are depicted in Figure \ref{fig:Andrade-RHEO}, using the \textit{Python} library \textit{corner} \cite{corner}. As we may see, the mean tidal viscosity of the mantle is strongly anti-correlated with the parameter $\zeta$: a tendency that will also be echoed by the more complex model. A small value of $\zeta$ reveals mantle deformation dominated by transient creep, which is, within the Andrade rheological model, also expected from a highly viscous continuum (with viscosities up to $\unit[10^{29}]{Pa\;s}$). A large value of $\zeta$ indicates mantle deforming preferentially by viscous creep, expected from lower values of $\eta_{\rm{m}}$ (down to $\unit[10^{20}]{Pa\;s}$). The posterior distribution of mantle viscosities and parameters $\zeta$ exhibit two regions of locally increased probability density: one at $\zeta\approx1$ and $\eta_{\rm{m}}\approx\unit[10^{22.5}]{Pa\;s}$, the other at $\zeta<0.1$ and $\eta_{\rm{m}}>\unit[10^{24}]{Pa\;s}$. Values of $\zeta$ greater than $100$ are less likely than values smaller than $100$.

If we compare the resulting Andrade parameter $\alpha=0.08^{+0.03}_{-0.02}$ with the typical values reported in the literature \cite<$0.1<\alpha<0.5$; see, e.g., the overview by>[]{castillo2011,Efroimsky2012a,Efroimsky2012b}, we may notice that it is unusually small. This discrepancy between our prediction and the laboratory data already indicates that although it is, in principle, possible to fit the lunar tidal response with a simple model assuming Andrade rheology in the mantle, the required parameters of this model might not be realistic. A similar point has been made by \citeA{khan2014} and used as an argument in favour of their interior model containing basal partial melt. Moreover, all samples of our Model 1 predict very low values of seismic $Q$ of the lunar mantle at the frequency of $\unit[1]{Hz}$ ($Q_{\rm{seis}}<100$), which is inconsistent with seismic measurements \cite{garcia2019}. Therefore, we will now focus our study on the more complex Sundberg-Cooper model.

\subsection{Lunar Mantle Governed by the Sundberg-Cooper Model} \label{subsec:sc_inversion}

In the present Subsection, as well as in Subsection \ref{subsec:melt_inversion}, we will explore lunar interior models that exhibit a second dissipation peak in the spectra of $k_2/Q_2$ and $Q_{2}^{-1}$. As in the previous inversion with Andrade mantle, we again employ the MCMC approach and seek the parameters of the Sundberg-Cooper model (Model 2 from Table \ref{tab:model_params}) fitting the empirical selenodetic parameters. Due to the greater dimension of the explored parameter space, the model only succeeded to converge after generating $1,440,000$ random samples, and we used a burn-in period of $28,800$ samples. The posterior distributions of the tidal quality factors demonstrate two peaks: a higher one with $Q_{2, \rm{monthly}}>Q_{2, \rm{annual}}$ and a lower one with $Q_{2, \rm{monthly}}<Q_{2, \rm{annual}}$. The latter generally presents a better fit to the observables considered.

\begin{figure}[htbp]
    \centering
    \includegraphics[width=\textwidth]{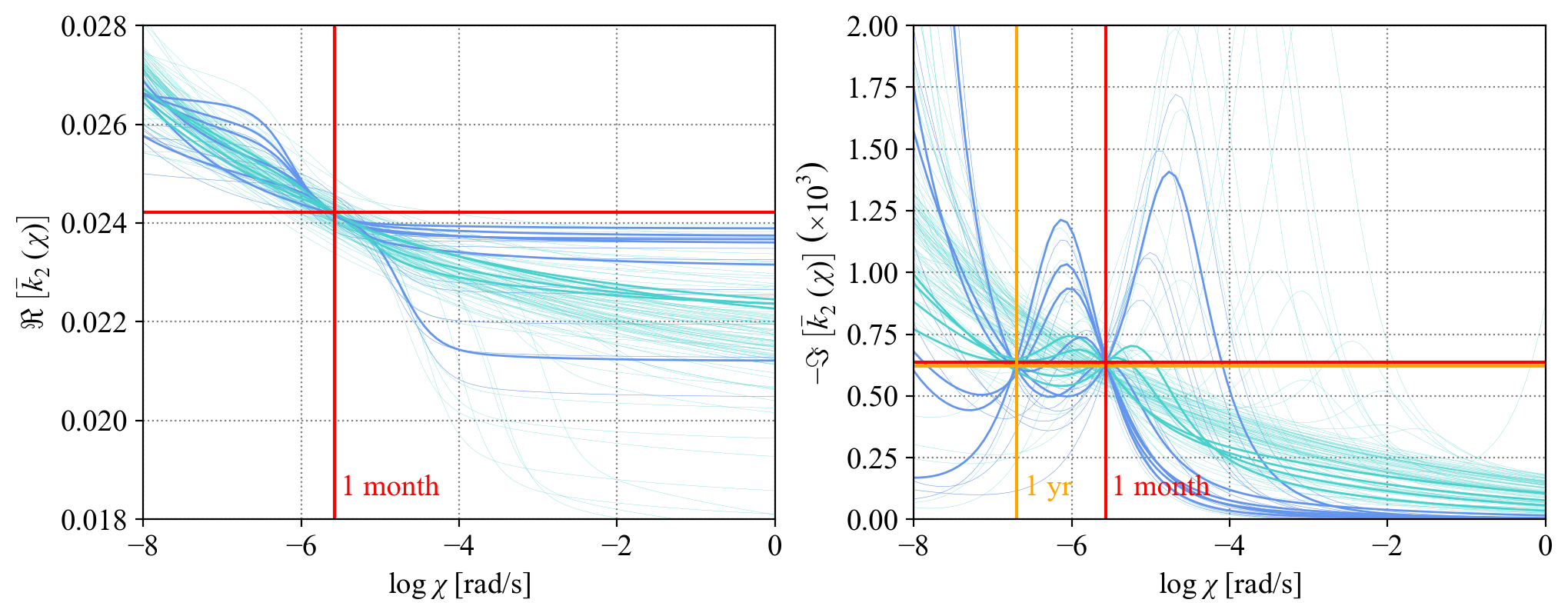}
    \caption{The real (left) and negative imaginary (right) parts of the complex Love number $\bar{k}_2$ as functions of frequency for $100$ randomly chosen samples from the posterior distribution (thin blue and turquoise lines) and for $10$ best-fitting samples (thick blue and turquoise lines). Samples plotted in turquoise only fit the geodetic constraints from our Table \ref{tab:observables}, samples plotted in blue are also consistent with mantle seismic $Q$ \cite<Table 3 of>{garcia2019}. The red and yellow lines indicate the values provided by \citeA{williams2015}. Model 2 with a mantle governed by the Sundberg-Cooper rheology.}
    \label{fig:SC-OVERVIEW}
\end{figure}

Figure \ref{fig:SC-OVERVIEW} illustrates the frequency dependence of the real and the imaginary part of the complex potential Love number $\bar{k}_2(\chi)$ for $100$ samples chosen randomly from the posterior distribution. The blue lines indicate samples that are also consistent with the mantle seismic $Q$ of $10^{3}-10^{5}$ \cite{nakamura1982,gillet2017,garcia2019}, the turquoise lines are samples that only fit the geodetic constraints from Table \ref{tab:observables}. Additionally, the thicker blue or turquoise lines show ten best-fitting samples, the parameters of which are listed in Table S2 of the SI. As we may see, for the best-fitting solutions, the tidal quality functions reported by \citeA{williams2015} at the monthly and the annual frequencies plot either on the different slopes of the secondary dissipation peak or they lie around the ``valley" between the primary and the secondary dissipation peak.

\begin{figure}[htbp]
    \centering
    \includegraphics[width=\textwidth]{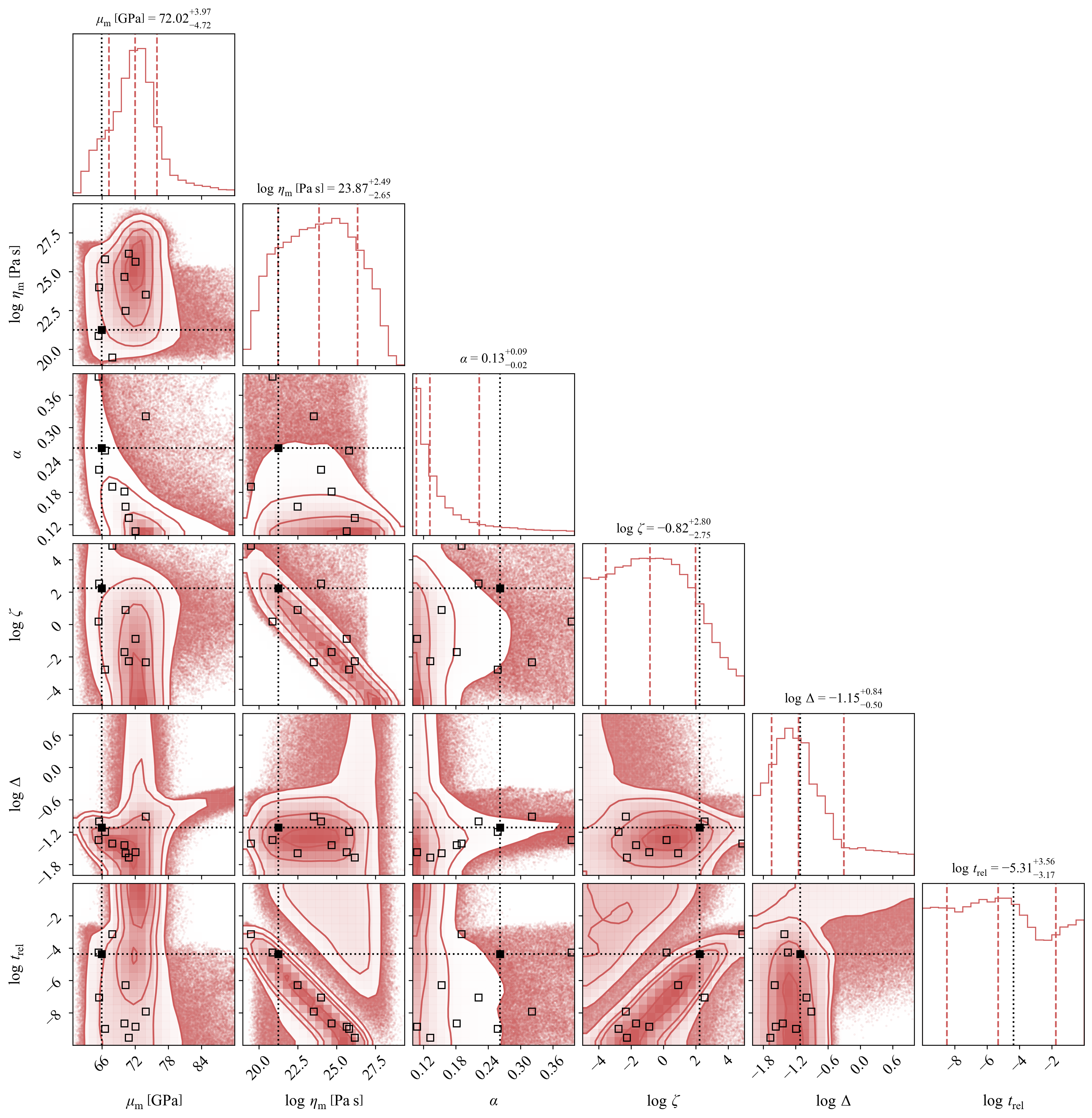}
    \caption{Same as Figure \ref{fig:Andrade-GEO}, but for the rheological parameters of the model with a Sundberg-Cooper mantle (Model 2).}
    \label{fig:SC-RHEO}
\end{figure}

For the interior structure of the lunar model, we find the same tendencies as in the previous subsection. Our prediction of the core size and the interior layers' densities remains unaffected by the change in the mantle's rheology. On the other hand, the range of predicted effective mantle rigidities becomes narrower and shifted to lower values ($\mu_{\rm{m}}=\unit[72.02^{+3.97}_{-4.72}]{GPa}$) within the Sundberg-Cooper model (Figure \ref{fig:SC-RHEO}). The trade-off between effective mantle viscosity and the Andrade parameter $\zeta$ is present, similar to Model 1, and the samples with the highest posterior probability density correspond to $\zeta<1$ and viscosities beyond $\unit[10^{23}]{Pa\;s}$. The Andrade parameter $\alpha$, which characterises the slope of the Andrade branch in the dissipation spectrum (i.e., at frequencies lower than the frequency of the secondary dissipation peak; the right panel of Figure \ref{fig:SC-OVERVIEW}), is preferentially at the lower bound of the considered range: around $\alpha=0.1$. This is a consequence of the model's tendency to fit the empirical quality function (equal to $-\Im{\left[\bar{k}_2(\chi)\right]}$) with the Andrade branch alone, making it as flat as possible. Among the ten best-fitting samples, listed in Table S2 of the SI and plotted as empty black squares in Figure \ref{fig:SC-RHEO}, are values from the entire range $[0.1, 0.4]$. Specifically, the best fit (the full black square in Figure \ref{fig:SC-RHEO}, with $\chi^2=1.39$) has $\alpha=0.26$.

The key ingredients of Model 2 are the parameters of the secondary (Debye) peak: the relaxation strength $\Delta$ and the relative relaxation time $t_{\rm{rel}}$. Figure \ref{fig:SC-RHEO} shows that these parameters attain different values in the samples that fit the tidal dissipation ($Q$ or $k_2/Q$) at the two considered frequencies (monthly and annual) with the secondary peak and different values in the samples fitting the tidal dissipation with the Andrade branch alone. The latter group, characterised by small $\alpha\approx0.1-0.2$, can reach any value of $\Delta$ and $t_{\rm{rel}}$ from the considered interval: this kind of fit is then equivalent to Model 1 (and inconsistent with the mantle's seismic $Q$). The former group, with $\alpha$ from the entire range of $[0.1,0.4]$, demonstrates a narrower range of $\Delta$ between $10^{-1.8}$ and $10^{-0.6}$. Furthermore, $\Delta$ in this second group is correlated with mantle rigidity. The relative relaxation time $t_{\rm{rel}}$ is anti-correlated with the effective mantle viscosity. Since the mantle viscosity determines the magnitude of the Maxwell time and because the tidal dissipation in the second group has a Debye peak in close vicinity of the monthly loading frequency (i.e., $\tau\sim\mbox{const.}$), $t_{\rm{rel}}=\tau/\tau_{\rm{_M}}$ has to decrease with increasing viscosity.

The relative relaxation time of samples with $\alpha>0.2$ is always smaller than $10^{-2}$. This result is consistent with the theoretical expectations, saying that $t_{\rm{rel}}\ll1$ \cite<e.g.,>{morris2009,lee2011,jackson2014}. Since $t_{\rm{rel}}$ is related to the grain size and the grain-boundary viscosity of the mantle material, it might enable us to evaluate whether the Sundberg-Cooper model is indeed applicable to the problem considered in this paper. We will discuss the implications of our $\Delta$ and $t_{\rm{rel}}$ estimates later in Subsection \ref{subsec:disc_sc}.

\subsection{Lunar Mantle with a Weak Basal Layer}\label{subsec:melt_inversion}

\begin{figure}[htbp]
    \centering
    \includegraphics[width=\textwidth]{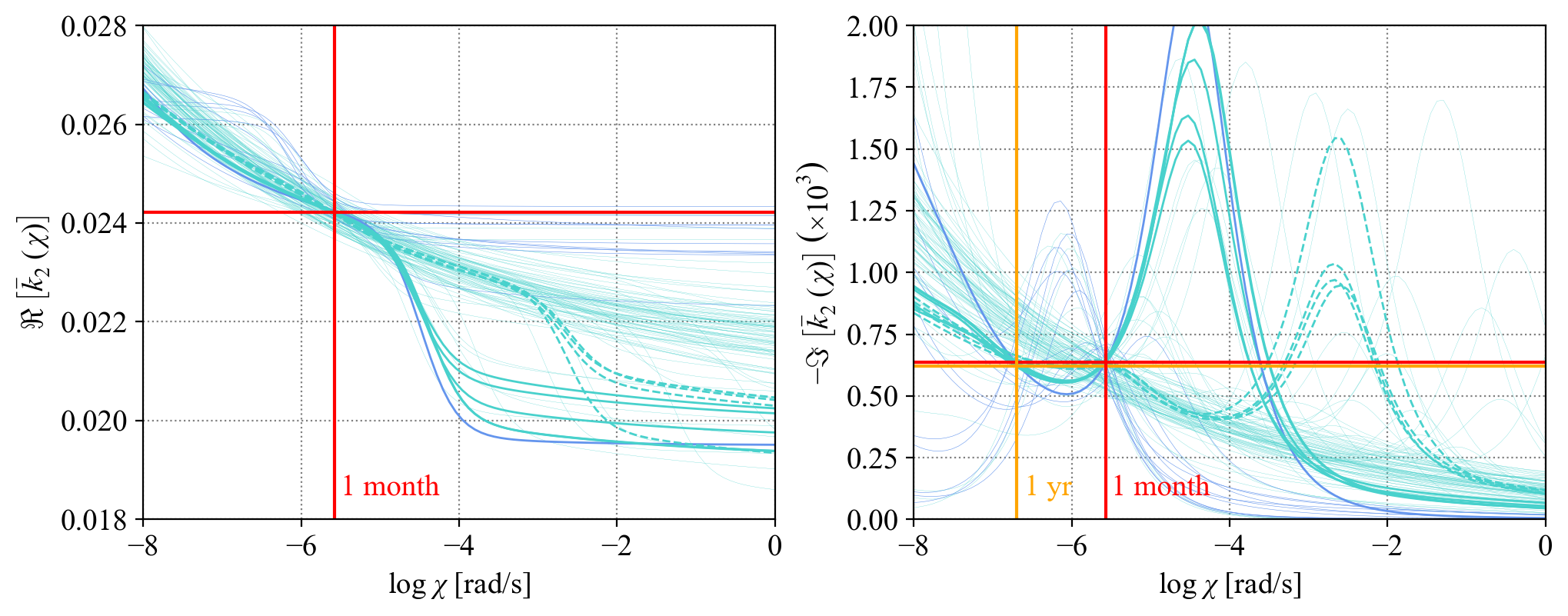}
    \caption{Same as Figure \ref{fig:SC-OVERVIEW}, but for Model 3 containing a basal layer. Dashed lines indicate best-fitting samples with $\eta_{\rm{LVZ}}\sim\unit[10^{13}]{Pa\;s}$.}
    \label{fig:Molten-OVERVIEW}
\end{figure}

The occurrence of the anomalous frequency dependence of lunar tidal $Q$ is often identified with the presence of a highly dissipative layer at the base of the lunar mantle. To compare the model assuming Sundberg-Cooper rheology with the more traditional interpretation of $Q$'s frequency dependence, we finally fitted the empirical constraint with Model 3, which consists of a core, a two-layered mantle, and a crust. Due to the higher dimensionality of the parameter space in Model 3 (see Table \ref{tab:model_params}), the inverse problem took longer to converge than the previous two models. We generated $4,258,000$ random samples and discarded the first $85,160$ samples. A randomly-chosen subgroup of samples from the posterior distribution is plotted in Figure \ref{fig:Molten-OVERVIEW}, along with the ten best-fitting parameter sets (tabulated in the SI, Table S3). Model 3 fits the considered observables better than Models 1 and 2, and $\chi^2$ of the best-fitting sample is $0.77$. However, nine of the ten best-fitting samples, indicated by the thick turquoise lines in Figure \ref{fig:Molten-OVERVIEW}, do not fall into the interval of expected seismic $Q$ of the mantle.

Figure \ref{fig:Molten-OVERVIEW} shows us that among the best-fitting parameter sets, there are two classes of models able to fit the anomalous frequency dependence of the tidal dissipation. Each of the two classes is associated with a different basal-layer viscosity. The first one is centered around $\eta_{\rm{LVZ}}\sim\unit[10^{15}]{Pa\;s}$ and fits the empirical values for the imaginary part of the tidal Love number (right panel of Figure \ref{fig:Molten-OVERVIEW}) with a ``valley" lying next to the basal layer's main dissipation peak and positioned between the loading frequencies $\chi=\unit[10^{-5}]{rad/s}$ and $\chi=\unit[10^{-4}]{rad/s}$. The second one, with $\eta_{\rm{LVZ}}\sim\unit[10^{13}]{Pa\;s}$, fits the dissipation data with a plateau lying next to a minor dissipation peak of the basal layer. This minor peak, corresponding to the same viscosity of the basal layer, is also present in Figure 2 of \citeA{harada2014}, although with a smaller magnitude. The difference in magnitude might be caused by the differences in the rheological model and parameters used in our study. Although the frequency dependence of the best-fitting samples in Model 3 generally follows a trend distinct from Model 2, a number of randomly chosen samples from the posterior distribution of Model 3 resemble those illustrated in Figure \ref{fig:SC-OVERVIEW}. Moreover, the samples with the basal layer's dissipation peak located between the monthly and the annual tidal frequencies tend to fit the mantle seismic $Q$ better than the other samples. The presence of a basal layer may thus mimic the Sundberg-Cooper mantle rheology---and \textit{vice versa}---as was indicated earlier in Figure \ref{fig:sc_vs_andrade}.

The rheological parameters of the overlying mantle are similar to those in the previous two models. Mantle viscosity is anti-correlated with the Andrade parameter $\zeta$, which is preferentially smaller than $100$. The Andrade parameter $\alpha$ tends to the lower bound of the considered interval, and nine of the ten best-fitting models have $\alpha<0.16$. A corner plot illustrating the rheological parameters is included in the SI.

\begin{figure}
    \centering
    \includegraphics[width=\textwidth]{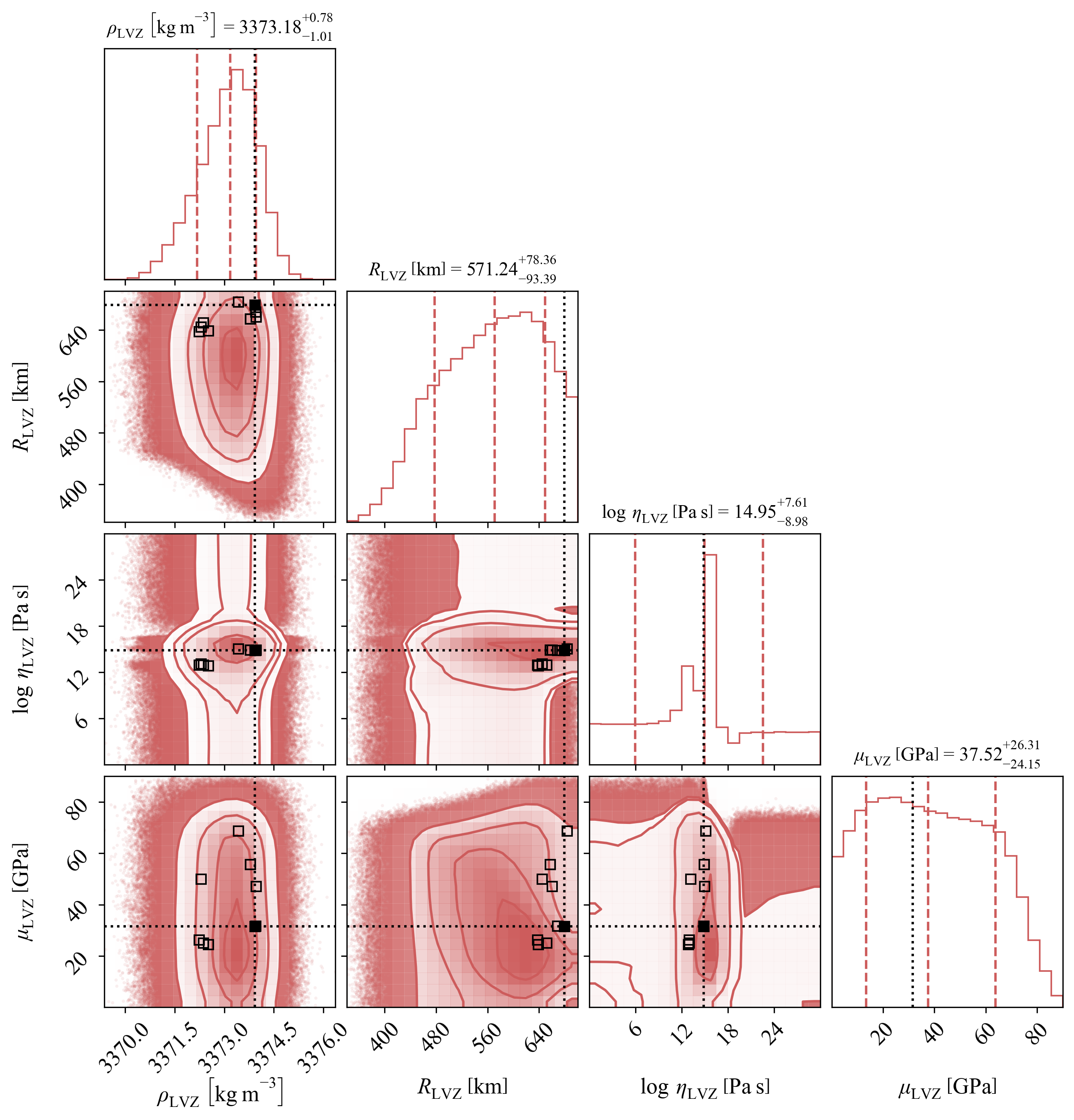}
    \caption{Same as Figure \ref{fig:Andrade-GEO}, but for the parameters of the basal layer (density $\rho_{\rm{LVZ}}$, outer radius $R_{\rm{LVZ}}$, viscosity $\eta_{\rm{LVZ}}$, and rigidity $\mu_{\rm{LVZ}}$) in Model 3.}
    \label{fig:Molten-LAYER}
\end{figure}

\begin{figure}
    \centering
   \includegraphics[width=\textwidth]{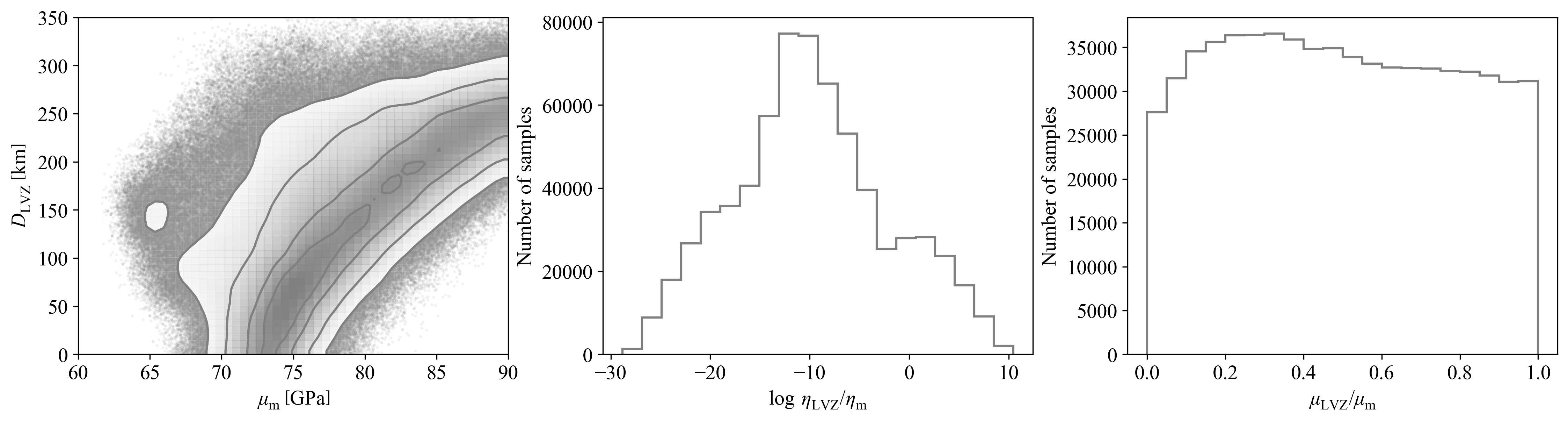}
    \caption{The marginal posterior distribution of mantle rigidity and basal layer's thickness (\textit{left}) and histograms of the viscosity contrast (\textit{middle}) and rigidity contrast (\textit{right}) between the basal layer and the overlying mantle in Model 3.}
    \label{fig:Molten-muM_Dlvz}
\end{figure}

We have already mentioned that the ten best-fitting samples of Model 3 fall into two distinct groups with different basal layer's viscosities. More specifically, even outside the small ensemble of best-fitting samples, the parameter sets with the lower possible basal layer's viscosity (around $\unit[10^{13}]{Pa\;s}$) always have $\alpha<0.24$ and preferentially bigger cores. The samples from the other category are more common and attain $\alpha$ from the entire studied interval. If we only consider the samples that also fit the mantle seismic $Q$, the preferred basal layer viscosity is $\sim\unit[10^{16}]{Pa\; s}$. An overview of all parameters of the basal layer is depicted in Figure \ref{fig:Molten-LAYER}. As we may see, the models with a maximum posterior probability density possess a basal layer with an outer radius of $\sim\unit[620]{km}$ and a rigidity of $\sim\unit[20]{GPa}$. The best-fitting samples typically have a basal layer extending to even greater radii. If we compare the layer's rigidities and viscosities to the rigidities and viscosities of the overlying mantle (Figure \ref{fig:Molten-muM_Dlvz}), we may find all possible ratios $\mu_{\rm{LVZ}}/\mu_{\rm{m}}$, with a very weak preference for values $<0.5$. Therefore, the rigidity contrast obtained from tidal data does not give a clear answer to the question of whether the basal layer can be partially molten.

On the other hand, the viscosity contrast between the basal layer and the overlying mantle is most often around ten orders of magnitude, and this is specifically true for the best-fitting models. Both the viscosity and the rigidity contrast might be indicative of the basal layer's composition and thermal state. We will discuss the implications of this result in more detail in Subsection \ref{subsec:disc_layer}. In addition to the contrasts, the left-most panel of Figure \ref{fig:Molten-muM_Dlvz} depicts the posterior distribution of mantle rigidities and basal layer's thicknesses. Since the low-viscosity basal layer increases the global deformability of the Moon, a thicker layer requires greater rigidity of the overlying mantle.

\begin{figure}
    \centering
    \includegraphics[width=\textwidth]{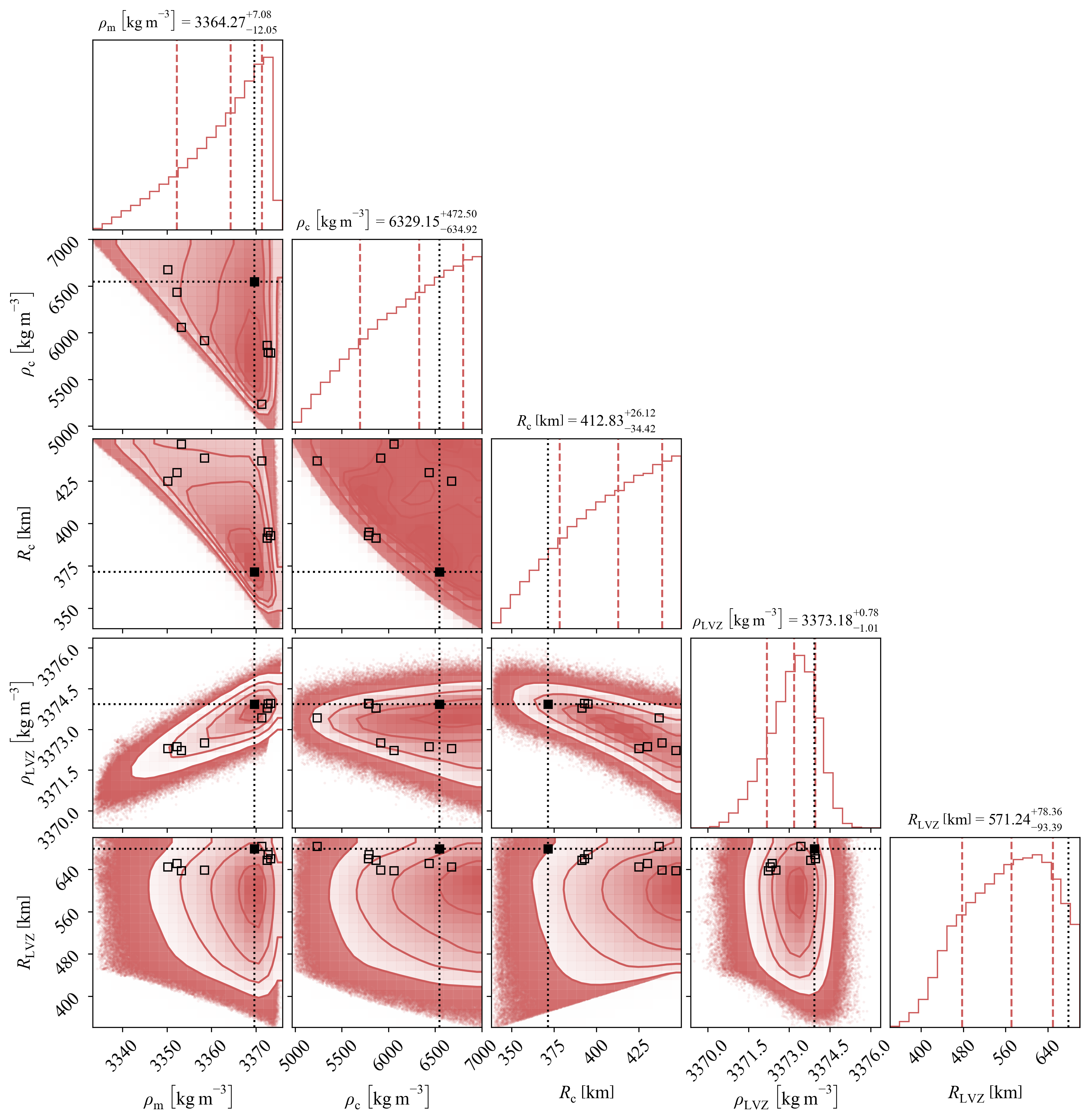}
    \caption{Same as Figure \ref{fig:Andrade-GEO}, but for the structural parameters (core density $\rho_{\rm{c}}$, mantle density $\rho_{\rm{m}}$, outer core radius $R_{\rm{c}}$, basal layer's density $\rho_{\rm{LVZ}}$, and basal layer's outer radius $R_{\rm{LVZ}}$) of Model 3.}
    \label{fig:Molten-GEO}
\end{figure}

Finally, Figure \ref{fig:Molten-GEO} shows the structural parameters of Model 3. With the inclusion of the basal layer, the characteristic trade-off between the core density and radius, known from Figure \ref{fig:Andrade-GEO}, disappears, or is absorbed by the variations in the densities of the other two layers. Similarly, the mantle density is less well-defined than in the previous two models. Instead, the model puts tight constraints on the density of the basal layer.

\section{Discussion} \label{sec:discussion}

In Section \ref{sec:moon}, we compared three different models of the lunar interior and presented the combinations of parameters required to fit the selenodetic constraints. Specifically, the more complex Models 2 and 3 were also able to fit the anomalous frequency dependence of lunar tidal $Q$ mentioned earlier. Now, we will discuss the implications of the two complex models and their fitted parameters for the lunar interior properties.

\subsection{Melt-free Lunar Interior} \label{subsec:disc_sc}

A model with a mantle governed by the Sundberg-Cooper rheology (Model 2) is able to fit the anomalous frequency dependence of lunar tidal $Q$ without the need to assume the existence of a highly dissipative layer at the mantle base. The frequency dependence is then simply explained by the presence of a Debye peak in the dissipation spectrum, associated with the elastically-accommodated GBS. Our best-fitting samples typically exhibit a relaxation timescale $\tau$ of this mechanism lying between $10^{4}$ and $\unit[10^{6}]{s}$, or $3$ and $300$ hours. How can these values be linked to the physical properties of the mantle?

Equation (\ref{eq:tau_gbs}), reprinted below for convenience, 
 \begin{linenomath*}
 \begin{equation*}
 \tau\;=\;\frac{\eta_{\rm{gb}}\,d}{\mu_{\rm{m}}\;\delta}\,\;,
 \end{equation*}
 \end{linenomath*}
gives us the relationship between $\tau$ and microphysical parameters. While $\mu_{\rm{m}}$ is obtained from the inversion of the seismic or tidal data and the grain boundary width $\delta$ is typically around $\unit[1]{nm}$ \cite{marquardt2018}, the other two parameters, namely the grain size $d$ and the grain boundary viscosity $\eta_{\rm{gb}}$, are largely unknown. For the range of $\tau$ obtained here, we predict $\eta_{\rm{gb}}\,d\sim10^6-10^8$. For micrometer to centimeter-sized grains, this implies a grain-boundary viscosity lying between $10^8$ and $\unit[10^{14}]{Pa\; s}$. To better illustrate the distributions of the microphysical parameters, Figure S5 in the SI shows the results of an MCMC inversion with an alternative version of Model 2. In this version, we did not vary the relative relaxation time $t_{\rm{rel}}$, but rather the grain size ($d=\unit[10^{-6}-10^{-2}]{m}$) and the grain-boundary viscosity ($\eta_{\rm{gb}}$ between $\unit[1]{Pa\; s}$ and $\eta_{\rm{m}}$).

\begin{figure}[htbp]
    \centering  
    \includegraphics[width=\textwidth]{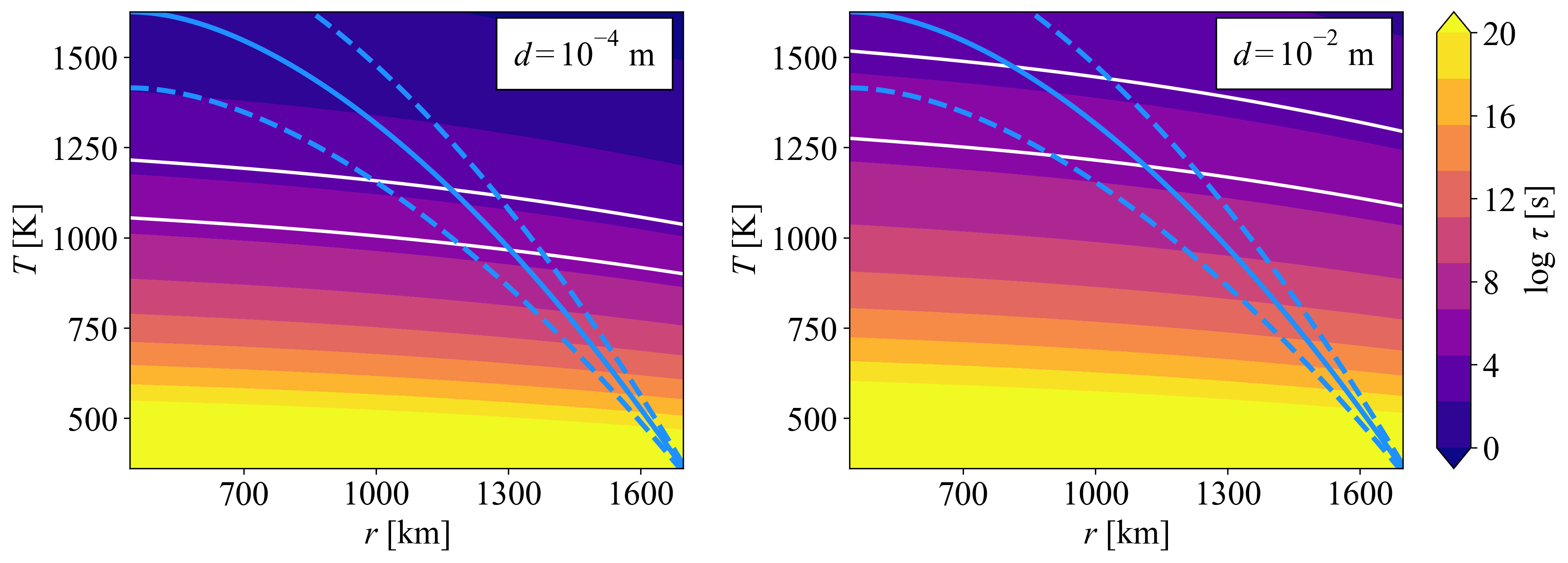}
    \caption{Relaxation time $\tau$ (colour-coded) of elastically accommodated GBS, as given by \citeA{jackson2014} and extrapolated to lunar interior conditions using the Arrhenian equation (\ref{eq:arrhenius}). White lines demarcate the relaxation times of $3$ and $300$ hours, resulting from our inversion. Blue lines indicate analytically-calculated conduction profiles proposed by \citeA{nimmo2012} for three different mantle heat productions ($8$, $9.5$, and $\unit[11]{nW\;m^{-3}}$), crustal heat production of $\unit[160]{nW\;m^{-3}}$ crustal thickness of $\unit[40]{km}$, and no heat exchange between core and mantle. Other parameters, such as the core size, core density, and mantle density, are adjusted to the best-fitting sample of Model 2. Grain sizes are given in the upper right corner of each plot.}
    \label{fig:tau_jackson}
\end{figure}

\citeA{jackson2014} presented results of laboratory experiments on fine-grained olivine subjected to torsional oscillations at high pressures ($P=\unit[200]{MPa}$) and relatively low temperatures ($T<\unit[900]{^{\circ}C}$), i.e., around the threshold between elastic response and elastically accommodated GBS. They found a GBS relaxation timescale of $\log\tau_{\rm{R}}=\unit[1.15\pm0.07]{s}$, where the subscript ``$R$" now stands for ``reference". Because the grain sizes of the samples studied by \citeA{jackson2014} were known, the estimate of $\tau_{\rm{R}}$ also served for the determination of $\eta_{\rm{gb}}=\unit[10^8]{Pa\;s}$, which is on the lower bound of the grain-boundary viscosities corresponding to our best-fitting samples. However, the viscosity, and consequently the relaxation timescale, depends on the pressure and temperature. Considering the reference temperature $T_{\rm{R}}=\unit[1173]{K}$, reference pressure $P_{\rm{R}}=\unit[200]{MPa}$, reference grain size $d_{\rm{R}}=\unit[10]{\mu m}$, activation volume $V^*=\unit[10]{cm^3\;mol^{-1}}$, and activation energy $E^*=\unit[259]{kJ\;mol^{-1}}$, as given by \citeA{jackson2014}, we can extrapolate their $\tau_{\rm{R}}$ to the conditions of the lunar mantle with the Arrhenius law \cite{jackson2010}:

\begin{linenomath*}
\begin{equation}
    \tau = \tau_{\rm{R}} \left(\frac{d}{d_{\rm{R}}}\right)^{m} \exp\left\{\frac{E^*}{R}\left(\frac{1}{T}-\frac{1}{T_{\rm{R}}}\right)\right\} \exp\left\{\frac{V^*}{R}\left(\frac{P}{T}-\frac{P_{\rm{R}}}{T_{\rm{R}}}\right)\right\}\; . \label{eq:arrhenius}
\end{equation}
\end{linenomath*}\\
In addition to the parameters introduced earlier, $m$ characterises the grain-size dependence of the relaxation process in question. We adopt the value $m=1.31$, found by \citeA{jackson2010} for anelastic processes. Figure \ref{fig:tau_jackson} illustrates the extrapolation of $\tau_{\rm{R}}$ of \citeA{jackson2014} to lunar interior conditions, considering the best-fitting parameter set of Model 2 and two depth-independent grain sizes. Over the colour-coded maps, we also plot the steady-state heat conduction profiles of \citeA{nimmo2012}. We note that the conduction profiles were only chosen for illustration purposes: the discussion of the thermal regime (conductive vs. convective) in the lunar mantle is beyond the scope of this paper.

The laboratory measurements of \citeA{jackson2014} were performed on a single sample of fine-grained polycrystalline olivine under constant pressure $P_{\rm{R}}$ and the Arrhenian extrapolation of $\tau$ was only tested for temperature dependence. Nevertheless, if we accept the assumption that these results are applicable to the Moon, Figure \ref{fig:tau_jackson} and the range of relaxation times able to fit the frequency dependence of tidal $Q$ ($\log\tau\in[4, 6]$) can help us to identify the minimum depth in which elastically accommodated GBS contributes to the tidal dissipation. For the smaller grain size ($d=\unit[0.1]{mm}$) and the reference profile of \citeA{nimmo2012} (solid line, mantle heat production of $\unit[9.5]{nW\;m^{-3}}$), we predict the minimum depth of $400$--$\unit[500]{km}$. For the larger grain size ($d=\unit[1]{cm}$), the minimum depth is $600$--$\unit[800]{km}$. A conductive profile corresponding to lower heat production than illustrated here would push the minimum depth to even greater values. The occurrence of elastically accommodated GBS in shallower depths would give rise to a relaxation peak (or to an onset of a relaxation band) at lower loading frequencies, which would not fit the observed annual and monthly tidal dissipation. Although the MCMC inversion from the previous section was performed for a model with a homogeneous mantle, i.e., assuming the occurrence of elastically-accommodated GBS at all depths from the surface down to the core, we also checked that a model described by the Andrade rheology above the derived depths and by the Sundberg-Cooper model below the derived depths might fit the considered observables with intermediate values of $\tau$ between $10^5$ and $10^{6}$. However, fitting of the observables with Sundberg-Cooper rheology only applicable to depths greater than $\unit[500]{km}$ (considering Andrade rheology at shallower depths) seems very challenging.

\begin{figure}[htbp]
    \centering
    \includegraphics[width=0.8\textwidth]{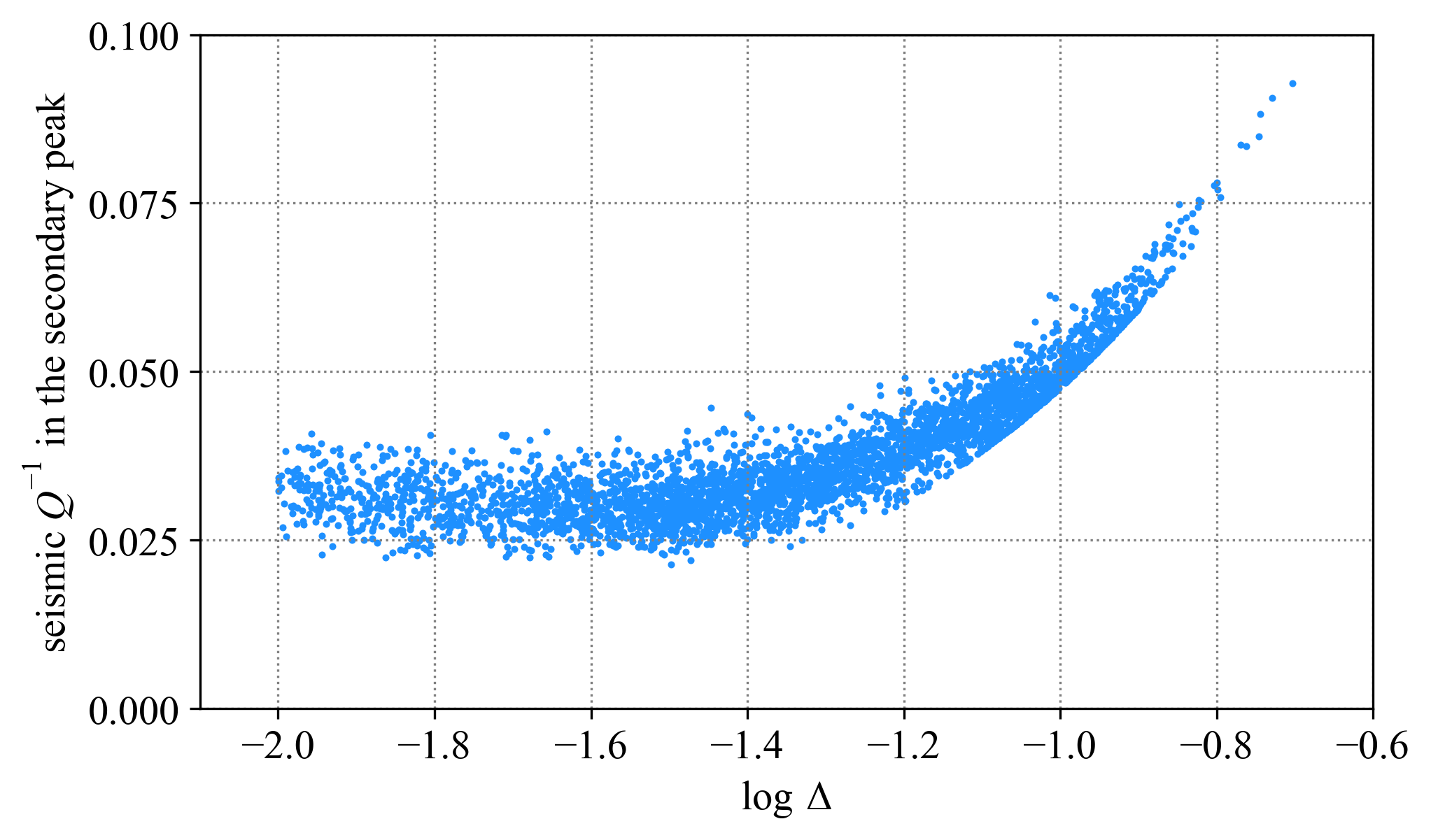}
    \caption{Peak value of the low-frequency seismic $Q^{-1}$ as a function of the relaxation strength $\Delta$ for $4,000$ randomly chosen samples of Model 2 exhibiting a Debye peak in the frequency interval ($\chi_{\rm{year}}$, $\unit[10^{-4}]{rad\;s^{-1}}$).}
    \label{fig:Delta_Q}
\end{figure}

Besides the timescale $\tau$, we have derived the relaxation strength of the hypothetical secondary peak. Considering only the group of samples fitting the anomalous frequency dependence of tidal $Q$, the relaxation strength falls into the interval $\log\Delta\in[-1.8, -0.6]$, or $\Delta\in[0.02, 0.25]$. Parameter $\Delta$ controls the height of the secondary dissipation peak in the Sundberg-Cooper model. Figure \ref{fig:Delta_Q} shows the dependence of the peak seismic $Q^{-1}$ at low, tidal frequencies on the relaxation strength $\Delta$ for randomly chosen $4,000$ samples of Model 2 that exhibit a Debye peak in the frequency range from $\chi_{\rm{year}}$ to $\unit[10^{-4}]{rad\;s^{-1}}$. Are these values consistent with theoretical prediction and laboratory data?

\citeA{Sundberg} reported relaxation strengths of polycrystalline olivine between $0.23$ and $1.91$, as found in different sources and under different assumptions on the grain shapes \cite{ke1947,raj1971,ghahremani1980}. Their own mechanical tests on peridotite (olivine-orthopyroxene) at temperatures between $1200$ and $\unit[1300]{^{\circ}C}$ were best fitted with $\Delta=0.43$ and the corresponding dissipation associated with elastically-accommodated GBS in their sample was $Q^{-1}=0.25-0.3$. On the other hand, \citeA{jackson2014}, who performed torsion oscillation experiments on olivine, found a relatively low dissipation peak with $Q^{-1}\le0.02$. Low secondary dissipation peaks with $Q^{-1}\sim10^{-2}$ were also predicted theoretically by \citeA{lee2010} for a grain boundary slope of $30^{\circ}$, while smaller slopes seem to allow $Q^{-1}$ exceeding $1$, especially when the individual grains are of comparable sizes and the grain boundary viscosity does not vary too much. Accordingly, \citeA{lee2011} note that $Q^{-1}$ in the secondary peak depends strongly on the slope of the grain boundaries.

The largest $\Delta$ predicted by our inversions and able to fit the frequency dependence of $Q$ lies on the lower bound of the range reported by \citeA{Sundberg}. At the same time, the small height of the Debye peak, observed by \citeA{jackson2014} and also found by \citeA{lee2010}, is only approximately consistent with $\log\;\Delta\lesssim-1.25$ (Figure \ref{fig:Delta_Q}). Following this brief discussion of dissipation arising due to elastically accommodated GBS, we conclude that the relaxation strength $\Delta$ (or $Q^{-1}$ in the secondary dissipation peak) is not well constrained and the values found in literature permit any of the $\Delta$s predicted in our Subsection \ref{subsec:sc_inversion}.

\subsection{Highly Dissipative Basal Layer} \label{subsec:disc_layer}

\begin{figure}[htbp]
    \centering
    \includegraphics[width=0.7\textwidth]{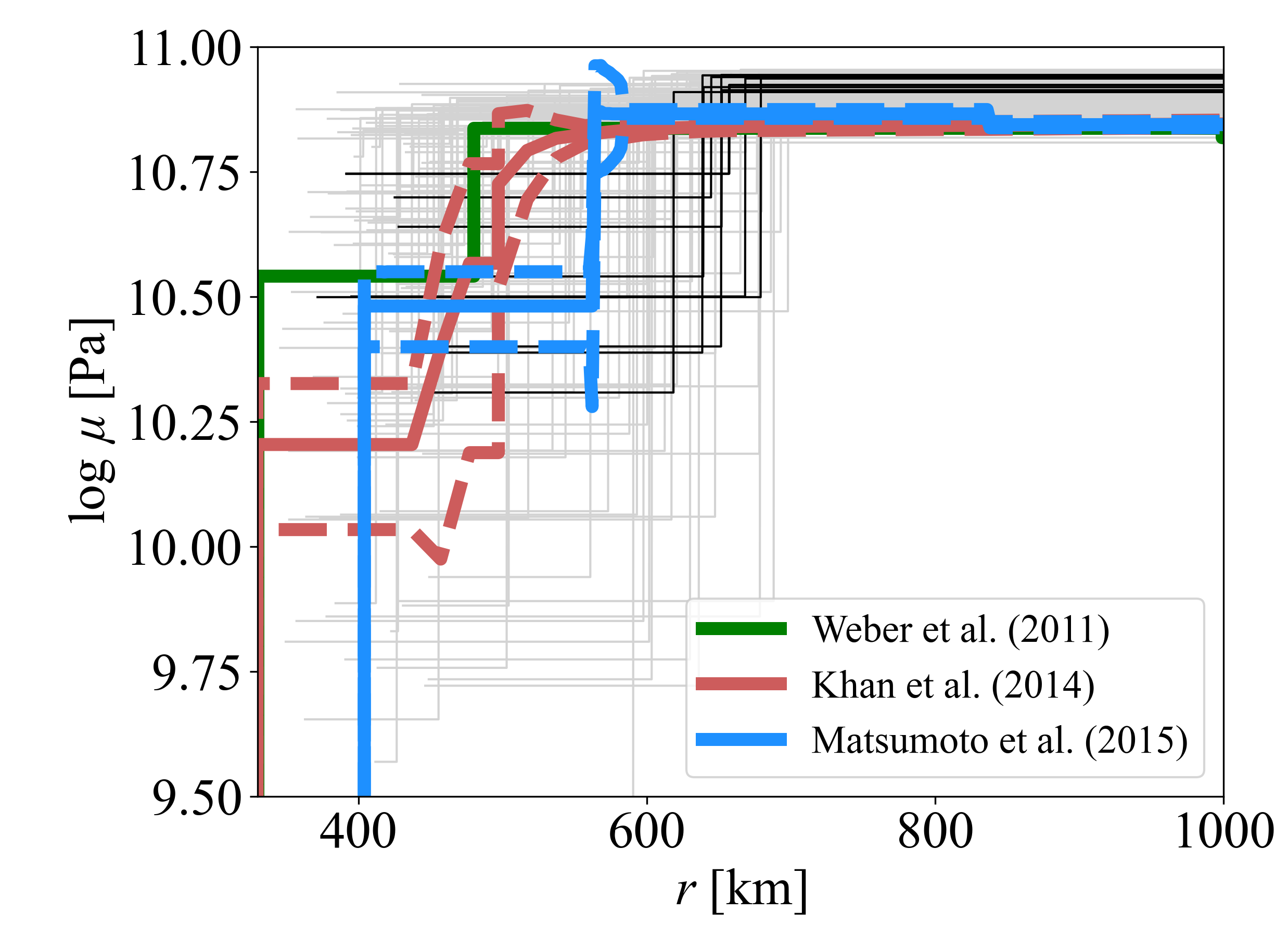}
    \caption{Rigidity prediction compared to seismic measurements. One hundred randomly chosen samples from the posterior distribution (light grey) and 10 best-fitting samples (black). Rigidity derived from seismic velocities and densities: green \cite{Weber}, red \cite{khan2014} and blue \cite{matsumoto2015}, dashed lines: errors. The data from the three studies are provided in \citeA{garcia2019}.}
    \label{fig:mu_basal}
\end{figure}

A highly dissipative layer located at any depth could also produce the desired secondary peak needed to explain the anomalous $Q$ dependence. (Note, however, that a presence of a highly dissipative layer at a shallow depth may lead to changes in the body's response to tides and might be incompatible with the measured values of the Love numbers.) Petrological considerations combined with an indication of a basal low-velocity zone place this anomalous layer in the deep interior. Therefore, as an alternative to the ``melt-free" Model 2, we tested the popular hypothesis of a putative highly dissipative layer at the base of the lunar mantle.

The derived rheological properties of the mantle and of the basal layer as well as the layer's thickness are poorly constrained and can be strongly biased. Firstly, the thickness $D_{\rm LVZ}$ of the basal layer is correlated with the value of the mantle rigidity $\mu_{\rm m}$ (Figure \ref{fig:Molten-GEO}); the thicker the basal layer, the larger mantle rigidity is required to satisfy the model constraints. The prediction of the mantle viscosity $\eta_{\rm m}$ is affected by the Andrade rheological parameters and is particularly anticorrelated with the parameter $\zeta$. On the other hand, the viscosity of the basal layer remains independent of the Andrade parameter $\alpha$, with the only exception that the solutions corresponding to the lower branch of the basal viscosity ($\eta_{\rm LVZ}=\unit[10^{13}]{Pa\,s}$) vanish for $\alpha>0.24$. The predicted contrast in viscosity between the two layers is therefore weakly dependent on the Andrade parameter $\zeta$ due to its anticorrelation with the mantle viscosity $\eta_{\rm m}$.

Secondly, the posterior distribution of the basal layer's rigidities ($\mu_{\rm LVZ}\le\mu_{\rm m}$) hints at a very weak anti-correlation with the outer radius of the basal layer $R_{\rm LVZ}$ (Figure~\ref{fig:Molten-LAYER}). However, the ten best-fitting models prefer a relatively large basal layer's outer radius independent of the rigidity. The predicted rigidities of a basal layer, especially for the best-fitting models, are consistent with seismic observations (Figure~\ref{fig:mu_basal}), including the rigidity decrease in the basal layer. These profiles are, however, obtained for a larger basal layer's outer radius compared to the seismic predictions. In general, the rigidity contrast between the basal layer and the overlying mantle is poorly constrained. Still, the models with the contrast in the range (0.1-0.5) are very weakly favoured (see the right-most panel of Figure~\ref{fig:Molten-muM_Dlvz} and Figure~\ref{fig:mu_basal}).
Lastly, there is no obvious correlation of the basal viscosity with the other considered parameters for any branch of the solutions (i.e., branches corresponding to viscosity $\sim\unit[10^{13}]{Pa\,s}$ and $\sim\unit[10^{15}]{Pa\,s}$). Low basal viscosity and large viscosity contrast are, therefore, the most robust results of the present inversion. 

\begin{figure}[htbp]
    \centering
    \includegraphics[width=\textwidth]{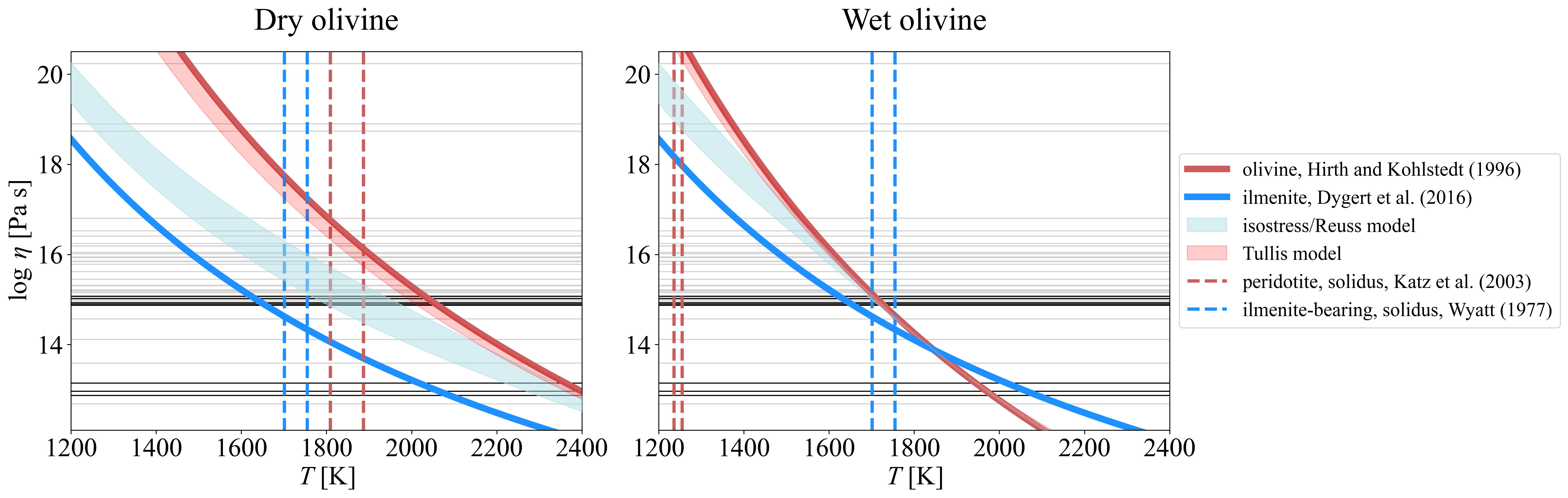}
    \caption{Basal viscosity prediction compared to rheological properties. One hundred randomly chosen samples from the posterior distribution (light grey) and 10 best-fitting samples (black). Over the predicted data is plotted the temperature dependence of viscosity of ilmenite \cite<blue,>[]{dygert2016}, dry olivine \cite<red,>[]{hirth1996}, and ilmenite-olivine aggregate ($\unit[2-16]{\%}$), the latter corresponding either to isostress (blue area, harmonic mean, suggested for high strain) or Tullis (red area, geometric mean, suggested for low strain) models. Errors of experimentally determined viscosities not included; ilmenite error factor is $\sim5$. Vertical lines delimit solidus temperatures for peridotite \cite{katz2003} and ilmenite-bearing material \cite{wyatt1977} at radii $\unit[330]{km}$ and $\unit[700]{km}$. Left panel: temperature dependence for the differential stress $\sigma_{\rm{D}}=\unit[1]{MPa}$, dry olivine.  Right panel: temperature dependence for $\sigma_{\rm{D}}=\unit[1]{MPa}$, wet olivine.}    
    \label{fig:eta_basal}
\end{figure}

Rigidity and viscosity magnitudes, and their contrast between the mantle and the basal layer, can be indicative of the variations in the composition, in the presence of melt, and in temperature. A stable partially molten zone in the lunar interior would pose strong constraints on the composition \cite{khan2014}. Given the absence of geologically recent volcanic activity, any melt residing in the deep lunar interior would have to be neutrally or negatively buoyant. Using an experimental approach on the synthetic equivalent of Moon samples, \citeA{vankanparker2012} concluded that the condition on the buoyancy below $\unit[1000]{km}$ is satisfied if high content of titanium dioxide is present in the melt. The presence of a partially molten layer is permitted at any depth below this neutral buoyancy level.

Moreover, evolutionary models suggest that high-density ilmenite-bearing cumulates enriched with TiO$_2$ and FeO are created towards the end of the shallow lunar magma ocean crystallisation, resulting in near-surface gravitational anomalies. This instability, combined with the low viscosity of those cumulates, might have eventually facilitated the mantle overturn, creating an ilmenite-rich layer at the base of the mantle \cite<e.g.,>[]{zhang2013,zhao2019,li2019}. Recently, \citeA{kraettli2022} suggested an alternative compositional model: a $\sim\unit[70]{km}$ thick layer of garnetite could have been created at the base of the mantle if two independently evolving melt reservoirs were present. The resulting high-density garnet, olivine, and FeTi-oxide assemblage is gravitationally stable and can contain a neutrally or negatively buoyant Fe-rich melt. The scenario of \citeA{kraettli2022} can also be accompanied by the mantle overturn, as suggested for the ilmenite-rich layer created at shallow depths.

Rheologically weak ilmenite combined with appropriate lower-mantle temperature can help to explain the low basal viscosity (Figure~\ref{fig:eta_basal}). Considering viscosities lower than $\eta_{\rm LVZ}\sim\unit[10^{16}]{Pa\,s}$, the basal layer would need to experience temperatures $\gtrsim\unit[1900]{K}$ if the lower mantle were only made of dry olivine. In contrast, for wet olivine, a temperature higher than $\unit[1600]{K}$ would be sufficient. Creep experiments \cite{dygert2016} conclude that the viscosity of ilmenite is more than three orders of magnitude lower than that of dry olivine. Consequently, a lower-mantle temperature higher than $\unit[1500]{K}$ might be acceptable to explain the predicted viscosities for pure ilmenite. Interestingly, if we consider viscosity $\sim\unit[10^{13}]{Pa\,}$ fitting the dissipation data with a plateau lying next to a minor dissipation peak of the basal layer, the temperature would have to be even higher ($\unit[2400]{K}$ for dry olivine, $\unit[2100]{K}$ for wet olivine, and $\unit[2000]{K}$ for ilmenite), i.e., it would need to attain values above the respective liquidi and critical porosities, where the melt presence would control the rheology. Melt content above the critical porosity would be inconsistent with only a small to moderate rigidity decrease. We will discuss the effect of the melt later in this Subsection.

The properties of ilmenite-olivine aggregates introduce yet another complexity. The viscosity of aggregates is suggested to depend on the value of the strain: it follows the Tullis model for low strain, whereas it tends to follow the lower bound on Figure~\ref{fig:eta_basal} (isostress model) for large strain \cite<see, e.g.,>[for a deeper discussion]{dygert2016}. The acceptable temperature range for olivine-ilmenite aggregate is close to the values for the pure olivine in the case of the Tullis model. The prediction for the isostress model (minimum bound, Reuss model) is consistent with temperature values larger than $\unit[1600]{K}$ considering viscosities $<\unit[10^{16}]{Pa\,s}$. Another obstacle in interpretation originates in the stress-sensitivity of the relevant creep. The viscosity can decrease by $\sim2.5$ orders of magnitude while decreasing the differential stress $\sigma_{\rm{D}}$ by one order of magnitude. In terms of acceptable thermal state, the temperature consistent with our prediction would decrease roughly by $\sim\unit[100]{K}$ considering two-fold higher differential stress and increase by the same value for two-fold lower stress, respectively.

Consequently, we find acceptable solutions both below and above the solidus. Our Model 3 thus cannot exclude or confirm a possible partial melt presence. An alternative explanation for the viscosity reduction can be the presence of water \cite<see also>[for a deeper discussion]{karato2013}, which would also reduce the solidus temperature and facilitate partial melting. Both the enrichment in ilmenite and elevated water content can lead to the desired value of viscosity at lower temperatures compared to the dry and/or ilmenite-free models (Figure~\ref{fig:eta_basal}).

Focusing now on the elastic properties, we note that the rigidities of olivine \cite<e.g.>[]{mao2015}, ilmenite \cite{jacobs2022}, and garnetite \cite{kraettli2022} are comparable. The temperature has only a limited impact on their value ($\unit[-0.01]{GPa/K}$ for olivine and ilmenite). Also, dependence on the water content (olivine-brucite) is only moderate \cite<$-1.3$ GPa/wt$\%$;>{jacobsen2008}. The magnitude of rigidity is, therefore, rather insensitive to possible constituents, temperature, and water content. The 84th percentile on Figure \ref{fig:Molten-LAYER}, corresponding to $\sim\unit[60]{GPa}$, fits the elastic properties of all considered minerals---ilmenite, olivine, and garnet. However, the 16th percentile ($\sim\unit[10]{GPa}$) would be difficult to explain by the changes in composition, high temperature, and/or water content alone.

\begin{figure}[htbp]
    \centering
    \includegraphics[width=\textwidth]{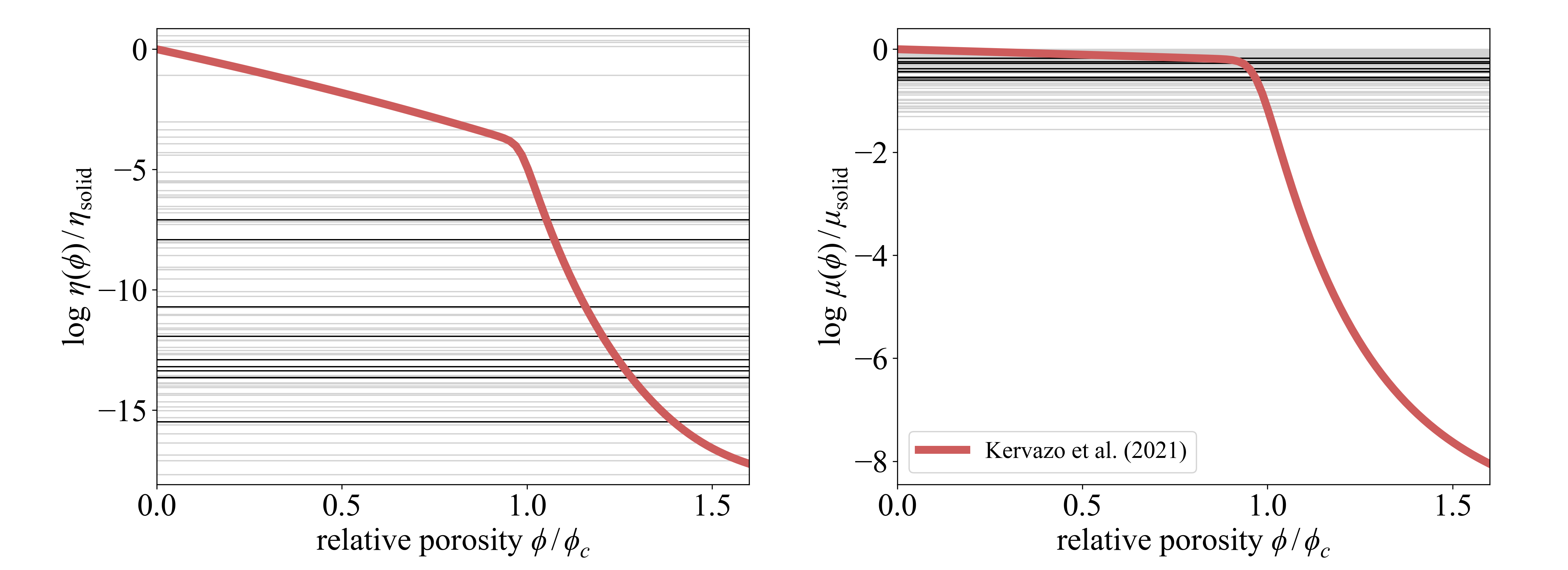}
    \caption{Impact of melt on the viscosity and rigidity contrast. The viscosity and rigidity contrast is expressed as a function of the $\phi/\phi_{\rm{c}}$ ($\phi$ denotes the porosity and $\phi_{\rm{c}}$ the critical porosity) and parameterised using \citeA{kervazo2021}; $\eta_{\rm solid}$ and $\mu_{\rm solid}$ represent values with no melt present at the solidus temperature; no change in composition is considered. The light grey and black horizontal lines depict the contrasts for 100 randomly chosen samples and 10 best-fitting samples, respectively.}
    \label{fig:melt_basal}
\end{figure}

The magnitude of rigidity (Figure~\ref{fig:melt_basal}) is, nevertheless, sensitive to the presence of melt around or above the disintegration point (characterised by the critical porosity $\phi_{\rm{c}}$), which describes the transition from the solid to liquid behaviour and its typical values lie between $25-40\%$ \cite<e.g.>{Renner2000}. Similarly, the viscosity value is very sensitive to the presence of melt for porosity higher than $\phi_{\rm{c}}$. For low porosities, it follows an exponential (Arrhenian) dependence. Figure~\ref{fig:melt_basal} suggests that the predicted rheological contrasts in Model 3 are consistent with $\phi\lesssim1.1\phi_{\rm{c}}$ for rigidity and with $\phi>1.1\phi_{\rm{c}}$ for the viscosity, considering best fitting samples. This apparent incompatibility may be accounted for by the presence of melt accompanied by the changes in the composition of the basal layer and by the susceptibility of viscosity to these changes. Consequently, the knowledge of the contrasts in both rheological parameters (rigidity and viscosity) could help tackle the trade-offs between porosity and composition or temperature. \\

\begin{figure}[htbp]
    \centering
    \includegraphics[width=\textwidth]{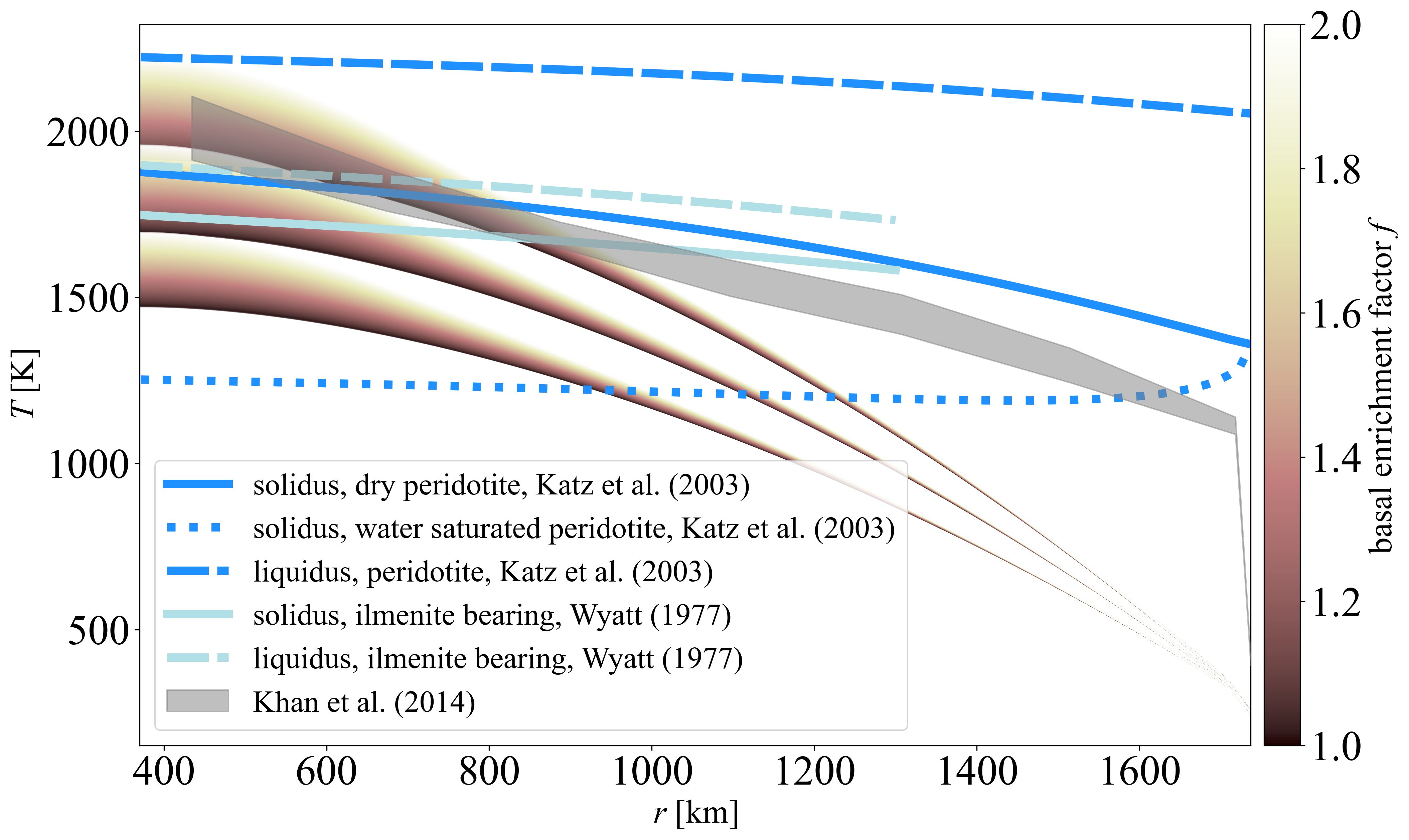}
    \caption{Comparison of temperature profiles for the best fitting sample of Model 3. Colour scale: conductive profile, calculated with the matrix propagator method; parameters as in Figure~\ref{fig:tau_jackson}. Individual branches correspond to average heating $8$, $9.5$ and $\unit[11]{nW/m^2}$ in the mantle. The coefficient $f$ denotes the enrichment in the radiogenic elements of the basal layer ($R_{\rm LVZ}=\unit[679]{km}$) compared to the rest of the mantle. Gray area is the temperature profile adapted from \citeA{khan2014}; darker blue lines: peridotite solidus (solid), water-saturated solidus (dotted), and liquidus (dashed) according to \citeA{katz2003}; light blue lines: clinopyroxene+ilmenite solidus (solid) and liquidus (dashed) according to \citeA{wyatt1977}.}
    \label{fig:temp_basal}
\end{figure}

The presence of a partially molten material would pose a strong constraint on the temperature and possible mode of the heat transfer in the lower mantle of the Moon, allowing only models that reach the temperature between the solidus and liquidus (Figure~\ref{fig:temp_basal}). The traditional advective models predict stagnant-lid mantle convection with a relatively thick lid at present \cite<e.g.>{zhang2013}. Below the stagnant lid, the temperature follows the adiabatic or, for large internal heating, sub-adiabatic gradient. We estimate the temperature increase across the entire mantle due to the adiabatic gradient to be bounded by $\unit[100]{K}$. Within those traditional models, it is plausible to reach solidus only in the lowermost thermal-compositional boundary layer. In the case of conductive models \cite<e.g.>[]{nimmo2012}, the temperature gradient is steeper than the solidus gradient and the solidus temperature can be reached in the entire basal layer, given appropriate internal heating (as demonstrated in Figure~\ref{fig:temp_basal}). Interestingly, the lunar selenotherm determined by the inversions of lunar geophysical data combined with phase-equilibrium computations \cite{khan2014} lies between the conductive and adiabatic gradients.

In the future, distinct sensitivity of rigidity, viscosity, and other transport properties to temperature, melt fraction, and composition may provide a way to separate the interior thermal and composition structure. At present, inversion errors and the uncertainties on material properties cannot confirm or rule out the existence of a partially molten basal layer. It therefore remains a valid hypothesis.

\subsection{Other Sources of Information}

The two models discussed in this section~--- one with a highly dissipative basal layer and the other with elastically-accommodated GBS in the mantle~--- cannot be distinguished from each other by the available selenodetic measurements. To answer the question stated in the title of our paper, one would need to resort to other types of empirical data. Among all geophysical methods devised for the exploration of planetary interiors, seismology is of foremost importance. Therefore, a question that cannot be solved by the interpretation of lunar tidal response might be answered by comparing the arrival times and the phases detected at individual seismic stations.

As we mentioned in Introduction, the Moon demonstrates a nearside-farside seismic asymmetry. Judging by the currently available seismic data collected on the near side, the deep interior of the far side is virtually aseismic or, alternatively, the seismic waves emanating from it are strongly attenuated or deflected. The existence of an aseismic area on the farside might not be entirely inconceivable. First, as pointed out by \citeA{Nakamura2005}, there are large zones with no located nests of deep moonquakes even on the nearside; and, in fact, most of the known deep seismic nests are part of an extended belt reaching from the south-west to the north-east of the lunar face. Second, there exists a pronounced dichotomy between the near side and far side of the Moon in terms of the crustal thickness, gravity field, and surface composition, which might point to a deeper, internal dichotomy as predicted by some evolutionary models \cite<e.g.,>[]{laneuville2013,zhu2019,jones2022}.\\

An obvious way to illuminate the lack of deep farside moonquakes detected by the Apollo seismic stations would be to place seismometers on the far side of the Moon. They would observe the far side activity, and record the known repeating nearside moonquakes or events determined from impact flash observations. The {{Farside Seismic Suite}} (FSS) mission, recently selected for flight as part of the NASA PRISM program and planned for launch in 2025, might provide such a measurement by delivering two seismometers to Schr\"odinger Crater \cite{Panning}. While this crater is far from the antipodes (in fact, close to the South pole), a seismometer residing in it should still be able to detect events from the far side, thereby addressing the hemispheric asymmetry in the Apollo observations. However, resolving polarisation of arrivals may be challenging for many moonquakes, meaning that many events will only have distance estimated, but not azimuth. (We are grateful to Mark P. Panning for an enlightening consultation on this topic.)

A better site for this science objective would be the far side Korolev crater residing by the equator, about 23 degrees from the antipodes (by which we understand the centre of the farside). It is now considered as one of the possible landing sites for the {{Lunar Geophysical Network}} (LGN) mission proposed to arrive on the Moon in 2030 and to deploy packages at four locations to enable geophysical measurements for 6 - 10 years \cite{Haviland}.

Still, having a station or even an array of seismic stations at or near the antipodes would be ideal. Observed by such a station or stations, all events at distances less than 90 degrees from the antipodes could be confidently assigned to the far side. So we would recommend the near-antipodes zone (that close to the centre of the farside) as a high-priority landing site for some future mission, a perfect area to monitor the seismic activity on the far side and, especially, to observe if and how seismic waves proliferate through the base of the mantle.\\

\begin{figure}[htbp]
    \centering
    \includegraphics[width=\textwidth]{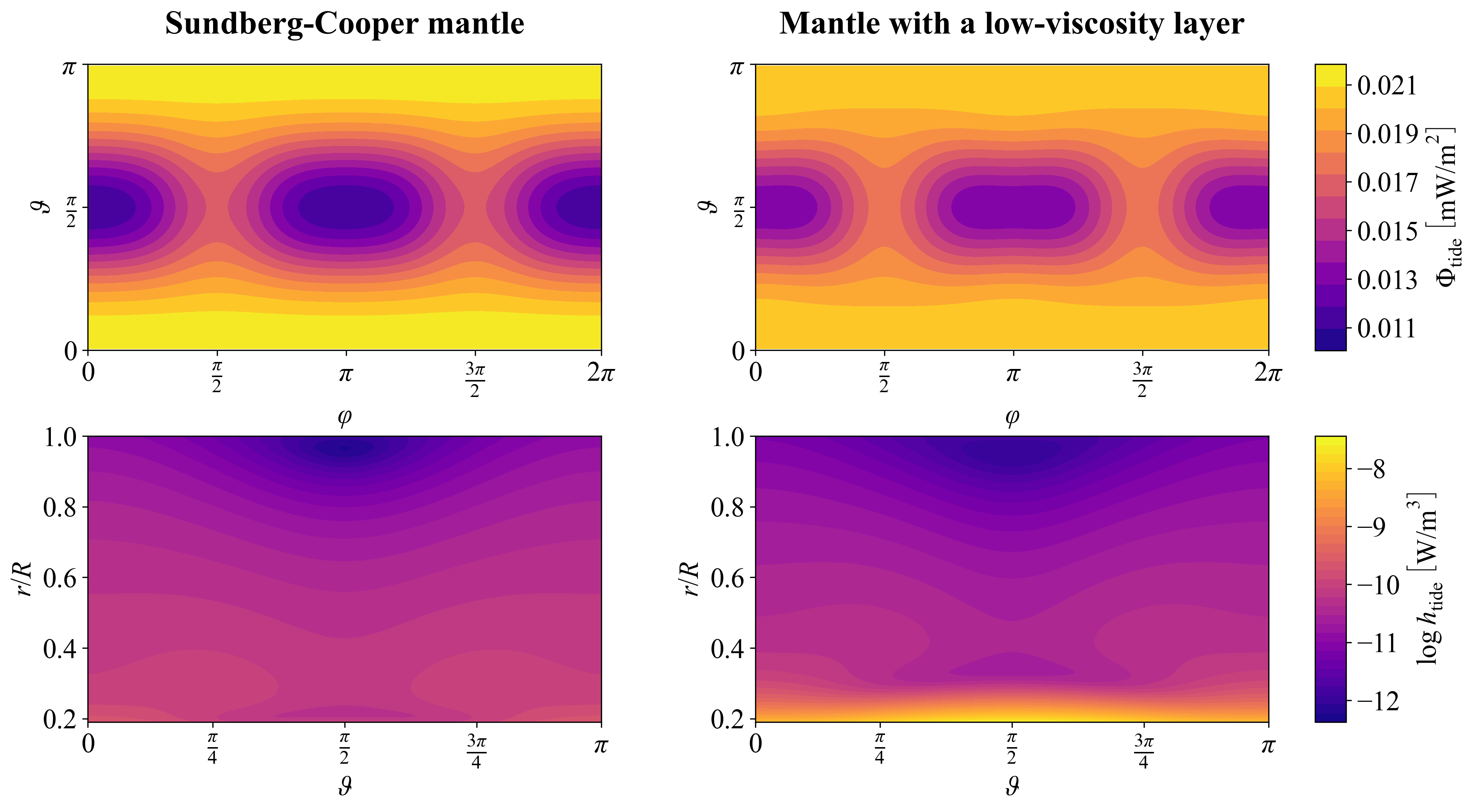}
    \caption{Average surface tidal heat flux (top) and volumetric tidal heating (bottom) for a specific realisation of each of the two models discussed in this work: the model considering elastically-accommodated GBS through the Sundberg-Cooper rheological model (Model 2, left) and the model with a basal low-viscosity zone (Model 3, right). In particular, the volumetric tidal heating is plotted as a function of relative radius $r/R$ and colatitude $\vartheta$ with longitude $\varphi$ equal to $0$.}
    \label{fig:heating_map}
\end{figure}

In addition to seismic measurements, and similarly to what is predicted for Jupiter's volcanic moon Io or for icy moons with subsurface oceans, the presence of a highly dissipative or a partially molten layer might be reflected in the tidal heating pattern on the lunar surface \cite<e.g.,>[]{segatz88,tobie05b}. However, as illustrated in the upper row of Figure \ref{fig:heating_map}, the positioning of the layer at the base of the mantle results in a very small difference between the surface heating patterns corresponding to the two alternative models. For samples with the same tidal response, both Models 2 and 3 show maxima of the average surface tidal heat flux $\Phi_{\rm{tide}}$ on the lunar poles and minima on the ``subterranean" point ($\varphi=0$) and its antipodes ($\varphi=\pi$). Moreover, the magnitude of $\Phi_{\rm{tide}}$ is generally very small, about three orders of magnitude lower than the flux produced by radiogenic heating of lunar interior \cite<e.g.,>[]{siegler2014}. The detection of any differences between the surface heat flux of the two models would be extremely challenging, if not impossible.

The lower row of Figure \ref{fig:heating_map} illustrates volumetric heat production due to tidal dissipation. As pointed out by \citeA{harada2014}, the presence of a low-viscosity zone at the base of the mantle results in a considerable local increase of tidal heating with respect to the rest of the mantle or to the model without the basal layer. While the tidal contribution to heat production in the high-viscosity parts of the mantle is around $\unit[10^{-11}]{W\;m^{-3}}$, the tidal heat production in the basal layer reaches $\sim\unit[10^{-8}]{W\;m^{-3}}$. For comparison, the global average of mantle heat production by all sources (radiogenic and tidal) is estimated to be $\unit[6.3\times10^{-9}]{W\;m^{-3}}$ \cite{siegler2014}. The predicted tidal dissipation in the basal layer can help to locally increase the temperature and exceed the solidus, especially if conductive heat transfer prevails in the lunar mantle. Combined with high enrichment of the basal layer in heat-producing elements, it may then contribute to maintaining the presence of melt.

Although virtually discarded at the beginning of this Subsection, let us nevertheless also discuss possible insights provided by future high-precision tidal measurements. At present, the tidal quality factor $Q$ and its frequency dependence are almost exclusively obtained from fitting the lunar physical librations, empirically determined by LLR. The only exception---to our knowledge---is the monthly $k_2/Q$ and $Q$ derived from the GRAIL data by \citeA{williams2015lpsc}. Future improvements in the satellite tracking \cite{dirkx2019,hu2022,stark2022} might provide new estimates of the tidal quality factors based on the lunar gravity field and help to further constrain their frequency dependence.

Among the quantities that we used in the inversion was degree-$3$ potential Love number $k_3$. This parameter is currently only known with a large error bar but its refinement would only help to discern between the two alternative models considered here if the elastically-accommodated GBS was contributing to the dissipation throughout the entire mantle (and not only in greater depths, as tentatively derived in Subsection \ref{subsec:disc_sc}). This is a consequence of a degree-dependent sensitivity of Love numbers to the interior structure. While degree-$2$ Love numbers and quality factors probe the lunar interior down to the core, higher-order quantities are only sensitive to shallower depths. The Love number $k_3$---or the quality factor $Q_3$---would thus not ``see" the basal low-viscosity layer, but it might sense complex tidal response in the upper mantle. As a result, the detection of the unexpected frequency dependence of tidal dissipation even in $Q_3$ (accompanied by a relatively high $k_3\sim0.01$) would clearly point at a mechanism acting in shallow depths.

Interestingly, the best-fitting samples of the two alternative models can be distinguished from each other relatively well. The main dissipation peaks associated with the basal layer in Model 3 emerge at high frequencies, beyond the monthly tidal frequency. Conversely, the Debye peak in Model 2 is, for most best-fitting samples, located between the monthly and the annual frequencies. This difference in the position of the secondary peak is also reflected in the magnitude of the elastic Love number $k_{2,\rm{e}}$, or the limit value of $k_{2}$ at high frequencies. For the best-fitting samples of Model 2, we see $k_{2,\rm{e}}=0.021-0.024$ at the frequency of $\unit[1]{Hz}$ (Figure \ref{fig:SC-OVERVIEW}). For the best-fitting samples of Model 3, it is $k_{2,\rm{e}}=0.0195-0.0205$ (Figure \ref{fig:Molten-OVERVIEW}). However, when considering all generated samples, Model 2 can attain much lower values of $k_{2,\rm{e}}$ than Model 3. For comparison, the value calculated by \citeA{Weber} for their seismic interior model is $k_{2,\rm{e}}=0.0232$. \citeA{williams2015} also derived lower bounds on the tidal $Q$ at the triennial and sexennial frequencies. When compared with the ensemble of our results for Models 2 and 3, only the samples with a dissipation peak between the monthly and the annual frequencies are permitted. For Model 3, this would imply a basal layer's viscosity of the order $\unit[10^{16}]{Pa\;s}$. Nevertheless, these constraints are model-dependent, and we chose not to use them to accept or reject samples.

Finally, we would like to note that any increase in the precision of $Q$ determination will greatly help in answering the question of whether any specific source of additional dissipation, be it a weak basal layer or elastic accommodation of strain at grain boundaries, is necessary in the first place. The existing empirical $Q$ or $k_2/Q$ at the monthly and the annual frequencies present an uncertainty between $10$ and $20\%$. Therefore, as we have also seen in Models 1 to 3, the tidal response of the Moon can still be fitted without the need for a secondary dissipation peak, although this often results in unusually small Andrade parameter $\alpha$ and unrealistically high attenuation of seismic waves in the lunar mantle.

\section{Conclusions} \label{sec:concl}

Tidal effects strongly depend not only on the interior density, viscosity, and rigidity profiles of celestial bodies, but also on the implied deformation mechanisms, which are reflected in the rheological models adopted. In this work, we attempted to illustrate that the unexpected frequency dependence of the tidal $Q$ measured by LLR \cite{williams2015} can be explained by lunar interior models both with and without a partially molten basal layer, and that each of the considered models leads to a different set of constraints on the interior properties.

As a first guess, we fitted the selenodetic parameters ($M$, MoIF; $k_2$, $k_3$, $h_2$, $Q$ at the monthly frequency, and $k_2/Q$ at the annual frequency) with a model consisting of a fluid core and a viscoelastic mantle governed by the Andrade rheology (Model 1). Within that model, we found a mantle viscosity of $\log\eta_{\rm{m}} [\unit{Pa\;s}] = 24.16^{+2.79}_{-2.82}$, mantle rigidity of $\mu_{\rm{m}} = \unit[80.30^{+6.37}_{-6.49}]{GPa}$, and the Andrade parameter $\alpha$ as low as $0.08^{+0.03}_{-0.02}$. The Andrade parameter $\zeta$ is anti-correlated with $\eta_{\rm{m}}$ and although it might attain a wide range of values, $\zeta<100$ seems more likely than $\zeta>100$. The predicted $\alpha$ is generally lower than reported in the literature \cite<0.1-0.4; e.g.,>[]{jackson2010,castillo2011,Efroimsky2012a,Efroimsky2012b}. This observation, along with seismological considerations, leads us to the conclusion that the tidal response of the Moon probably cannot be explained by the Andrade model alone and requires either a basal low-viscosity zone \cite<in line with the conclusion of>[]{khan2014} or an additional dissipation mechanism in the mantle \cite<similar to>[]{nimmo2012}.

Therefore, we fitted the selenodetic data with two more complex models and paid special attention to the best-fitting samples that exhibited a dissipation peak close to the monthly frequency. Both models are able to produce the same frequency dependence of the tidal parameters. In Model 2, consisting of a liquid core, an elastic crust, and a Sundberg-Cooper mantle, the fitting of the lunar tidal dissipation requires the relaxation time $\tau$ associated with elastically-accommodated GBS to be in the range from $3$ to $300$ hours, corresponding to a grain boundary viscosity between $10^{8}$ and $\unit[10^{14}]{Pa\; s}$ (the exact value depends on the grain size, which follows a uniform distribution). The relaxation strength $\Delta$ is then predicted to lie in the interval $[0.02, 0.25]$. For the Andrade parameter $\alpha$, all values in the considered interval $[0.1, 0.4]$ can be attained, and $\zeta$ follows a tendency similar to Model 1. We further obtain a mantle viscosity of $\log\eta_{\rm{m}} [\unit{Pa\;s}] = 23.87^{+2.49}_{-2.65}$ and a mantle rigidity $\mu_{\rm{m}}=\unit[72.02^{+3.97}_{-4.72}]{GPa}$. 

In Model 3, containing a liquid core, a low-rigidity basal layer, an Andrade mantle, and an elastic crust, the tidal parameters permit a wide range of basal layer thicknesses $D_{\rm{LVZ}}\in[0,350]\; \unit{km}$ and rigidities $\mu_{\rm{LVZ}}\in[0, \mu_{\rm{m}}]$. The predicted values of $\mu_{\rm{LVZ}}$ are consistent with elastic properties of all considered minerals (olivine, ilmenite, granite) and with a wide range of lower-mantle temperatures. For the basal layer viscosity $\eta_{\rm{LVZ}}$, we find two categories of samples providing the best fit to the observed frequency dependence of the tidal dissipation, along with the other selenodetic parameters: one with $\eta_{\rm{LVZ}}\sim\unit[10^{13}]{Pa\;s}$ and the other, preferred, with $\eta_{\rm{LVZ}}\sim\unit[10^{15}]{Pa\;s}$. We note that this result was obtained by fitting $Q$ at the monthly frequency and $k_2/Q$ at the annual frequency. Therefore, the derived basal layer viscosity in the second category is one order of magnitude smaller than reported by \citeA{Efroimsky2012a,Efroimsky2012b,harada2014,harada2016,matsumoto2015,tan2021}, and \citeA{kronrod2022}, who fitted $Q$ at both frequencies. A solution with $\eta_{\rm{LVZ}}\sim\unit[10^{16}]{Pa\;s}$ is, however, also possible, and it would be preferred if we also constrained our models by $Q$ at the triennial and sexennial frequencies or by the mantle seismic $Q$ at the frequency of $\unit[1]{Hz}$. The first category of the best-fitting samples, with $\eta_{\rm{LVZ}}\sim\unit[10^{13}]{Pa\;s}$, is three orders of magnitude smaller and results from the emergence of multiple peaks in a multilayered body. Nevertheless, none of these basal-layer viscosities is able to pose strong constraints on the lower-mantle temperature, owing to the large uncertainties on the rheological properties of lunar minerals. For the viscosity and rigidity of the overlying mantle, we get $\log\eta_{\rm{m}} [\unit{Pa\;s}] = 24.08^{+2.73}_{-2.77}$ and $\mu_{\rm{m}}=\unit[78.03^{+7.15}_{-5.85}]{GPa}$. As in the other two models, the exact value of viscosity depends on the Andrade parameter $\zeta$, which is likely smaller than $100$. Finally, the Andrade parameter $\alpha$ in Model 3 is typically small and almost all best-fitting samples have $\alpha<0.16$, although more realistic values are also possible.

The existence of a basal weak or possibly semi-molten layer in the mantles of terrestrial bodies has been recently also suggested for Mercury \cite{steinbruegge2021} and for Mars \cite{samuel2021}. In the case of Mercury, a lower mantle viscosity as low as $\unit[10^{13}]{Pa\; s}$ was proposed to match the latest measurements of the moment of inertia and of $k_2$; although this result was later critically reassessed by \citeA{goossens2022}, who showed that more realistic values around $\unit[10^{18}]{Pa\; s}$ might still explain the observations. In the case of Mars, the putative basal semi-molten layer was introduced by \citeA{samuel2021} to provide an alternative fit to seismic data which would not require the existence of a large core with an unexpectedly high concentration of light elements \cite<reported in>[]{staehler2021}. Lastly, large provinces of decreased shear seismic velocities also exist at the base of the Earth's mantle. These zones form a heterogeneous pattern in the deep terrestrial interior; however, according to numerical models, the formation of a continuous layer right above the core-mantle boundary is also possible for some values of model parameters \cite<e.g.,>[]{dannberg2021}. A new question thus arises: is a weak basal layer something common among terrestrial planet's mantles? Is it a natural and widely present outcome of magma ocean solidification and subsequent dynamical processes? Or is it merely a popular explanation of the data available?

Since the available tidal parameters were deemed insufficient to distinguish a weak basal layer above the lunar core from the manifestation of elastically accommodated GBS in the mantle, we conclude that an answer to the question stated in the title of our paper awaits future lunar seismic experiments (ideally with a uniform distribution of seismometers across the lunar surface) as well as a better understanding of elastic parameters of olivine-ilmenite assemblages near their melting point. Additionally, a tighter bound on the hypothetical basal layer parameters or on the strength and position of the secondary Debye peak in the alternative, Sundberg-Cooper model might be given by updated values of tidal $Q$ at multiple frequencies or by an independent inference of interior dissipation from the tidal phase lag and frequency-dependent $k_2$, theoretically measurable by laser altimetry or orbital tracking data \cite{dirkx2019,hu2022,stark2022}. A combination of all those sources of information will probably still not provide a bright picture of the deep lunar interior; however, it will help us to refute at least some of the many possible interior models.

\section*{Open Research}

The software developed for the calculation of selenodetic parameters of multi-layered bodies, the Python interface for running the MCMC inversion, and the plotting tools used for the figures presented in this study are available in \citeA{walterova2023data}.







\acknowledgments
We would like to thank the editor Laurent Mont\'{e}si and the two anonymous reviewers for their many constructive comments that were essential in shaping the present manuscript. Furthermore, our special gratitude goes to Amir Khan for reading the entire manuscript and providing numerous suggestions and comments, which helped us to improve the quality of the paper by reformulating the introduction and considering the additional constraint on mantle seismic $Q$. We are also grateful to James G. Williams, Mark P. Panning, and Alexander S. Konopliv for extremely helpful conversations on various aspects of lunar science. M.W. thanks Ana-Catalina Plesa and Martin Knapmeyer for discussions about the lunar interior and Philipp A. Baumeister for introducing her to the Python libraries used in this work. She also gratefully acknowledges the financial support and endorsement from the DLR Management Board Young Research Group Leader Program and the Executive Board Member for Space Research and Technology. M.B. received funding from Czech Science Foundation grant no. 22-20388S.


%
%



\nocite{williams2012}
\nocite{goossens2011}
\nocite{yan2012}
\nocite{frohlich2009}
\nocite{qin2012}

\bibliography{references}

\begin{thebibliography}{}

\bibitem [\protect \citeauthoryear {%
{Andrade}%
}{%
{Andrade}%
}{%
{\protect \APACyear {1910}}%
}]{%
andrade1910}
\APACinsertmetastar {%
andrade1910}%
\begin{APACrefauthors}%
{Andrade}, E\BPBI N\BPBI D\BPBI C.%
\end{APACrefauthors}%
\unskip\
\newblock
\APACrefYearMonthDay{1910}{{\APACmonth{06}}}{}.
\newblock
{\BBOQ}\APACrefatitle {{On the Viscous Flow in Metals, and Allied Phenomena}}
  {{On the Viscous Flow in Metals, and Allied Phenomena}}.{\BBCQ}
\newblock
\APACjournalVolNumPages{Proceedings of the Royal Society of London Series
  A}{84}{567}{1-12}.
\PrintBackRefs{\CurrentBib}

\bibitem [\protect \citeauthoryear {%
{Bagheri}%
\ \protect \BOthers {.}}{%
{Bagheri}%
\ \protect \BOthers {.}}{%
{\protect \APACyear {2022}}%
}]{%
TidalReview}
\APACinsertmetastar {%
TidalReview}%
\begin{APACrefauthors}%
{Bagheri}, A.%
, {Efroimsky}, M.%
, {Castillo-Rogez}, J.%
, {Goossens}, S.%
, {Plesa}, A\BHBI C.%
, {Rambaux}, N.%
\BDBL {}{Giardini}, D.%
\end{APACrefauthors}%
\unskip\
\newblock
\APACrefYearMonthDay{2022}{{\APACmonth{08}}}{}.
\newblock
{\BBOQ}\APACrefatitle {{Tidal insights into rocky and icy bodies: An
  introduction and overview}} {{Tidal insights into rocky and icy bodies: An
  introduction and overview}}.{\BBCQ}
\newblock
\APACjournalVolNumPages{Advances in Geophysics}{63}{}{231-320}.
\PrintBackRefs{\CurrentBib}

\bibitem [\protect \citeauthoryear {%
Bagheri%
, Khan%
, Al-Attar%
, Crawford%
\BCBL {}\ \BBA {} Giardini%
}{%
Bagheri%
\ \protect \BOthers {.}}{%
{\protect \APACyear {2019}}%
}]{%
Mars}
\APACinsertmetastar {%
Mars}%
\begin{APACrefauthors}%
Bagheri, A.%
, Khan, A.%
, Al-Attar, D.%
, Crawford, O.%
\BCBL {}\ \BBA {} Giardini, D.%
\end{APACrefauthors}%
\unskip\
\newblock
\APACrefYearMonthDay{2019}{}{}.
\newblock
{\BBOQ}\APACrefatitle {Tidal Response of Mars Constrained From Laboratory-Based
  Viscoelastic Dissipation Models and Geophysical Data} {Tidal response of mars
  constrained from laboratory-based viscoelastic dissipation models and
  geophysical data}.{\BBCQ}
\newblock
\APACjournalVolNumPages{Journal of Geophysical Research:
  Planets}{124}{11}{2703-2727}.
\newblock
\begin{APACrefURL}
  \url{https://agupubs.onlinelibrary.wiley.com/doi/abs/10.1029/2019JE006015}
  \end{APACrefURL}
\newblock
\begin{APACrefDOI} \doi{https://doi.org/10.1029/2019JE006015} \end{APACrefDOI}
\PrintBackRefs{\CurrentBib}

\bibitem [\protect \citeauthoryear {%
Bagheri%
\ \protect \BOthers {.}}{%
Bagheri%
\ \protect \BOthers {.}}{%
{\protect \APACyear {2022}}%
}]{%
BAGHERI2022114871}
\APACinsertmetastar {%
BAGHERI2022114871}%
\begin{APACrefauthors}%
Bagheri, A.%
, Khan, A.%
, Deschamps, F.%
, Samuel, H.%
, Kruglyakov, M.%
\BCBL {}\ \BBA {} Giardini, D.%
\end{APACrefauthors}%
\unskip\
\newblock
\APACrefYearMonthDay{2022}{}{}.
\newblock
{\BBOQ}\APACrefatitle {The tidal-thermal evolution of the Pluto-Charon system}
  {The tidal-thermal evolution of the pluto-charon system}.{\BBCQ}
\newblock
\APACjournalVolNumPages{Icarus}{376}{}{114871}.
\newblock
\begin{APACrefURL}
  \url{https://www.sciencedirect.com/science/article/pii/S001910352100508X}
  \end{APACrefURL}
\newblock
\begin{APACrefDOI} \doi{https://doi.org/10.1016/j.icarus.2021.114871}
  \end{APACrefDOI}
\PrintBackRefs{\CurrentBib}

\bibitem [\protect \citeauthoryear {%
{Barnhoorn}%
, {Jackson}%
, {Fitz Gerald}%
, {Kishimoto}%
\BCBL {}\ \BBA {} {Itatani}%
}{%
{Barnhoorn}%
\ \protect \BOthers {.}}{%
{\protect \APACyear {2016}}%
}]{%
barnhoorn2016}
\APACinsertmetastar {%
barnhoorn2016}%
\begin{APACrefauthors}%
{Barnhoorn}, A.%
, {Jackson}, I.%
, {Fitz Gerald}, J\BPBI D.%
, {Kishimoto}, A.%
\BCBL {}\ \BBA {} {Itatani}, K.%
\end{APACrefauthors}%
\unskip\
\newblock
\APACrefYearMonthDay{2016}{{\APACmonth{07}}}{}.
\newblock
{\BBOQ}\APACrefatitle {{Grain size-sensitive viscoelastic relaxation and
  seismic properties of polycrystalline MgO}} {{Grain size-sensitive
  viscoelastic relaxation and seismic properties of polycrystalline
  MgO}}.{\BBCQ}
\newblock
\APACjournalVolNumPages{Journal of Geophysical Research (Solid
  Earth)}{121}{7}{4955-4976}.
\newblock
\begin{APACrefDOI} \doi{10.1002/2016JB013126} \end{APACrefDOI}
\PrintBackRefs{\CurrentBib}

\bibitem [\protect \citeauthoryear {%
Biot%
}{%
Biot%
}{%
{\protect \APACyear {1954}}%
}]{%
biot}
\APACinsertmetastar {%
biot}%
\begin{APACrefauthors}%
Biot, M\BPBI A.%
\end{APACrefauthors}%
\unskip\
\newblock
\APACrefYearMonthDay{1954}{}{}.
\newblock
{\BBOQ}\APACrefatitle {Theory of stress-strain relations in anisotropic
  viscoelasticity and relaxation phenomena} {Theory of stress-strain relations
  in anisotropic viscoelasticity and relaxation phenomena}.{\BBCQ}
\newblock
\APACjournalVolNumPages{Journal of Applied Physics}{25}{11}{1385--1391}.
\PrintBackRefs{\CurrentBib}

\bibitem [\protect \citeauthoryear {%
{Bolmont}%
\ \protect \BOthers {.}}{%
{Bolmont}%
\ \protect \BOthers {.}}{%
{\protect \APACyear {2020}}%
}]{%
bolmont2020}
\APACinsertmetastar {%
bolmont2020}%
\begin{APACrefauthors}%
{Bolmont}, E.%
, {Breton}, S\BPBI N.%
, {Tobie}, G.%
, {Dumoulin}, C.%
, {Mathis}, S.%
\BCBL {}\ \BBA {} {Grasset}, O.%
\end{APACrefauthors}%
\unskip\
\newblock
\APACrefYearMonthDay{2020}{{\APACmonth{12}}}{}.
\newblock
{\BBOQ}\APACrefatitle {{Solid tidal friction in multi-layer planets:
  Application to Earth, Venus, a Super Earth and the TRAPPIST-1 planets.
  Potential approximation of a multi-layer planet as a homogeneous body}}
  {{Solid tidal friction in multi-layer planets: Application to Earth, Venus, a
  Super Earth and the TRAPPIST-1 planets. Potential approximation of a
  multi-layer planet as a homogeneous body}}.{\BBCQ}
\newblock
\APACjournalVolNumPages{Astronomy \& Astrophysics}{644}{}{A165}.
\newblock
\begin{APACrefDOI} \doi{10.1051/0004-6361/202038204} \end{APACrefDOI}
\PrintBackRefs{\CurrentBib}

\bibitem [\protect \citeauthoryear {%
{Bou{\'e}}%
\ \BBA {} {Efroimsky}%
}{%
{Bou{\'e}}%
\ \BBA {} {Efroimsky}%
}{%
{\protect \APACyear {2019}}%
}]{%
BoueEfroimsky2019}
\APACinsertmetastar {%
BoueEfroimsky2019}%
\begin{APACrefauthors}%
{Bou{\'e}}, G.%
\BCBT {}\ \BBA {} {Efroimsky}, M.%
\end{APACrefauthors}%
\unskip\
\newblock
\APACrefYearMonthDay{2019}{}{}.
\newblock
{\BBOQ}\APACrefatitle {Tidal evolution of the Keplerian elements} {Tidal
  evolution of the keplerian elements}.{\BBCQ}
\newblock
\APACjournalVolNumPages{Celestial Mechanics and Dynamical
  Astronomy}{131}{}{30}.
\newblock
\begin{APACrefDOI} \doi{10.1007/s10569-019-9908-2} \end{APACrefDOI}
\PrintBackRefs{\CurrentBib}

\bibitem [\protect \citeauthoryear {%
{Briaud}%
, {Fienga}%
\BCBL {}\ \protect \BOthers {.}}{%
{Briaud}%
, {Fienga}%
\BCBL {}\ \protect \BOthers {.}}{%
{\protect \APACyear {2023}}%
}]{%
briaud2023}
\APACinsertmetastar {%
briaud2023}%
\begin{APACrefauthors}%
{Briaud}, A.%
, {Fienga}, A.%
, {Melini}, D.%
, {Rambaux}, N.%
, {M{\'e}min}, A.%
, {Spada}, G.%
\BDBL {}{Baguet}, D.%
\end{APACrefauthors}%
\unskip\
\newblock
\APACrefYearMonthDay{2023}{{\APACmonth{04}}}{}.
\newblock
{\BBOQ}\APACrefatitle {{Constraints on the lunar core viscosity from tidal
  deformation}} {{Constraints on the lunar core viscosity from tidal
  deformation}}.{\BBCQ}
\newblock
\APACjournalVolNumPages{\icarus}{394}{}{115426}.
\newblock
\begin{APACrefDOI} \doi{10.1016/j.icarus.2023.115426} \end{APACrefDOI}
\PrintBackRefs{\CurrentBib}

\bibitem [\protect \citeauthoryear {%
{Briaud}%
, {Ganino}%
, {Fienga}%
, {M{\'e}min}%
\BCBL {}\ \BBA {} {Rambaux}%
}{%
{Briaud}%
, {Ganino}%
\BCBL {}\ \protect \BOthers {.}}{%
{\protect \APACyear {2023}}%
}]{%
briaud2023nat}
\APACinsertmetastar {%
briaud2023nat}%
\begin{APACrefauthors}%
{Briaud}, A.%
, {Ganino}, C.%
, {Fienga}, A.%
, {M{\'e}min}, A.%
\BCBL {}\ \BBA {} {Rambaux}, N.%
\end{APACrefauthors}%
\unskip\
\newblock
\APACrefYearMonthDay{2023}{{\APACmonth{05}}}{}.
\newblock
{\BBOQ}\APACrefatitle {{The lunar solid inner core and the mantle overturn}}
  {{The lunar solid inner core and the mantle overturn}}.{\BBCQ}
\newblock
\APACjournalVolNumPages{\nat}{617}{7962}{743-746}.
\newblock
\begin{APACrefDOI} \doi{10.1038/s41586-023-05935-7} \end{APACrefDOI}
\PrintBackRefs{\CurrentBib}

\bibitem [\protect \citeauthoryear {%
{Castillo-Rogez}%
, {Efroimsky}%
\BCBL {}\ \BBA {} {Lainey}%
}{%
{Castillo-Rogez}%
\ \protect \BOthers {.}}{%
{\protect \APACyear {2011}}%
}]{%
castillo2011}
\APACinsertmetastar {%
castillo2011}%
\begin{APACrefauthors}%
{Castillo-Rogez}, J\BPBI C.%
, {Efroimsky}, M.%
\BCBL {}\ \BBA {} {Lainey}, V.%
\end{APACrefauthors}%
\unskip\
\newblock
\APACrefYearMonthDay{2011}{{\APACmonth{09}}}{}.
\newblock
{\BBOQ}\APACrefatitle {{The tidal history of Iapetus: Spin dynamics in the
  light of a refined dissipation model}} {{The tidal history of Iapetus: Spin
  dynamics in the light of a refined dissipation model}}.{\BBCQ}
\newblock
\APACjournalVolNumPages{Journal of Geophysical Research
  (Planets)}{116}{E9}{E09008}.
\newblock
\begin{APACrefDOI} \doi{10.1029/2010JE003664} \end{APACrefDOI}
\PrintBackRefs{\CurrentBib}

\bibitem [\protect \citeauthoryear {%
{Dannberg}%
, {Myhill}%
, {Gassm{\"o}ller}%
\BCBL {}\ \BBA {} {Cottaar}%
}{%
{Dannberg}%
\ \protect \BOthers {.}}{%
{\protect \APACyear {2021}}%
}]{%
dannberg2021}
\APACinsertmetastar {%
dannberg2021}%
\begin{APACrefauthors}%
{Dannberg}, J.%
, {Myhill}, R.%
, {Gassm{\"o}ller}, R.%
\BCBL {}\ \BBA {} {Cottaar}, S.%
\end{APACrefauthors}%
\unskip\
\newblock
\APACrefYearMonthDay{2021}{{\APACmonth{11}}}{}.
\newblock
{\BBOQ}\APACrefatitle {{The morphology, evolution and seismic visibility of
  partial melt at the core-mantle boundary: implications for ULVZs}} {{The
  morphology, evolution and seismic visibility of partial melt at the
  core-mantle boundary: implications for ULVZs}}.{\BBCQ}
\newblock
\APACjournalVolNumPages{Geophysical Journal International}{227}{2}{1028-1059}.
\newblock
\begin{APACrefDOI} \doi{10.1093/gji/ggab242} \end{APACrefDOI}
\PrintBackRefs{\CurrentBib}

\bibitem [\protect \citeauthoryear {%
{Darwin}%
}{%
{Darwin}%
}{%
{\protect \APACyear {1879}}%
}]{%
darwin}
\APACinsertmetastar {%
darwin}%
\begin{APACrefauthors}%
{Darwin}, G\BPBI H.%
\end{APACrefauthors}%
\unskip\
\newblock
\APACrefYearMonthDay{1879}{{\APACmonth{01}}}{}.
\newblock
{\BBOQ}\APACrefatitle {{On the Analytical Expressions Which Give the History of
  a Fluid Planet of Small Viscosity, Attended by a Single Satellite}} {{On the
  Analytical Expressions Which Give the History of a Fluid Planet of Small
  Viscosity, Attended by a Single Satellite}}.{\BBCQ}
\newblock
\APACjournalVolNumPages{Proceedings of the Royal Society of London Series
  I}{30}{}{255-278}.
\PrintBackRefs{\CurrentBib}

\bibitem [\protect \citeauthoryear {%
{Dirkx}%
\ \protect \BOthers {.}}{%
{Dirkx}%
\ \protect \BOthers {.}}{%
{\protect \APACyear {2019}}%
}]{%
dirkx2019}
\APACinsertmetastar {%
dirkx2019}%
\begin{APACrefauthors}%
{Dirkx}, D.%
, {Prochazka}, I.%
, {Bauer}, S.%
, {Visser}, P.%
, {Noomen}, R.%
, {Gurvits}, L\BPBI I.%
\BCBL {}\ \BBA {} {Vermeersen}, B.%
\end{APACrefauthors}%
\unskip\
\newblock
\APACrefYearMonthDay{2019}{{\APACmonth{11}}}{}.
\newblock
{\BBOQ}\APACrefatitle {{Laser and radio tracking for planetary science
  missions{\textemdash}a comparison}} {{Laser and radio tracking for planetary
  science missions{\textemdash}a comparison}}.{\BBCQ}
\newblock
\APACjournalVolNumPages{Journal of Geodesy}{93}{11}{2405-2420}.
\newblock
\begin{APACrefDOI} \doi{10.1007/s00190-018-1171-x} \end{APACrefDOI}
\PrintBackRefs{\CurrentBib}

\bibitem [\protect \citeauthoryear {%
{Dumoulin}%
, {Tobie}%
, {Verhoeven}%
, {Rosenblatt}%
\BCBL {}\ \BBA {} {Rambaux}%
}{%
{Dumoulin}%
\ \protect \BOthers {.}}{%
{\protect \APACyear {2017}}%
}]{%
dumoulin2017}
\APACinsertmetastar {%
dumoulin2017}%
\begin{APACrefauthors}%
{Dumoulin}, C.%
, {Tobie}, G.%
, {Verhoeven}, O.%
, {Rosenblatt}, P.%
\BCBL {}\ \BBA {} {Rambaux}, N.%
\end{APACrefauthors}%
\unskip\
\newblock
\APACrefYearMonthDay{2017}{{\APACmonth{06}}}{}.
\newblock
{\BBOQ}\APACrefatitle {{Tidal constraints on the interior of Venus}} {{Tidal
  constraints on the interior of Venus}}.{\BBCQ}
\newblock
\APACjournalVolNumPages{Journal of Geophysical Research
  (Planets)}{122}{6}{1338-1352}.
\newblock
\begin{APACrefDOI} \doi{10.1002/2016JE005249} \end{APACrefDOI}
\PrintBackRefs{\CurrentBib}

\bibitem [\protect \citeauthoryear {%
Dygert%
, Hirth%
\BCBL {}\ \BBA {} Liang%
}{%
Dygert%
\ \protect \BOthers {.}}{%
{\protect \APACyear {2016}}%
}]{%
dygert2016}
\APACinsertmetastar {%
dygert2016}%
\begin{APACrefauthors}%
Dygert, N.%
, Hirth, G.%
\BCBL {}\ \BBA {} Liang, Y.%
\end{APACrefauthors}%
\unskip\
\newblock
\APACrefYearMonthDay{2016}{}{}.
\newblock
{\BBOQ}\APACrefatitle {A flow law for ilmenite in dislocation creep:
  Implications for lunar cumulate mantle overturn} {A flow law for ilmenite in
  dislocation creep: Implications for lunar cumulate mantle overturn}.{\BBCQ}
\newblock
\APACjournalVolNumPages{Geophysical Research Letters}{43}{2}{532-540}.
\newblock
\begin{APACrefURL}
  \url{https://agupubs.onlinelibrary.wiley.com/doi/abs/10.1002/2015GL066546}
  \end{APACrefURL}
\newblock
\begin{APACrefDOI} \doi{https://doi.org/10.1002/2015GL066546} \end{APACrefDOI}
\PrintBackRefs{\CurrentBib}

\bibitem [\protect \citeauthoryear {%
{Efroimsky}%
}{%
{Efroimsky}%
}{%
{\protect \APACyear {2012}}%
{\protect \APACexlab {{\protect \BCnt {1}}}}}]{%
Efroimsky2012a}
\APACinsertmetastar {%
Efroimsky2012a}%
\begin{APACrefauthors}%
{Efroimsky}, M.%
\end{APACrefauthors}%
\unskip\
\newblock
\APACrefYearMonthDay{2012{\protect \BCnt {1}}}{{\APACmonth{03}}}{}.
\newblock
{\BBOQ}\APACrefatitle {{Bodily tides near spin-orbit resonances}} {{Bodily
  tides near spin-orbit resonances}}.{\BBCQ}
\newblock
\APACjournalVolNumPages{Celestial Mechanics and Dynamical
  Astronomy}{112}{3}{283-330}.
\newblock
\begin{APACrefDOI} \doi{10.1007/s10569-011-9397-4} \end{APACrefDOI}
\PrintBackRefs{\CurrentBib}

\bibitem [\protect \citeauthoryear {%
{Efroimsky}%
}{%
{Efroimsky}%
}{%
{\protect \APACyear {2012}}%
{\protect \APACexlab {{\protect \BCnt {2}}}}}]{%
Efroimsky2012b}
\APACinsertmetastar {%
Efroimsky2012b}%
\begin{APACrefauthors}%
{Efroimsky}, M.%
\end{APACrefauthors}%
\unskip\
\newblock
\APACrefYearMonthDay{2012{\protect \BCnt {2}}}{{\APACmonth{02}}}{}.
\newblock
{\BBOQ}\APACrefatitle {{Tidal Dissipation Compared to Seismic Dissipation: In
  Small Bodies, Earths, and Super-Earths}} {{Tidal Dissipation Compared to
  Seismic Dissipation: In Small Bodies, Earths, and Super-Earths}}.{\BBCQ}
\newblock
\APACjournalVolNumPages{The Astrophysical Journal}{746}{2}{150}.
\newblock
\begin{APACrefDOI} \doi{10.1088/0004-637X/746/2/150} \end{APACrefDOI}
\PrintBackRefs{\CurrentBib}

\bibitem [\protect \citeauthoryear {%
{Efroimsky}%
}{%
{Efroimsky}%
}{%
{\protect \APACyear {2015}}%
}]{%
Efroimsky2015}
\APACinsertmetastar {%
Efroimsky2015}%
\begin{APACrefauthors}%
{Efroimsky}, M.%
\end{APACrefauthors}%
\unskip\
\newblock
\APACrefYearMonthDay{2015}{{\APACmonth{10}}}{}.
\newblock
{\BBOQ}\APACrefatitle {{Tidal Evolution of Asteroidal Binaries. Ruled by
  Viscosity. Ignorant of Rigidity.}} {{Tidal Evolution of Asteroidal Binaries.
  Ruled by Viscosity. Ignorant of Rigidity.}}{\BBCQ}
\newblock
\APACjournalVolNumPages{The Astronomical Journal}{150}{4}{98}.
\newblock
\begin{APACrefDOI} \doi{10.1088/0004-6256/150/4/98} \end{APACrefDOI}
\PrintBackRefs{\CurrentBib}

\bibitem [\protect \citeauthoryear {%
{Efroimsky}%
\ \BBA {} {Makarov}%
}{%
{Efroimsky}%
\ \BBA {} {Makarov}%
}{%
{\protect \APACyear {2013}}%
}]{%
EfroimskyMakarov2013}
\APACinsertmetastar {%
EfroimskyMakarov2013}%
\begin{APACrefauthors}%
{Efroimsky}, M.%
\BCBT {}\ \BBA {} {Makarov}, V\BPBI V.%
\end{APACrefauthors}%
\unskip\
\newblock
\APACrefYearMonthDay{2013}{}{}.
\newblock
{\BBOQ}\APACrefatitle {Tidal Friction and Tidal Lagging. Applicability
  Limitations of a Popular Formula for the Tidal Torque} {Tidal friction and
  tidal lagging. applicability limitations of a popular formula for the tidal
  torque}.{\BBCQ}
\newblock
\APACjournalVolNumPages{The Astrophysical Journal}{764}{}{26}.
\newblock
\begin{APACrefDOI} \doi{10.1088/0004-637X/764/1/26} \end{APACrefDOI}
\PrintBackRefs{\CurrentBib}

\bibitem [\protect \citeauthoryear {%
{Efroimsky}%
\ \BBA {} {Makarov}%
}{%
{Efroimsky}%
\ \BBA {} {Makarov}%
}{%
{\protect \APACyear {2014}}%
}]{%
EfroimskyMakarov2014}
\APACinsertmetastar {%
EfroimskyMakarov2014}%
\begin{APACrefauthors}%
{Efroimsky}, M.%
\BCBT {}\ \BBA {} {Makarov}, V\BPBI V.%
\end{APACrefauthors}%
\unskip\
\newblock
\APACrefYearMonthDay{2014}{{\APACmonth{11}}}{}.
\newblock
{\BBOQ}\APACrefatitle {{Tidal Dissipation in a Homogeneous Spherical Body. I.
  Methods}} {{Tidal Dissipation in a Homogeneous Spherical Body. I.
  Methods}}.{\BBCQ}
\newblock
\APACjournalVolNumPages{The Astrophysical Journal}{795}{1}{6}.
\newblock
\begin{APACrefDOI} \doi{10.1088/0004-637X/795/1/6} \end{APACrefDOI}
\PrintBackRefs{\CurrentBib}

\bibitem [\protect \citeauthoryear {%
{Foreman-Mackey}%
, {Hogg}%
, {Lang}%
\BCBL {}\ \BBA {} {Goodman}%
}{%
{Foreman-Mackey}%
\ \protect \BOthers {.}}{%
{\protect \APACyear {2013}}%
}]{%
foreman2013}
\APACinsertmetastar {%
foreman2013}%
\begin{APACrefauthors}%
{Foreman-Mackey}, D.%
, {Hogg}, D\BPBI W.%
, {Lang}, D.%
\BCBL {}\ \BBA {} {Goodman}, J.%
\end{APACrefauthors}%
\unskip\
\newblock
\APACrefYearMonthDay{2013}{{\APACmonth{03}}}{}.
\newblock
{\BBOQ}\APACrefatitle {{emcee: The MCMC Hammer}} {{emcee: The MCMC
  Hammer}}.{\BBCQ}
\newblock
\APACjournalVolNumPages{Publications of the Astronomical Society of the
  Pacific}{125}{925}{306}.
\newblock
\begin{APACrefDOI} \doi{10.1086/670067} \end{APACrefDOI}
\PrintBackRefs{\CurrentBib}

\bibitem [\protect \citeauthoryear {%
Foreman-Mackey%
}{%
Foreman-Mackey%
}{%
{\protect \APACyear {2016}}%
}]{%
corner}
\APACinsertmetastar {%
corner}%
\begin{APACrefauthors}%
Foreman-Mackey, D.%
\end{APACrefauthors}%
\unskip\
\newblock
\APACrefYearMonthDay{2016}{jun}{}.
\newblock
{\BBOQ}\APACrefatitle {corner.py: Scatterplot matrices in Python} {corner.py:
  Scatterplot matrices in python}.{\BBCQ}
\newblock
\APACjournalVolNumPages{The Journal of Open Source Software}{1}{2}{24}.
\newblock
\begin{APACrefURL} \url{https://doi.org/10.21105/joss.00024} \end{APACrefURL}
\newblock
\begin{APACrefDOI} \doi{10.21105/joss.00024} \end{APACrefDOI}
\PrintBackRefs{\CurrentBib}

\bibitem [\protect \citeauthoryear {%
{Frohlich}%
\ \BBA {} {Nakamura}%
}{%
{Frohlich}%
\ \BBA {} {Nakamura}%
}{%
{\protect \APACyear {2009}}%
}]{%
frohlich2009}
\APACinsertmetastar {%
frohlich2009}%
\begin{APACrefauthors}%
{Frohlich}, C.%
\BCBT {}\ \BBA {} {Nakamura}, Y.%
\end{APACrefauthors}%
\unskip\
\newblock
\APACrefYearMonthDay{2009}{{\APACmonth{04}}}{}.
\newblock
{\BBOQ}\APACrefatitle {{The physical mechanisms of deep moonquakes and
  intermediate-depth earthquakes: How similar and how different?}} {{The
  physical mechanisms of deep moonquakes and intermediate-depth earthquakes:
  How similar and how different?}}{\BBCQ}
\newblock
\APACjournalVolNumPages{Physics of the Earth and Planetary
  Interiors}{173}{3-4}{365-374}.
\newblock
\begin{APACrefDOI} \doi{10.1016/j.pepi.2009.02.004} \end{APACrefDOI}
\PrintBackRefs{\CurrentBib}

\bibitem [\protect \citeauthoryear {%
{Fuqua Haviland}%
\ \protect \BOthers {.}}{%
{Fuqua Haviland}%
\ \protect \BOthers {.}}{%
{\protect \APACyear {2022}}%
}]{%
Haviland}
\APACinsertmetastar {%
Haviland}%
\begin{APACrefauthors}%
{Fuqua Haviland}, H.%
, {Weber}, R\BPBI C.%
, {Neal}, C\BPBI R.%
, {Lognonn{\'e}}, P.%
, {Garcia}, R\BPBI F.%
, {Schmerr}, N.%
\BDBL {}{Bremner}, P\BPBI M.%
\end{APACrefauthors}%
\unskip\
\newblock
\APACrefYearMonthDay{2022}{{\APACmonth{02}}}{}.
\newblock
{\BBOQ}\APACrefatitle {{The Lunar Geophysical Network Landing Sites Science
  Rationale}} {{The Lunar Geophysical Network Landing Sites Science
  Rationale}}.{\BBCQ}
\newblock
\APACjournalVolNumPages{The Planetary Science Journal}{3}{2}{40}.
\newblock
\begin{APACrefDOI} \doi{10.3847/PSJ/ac0f82} \end{APACrefDOI}
\PrintBackRefs{\CurrentBib}

\bibitem [\protect \citeauthoryear {%
{Garcia}%
, {Gagnepain-Beyneix}%
, {Chevrot}%
\BCBL {}\ \BBA {} {Lognonn{\'e}}%
}{%
{Garcia}%
\ \protect \BOthers {.}}{%
{\protect \APACyear {2011}}%
}]{%
garcia2011}
\APACinsertmetastar {%
garcia2011}%
\begin{APACrefauthors}%
{Garcia}, R\BPBI F.%
, {Gagnepain-Beyneix}, J.%
, {Chevrot}, S.%
\BCBL {}\ \BBA {} {Lognonn{\'e}}, P.%
\end{APACrefauthors}%
\unskip\
\newblock
\APACrefYearMonthDay{2011}{{\APACmonth{09}}}{}.
\newblock
{\BBOQ}\APACrefatitle {{Very preliminary reference Moon model}} {{Very
  preliminary reference Moon model}}.{\BBCQ}
\newblock
\APACjournalVolNumPages{Physics of the Earth and Planetary
  Interiors}{188}{1}{96-113}.
\newblock
\begin{APACrefDOI} \doi{10.1016/j.pepi.2011.06.015} \end{APACrefDOI}
\PrintBackRefs{\CurrentBib}

\bibitem [\protect \citeauthoryear {%
{Garcia}%
\ \protect \BOthers {.}}{%
{Garcia}%
\ \protect \BOthers {.}}{%
{\protect \APACyear {2019}}%
}]{%
garcia2019}
\APACinsertmetastar {%
garcia2019}%
\begin{APACrefauthors}%
{Garcia}, R\BPBI F.%
, {Khan}, A.%
, {Drilleau}, M.%
, {Margerin}, L.%
, {Kawamura}, T.%
, {Sun}, D.%
\BDBL {}{Zhu}, P.%
\end{APACrefauthors}%
\unskip\
\newblock
\APACrefYearMonthDay{2019}{{\APACmonth{11}}}{}.
\newblock
{\BBOQ}\APACrefatitle {{Lunar Seismology: An Update on Interior Structure
  Models}} {{Lunar Seismology: An Update on Interior Structure Models}}.{\BBCQ}
\newblock
\APACjournalVolNumPages{Space Science Reviews}{215}{8}{50}.
\newblock
\begin{APACrefDOI} \doi{10.1007/s11214-019-0613-y} \end{APACrefDOI}
\PrintBackRefs{\CurrentBib}

\bibitem [\protect \citeauthoryear {%
{Gerstenkorn}%
}{%
{Gerstenkorn}%
}{%
{\protect \APACyear {1967}}%
}]{%
Gerstenkorn}
\APACinsertmetastar {%
Gerstenkorn}%
\begin{APACrefauthors}%
{Gerstenkorn}, H.%
\end{APACrefauthors}%
\unskip\
\newblock
\APACrefYearMonthDay{1967}{{\APACmonth{01}}}{}.
\newblock
{\BBOQ}\APACrefatitle {{The Earth as a Maxwell body}} {{The Earth as a Maxwell
  body}}.{\BBCQ}
\newblock
\APACjournalVolNumPages{\icarus}{6}{1-3}{92-99}.
\newblock
\begin{APACrefDOI} \doi{10.1016/0019-1035(67)90006-1} \end{APACrefDOI}
\PrintBackRefs{\CurrentBib}

\bibitem [\protect \citeauthoryear {%
{Gevorgyan}%
}{%
{Gevorgyan}%
}{%
{\protect \APACyear {2021}}%
}]{%
gevorgyan2021}
\APACinsertmetastar {%
gevorgyan2021}%
\begin{APACrefauthors}%
{Gevorgyan}, Y.%
\end{APACrefauthors}%
\unskip\
\newblock
\APACrefYearMonthDay{2021}{{\APACmonth{06}}}{}.
\newblock
{\BBOQ}\APACrefatitle {{Homogeneous model for the TRAPPIST-1e planet with an
  icy layer}} {{Homogeneous model for the TRAPPIST-1e planet with an icy
  layer}}.{\BBCQ}
\newblock
\APACjournalVolNumPages{Astronomy \& Astrophysics}{650}{}{A141}.
\newblock
\begin{APACrefDOI} \doi{10.1051/0004-6361/202140736} \end{APACrefDOI}
\PrintBackRefs{\CurrentBib}

\bibitem [\protect \citeauthoryear {%
{Gevorgyan}%
, {Matsuyama}%
\BCBL {}\ \BBA {} {Ragazzo}%
}{%
{Gevorgyan}%
\ \protect \BOthers {.}}{%
{\protect \APACyear {2023}}%
}]{%
gevorgyan2023}
\APACinsertmetastar {%
gevorgyan2023}%
\begin{APACrefauthors}%
{Gevorgyan}, Y.%
, {Matsuyama}, I.%
\BCBL {}\ \BBA {} {Ragazzo}, C.%
\end{APACrefauthors}%
\unskip\
\newblock
\APACrefYearMonthDay{2023}{{\APACmonth{03}}}{}.
\newblock
{\BBOQ}\APACrefatitle {{Equivalence between simple multilayered and homogeneous
  laboratory-based rheological models in planetary science}} {{Equivalence
  between simple multilayered and homogeneous laboratory-based rheological
  models in planetary science}}.{\BBCQ}
\newblock
\APACjournalVolNumPages{arXiv e-prints}{}{}{arXiv:2303.05253}.
\newblock
\begin{APACrefDOI} \doi{10.48550/arXiv.2303.05253} \end{APACrefDOI}
\PrintBackRefs{\CurrentBib}

\bibitem [\protect \citeauthoryear {%
Ghahremani%
}{%
Ghahremani%
}{%
{\protect \APACyear {1980}}%
}]{%
ghahremani1980}
\APACinsertmetastar {%
ghahremani1980}%
\begin{APACrefauthors}%
Ghahremani, F.%
\end{APACrefauthors}%
\unskip\
\newblock
\APACrefYearMonthDay{1980}{}{}.
\newblock
{\BBOQ}\APACrefatitle {Effect of grain boundary sliding on anelasticity of
  polycrystals} {Effect of grain boundary sliding on anelasticity of
  polycrystals}.{\BBCQ}
\newblock
\APACjournalVolNumPages{International Journal of Solids and
  Structures}{16}{9}{825-845}.
\newblock
\begin{APACrefURL}
  \url{https://www.sciencedirect.com/science/article/pii/0020768380900529}
  \end{APACrefURL}
\newblock
\begin{APACrefDOI} \doi{https://doi.org/10.1016/0020-7683(80)90052-9}
  \end{APACrefDOI}
\PrintBackRefs{\CurrentBib}

\bibitem [\protect \citeauthoryear {%
{Gillet}%
, {Margerin}%
, {Calvet}%
\BCBL {}\ \BBA {} {Monnereau}%
}{%
{Gillet}%
\ \protect \BOthers {.}}{%
{\protect \APACyear {2017}}%
}]{%
gillet2017}
\APACinsertmetastar {%
gillet2017}%
\begin{APACrefauthors}%
{Gillet}, K.%
, {Margerin}, L.%
, {Calvet}, M.%
\BCBL {}\ \BBA {} {Monnereau}, M.%
\end{APACrefauthors}%
\unskip\
\newblock
\APACrefYearMonthDay{2017}{{\APACmonth{01}}}{}.
\newblock
{\BBOQ}\APACrefatitle {{Scattering attenuation profile of the Moon:
  Implications for shallow moonquakes and the structure of the megaregolith}}
  {{Scattering attenuation profile of the Moon: Implications for shallow
  moonquakes and the structure of the megaregolith}}.{\BBCQ}
\newblock
\APACjournalVolNumPages{Physics of the Earth and Planetary
  Interiors}{262}{}{28-40}.
\newblock
\begin{APACrefDOI} \doi{10.1016/j.pepi.2016.11.001} \end{APACrefDOI}
\PrintBackRefs{\CurrentBib}

\bibitem [\protect \citeauthoryear {%
{Goodman}%
\ \BBA {} {Weare}%
}{%
{Goodman}%
\ \BBA {} {Weare}%
}{%
{\protect \APACyear {2010}}%
}]{%
goodman2010}
\APACinsertmetastar {%
goodman2010}%
\begin{APACrefauthors}%
{Goodman}, J.%
\BCBT {}\ \BBA {} {Weare}, J.%
\end{APACrefauthors}%
\unskip\
\newblock
\APACrefYearMonthDay{2010}{{\APACmonth{01}}}{}.
\newblock
{\BBOQ}\APACrefatitle {{Ensemble samplers with affine invariance}} {{Ensemble
  samplers with affine invariance}}.{\BBCQ}
\newblock
\APACjournalVolNumPages{Communications in Applied Mathematics and Computational
  Science}{5}{1}{65-80}.
\newblock
\begin{APACrefDOI} \doi{10.2140/camcos.2010.5.65} \end{APACrefDOI}
\PrintBackRefs{\CurrentBib}

\bibitem [\protect \citeauthoryear {%
{Goossens}%
\ \protect \BOthers {.}}{%
{Goossens}%
\ \protect \BOthers {.}}{%
{\protect \APACyear {2011}}%
}]{%
goossens2011}
\APACinsertmetastar {%
goossens2011}%
\begin{APACrefauthors}%
{Goossens}, S.%
, {Matsumoto}, K.%
, {Liu}, Q.%
, {Kikuchi}, F.%
, {Sato}, K.%
, {Hanada}, H.%
\BDBL {}{Chen}, M.%
\end{APACrefauthors}%
\unskip\
\newblock
\APACrefYearMonthDay{2011}{{\APACmonth{04}}}{}.
\newblock
{\BBOQ}\APACrefatitle {{Lunar gravity field determination using SELENE
  same-beam differential VLBI tracking data}} {{Lunar gravity field
  determination using SELENE same-beam differential VLBI tracking
  data}}.{\BBCQ}
\newblock
\APACjournalVolNumPages{Journal of Geodesy}{85}{4}{205-228}.
\newblock
\begin{APACrefDOI} \doi{10.1007/s00190-010-0430-2} \end{APACrefDOI}
\PrintBackRefs{\CurrentBib}

\bibitem [\protect \citeauthoryear {%
{Goossens}%
\ \protect \BOthers {.}}{%
{Goossens}%
\ \protect \BOthers {.}}{%
{\protect \APACyear {2022}}%
}]{%
goossens2022}
\APACinsertmetastar {%
goossens2022}%
\begin{APACrefauthors}%
{Goossens}, S.%
, {Renaud}, J\BPBI P.%
, {Henning}, W\BPBI G.%
, {Mazarico}, E.%
, {Bertone}, S.%
\BCBL {}\ \BBA {} {Genova}, A.%
\end{APACrefauthors}%
\unskip\
\newblock
\APACrefYearMonthDay{2022}{{\APACmonth{02}}}{}.
\newblock
{\BBOQ}\APACrefatitle {{Evaluation of Recent Measurements of Mercury's Moments
  of Inertia and Tides Using a Comprehensive Markov Chain Monte Carlo Method}}
  {{Evaluation of Recent Measurements of Mercury's Moments of Inertia and Tides
  Using a Comprehensive Markov Chain Monte Carlo Method}}.{\BBCQ}
\newblock
\APACjournalVolNumPages{The Planetary Science Journal}{3}{2}{37}.
\newblock
\begin{APACrefDOI} \doi{10.3847/PSJ/ac4bb8} \end{APACrefDOI}
\PrintBackRefs{\CurrentBib}

\bibitem [\protect \citeauthoryear {%
{Harada}%
\ \protect \BOthers {.}}{%
{Harada}%
\ \protect \BOthers {.}}{%
{\protect \APACyear {2014}}%
}]{%
harada2014}
\APACinsertmetastar {%
harada2014}%
\begin{APACrefauthors}%
{Harada}, Y.%
, {Goossens}, S.%
, {Matsumoto}, K.%
, {Yan}, J.%
, {Ping}, J.%
, {Noda}, H.%
\BCBL {}\ \BBA {} {Haruyama}, J.%
\end{APACrefauthors}%
\unskip\
\newblock
\APACrefYearMonthDay{2014}{{\APACmonth{08}}}{}.
\newblock
{\BBOQ}\APACrefatitle {{Strong tidal heating in an ultralow-viscosity zone at
  the core-mantle boundary of the Moon}} {{Strong tidal heating in an
  ultralow-viscosity zone at the core-mantle boundary of the Moon}}.{\BBCQ}
\newblock
\APACjournalVolNumPages{Nature Geoscience}{7}{8}{569-572}.
\newblock
\begin{APACrefDOI} \doi{10.1038/ngeo2211} \end{APACrefDOI}
\PrintBackRefs{\CurrentBib}

\bibitem [\protect \citeauthoryear {%
{Harada}%
\ \protect \BOthers {.}}{%
{Harada}%
\ \protect \BOthers {.}}{%
{\protect \APACyear {2016}}%
}]{%
harada2016}
\APACinsertmetastar {%
harada2016}%
\begin{APACrefauthors}%
{Harada}, Y.%
, {Goossens}, S.%
, {Matsumoto}, K.%
, {Yan}, J.%
, {Ping}, J.%
, {Noda}, H.%
\BCBL {}\ \BBA {} {Haruyama}, J.%
\end{APACrefauthors}%
\unskip\
\newblock
\APACrefYearMonthDay{2016}{{\APACmonth{09}}}{}.
\newblock
{\BBOQ}\APACrefatitle {{The deep lunar interior with a low-viscosity zone:
  Revised constraints from recent geodetic parameters on the tidal response of
  the Moon}} {{The deep lunar interior with a low-viscosity zone: Revised
  constraints from recent geodetic parameters on the tidal response of the
  Moon}}.{\BBCQ}
\newblock
\APACjournalVolNumPages{Icarus}{276}{}{96-101}.
\newblock
\begin{APACrefDOI} \doi{10.1016/j.icarus.2016.04.021} \end{APACrefDOI}
\PrintBackRefs{\CurrentBib}

\bibitem [\protect \citeauthoryear {%
{Hirth}%
\ \BBA {} {Kohlstedt}%
}{%
{Hirth}%
\ \BBA {} {Kohlstedt}%
}{%
{\protect \APACyear {1996}}%
}]{%
hirth1996}
\APACinsertmetastar {%
hirth1996}%
\begin{APACrefauthors}%
{Hirth}, G.%
\BCBT {}\ \BBA {} {Kohlstedt}, D\BPBI L.%
\end{APACrefauthors}%
\unskip\
\newblock
\APACrefYearMonthDay{1996}{{\APACmonth{10}}}{}.
\newblock
{\BBOQ}\APACrefatitle {{Water in the oceanic upper mantle: implications for
  rheology, melt extraction and the evolution of the lithosphere}} {{Water in
  the oceanic upper mantle: implications for rheology, melt extraction and the
  evolution of the lithosphere}}.{\BBCQ}
\newblock
\APACjournalVolNumPages{Earth and Planetary Science Letters}{144}{1-2}{93-108}.
\newblock
\begin{APACrefDOI} \doi{10.1016/0012-821X(96)00154-9} \end{APACrefDOI}
\PrintBackRefs{\CurrentBib}

\bibitem [\protect \citeauthoryear {%
{Hu}%
\ \protect \BOthers {.}}{%
{Hu}%
\ \protect \BOthers {.}}{%
{\protect \APACyear {2022}}%
}]{%
hu2022}
\APACinsertmetastar {%
hu2022}%
\begin{APACrefauthors}%
{Hu}, X.%
, {Stark}, A.%
, {Dirkx}, D.%
, {Hussmann}, H.%
, {Fienga}, A.%
, {Fayolle-Chambe}, M.%
\BDBL {}{Oberst}, J.%
\end{APACrefauthors}%
\unskip\
\newblock
\APACrefYearMonthDay{2022}{{\APACmonth{05}}}{}.
\newblock
{\BBOQ}\APACrefatitle {{Sensitivity analysis of frequency-dependent
  visco-elastic effects on lunar orbiters}} {{Sensitivity analysis of
  frequency-dependent visco-elastic effects on lunar orbiters}}.{\BBCQ}
\newblock
\BIn{} \APACrefbtitle {EGU General Assembly Conference Abstracts} {Egu general
  assembly conference abstracts}\ (\BPG~EGU22-9722).
\newblock
\begin{APACrefDOI} \doi{10.5194/egusphere-egu22-9722} \end{APACrefDOI}
\PrintBackRefs{\CurrentBib}

\bibitem [\protect \citeauthoryear {%
{Jackson}%
, {Faul}%
\BCBL {}\ \BBA {} {Skelton}%
}{%
{Jackson}%
\ \protect \BOthers {.}}{%
{\protect \APACyear {2014}}%
}]{%
jackson2014}
\APACinsertmetastar {%
jackson2014}%
\begin{APACrefauthors}%
{Jackson}, I.%
, {Faul}, U\BPBI H.%
\BCBL {}\ \BBA {} {Skelton}, R.%
\end{APACrefauthors}%
\unskip\
\newblock
\APACrefYearMonthDay{2014}{{\APACmonth{03}}}{}.
\newblock
{\BBOQ}\APACrefatitle {{Elastically accommodated grain-boundary sliding: New
  insights from experiment and modeling}} {{Elastically accommodated
  grain-boundary sliding: New insights from experiment and modeling}}.{\BBCQ}
\newblock
\APACjournalVolNumPages{Physics of the Earth and Planetary
  Interiors}{228}{}{203-210}.
\newblock
\begin{APACrefDOI} \doi{10.1016/j.pepi.2013.11.014} \end{APACrefDOI}
\PrintBackRefs{\CurrentBib}

\bibitem [\protect \citeauthoryear {%
{Jackson}%
\ \protect \BOthers {.}}{%
{Jackson}%
\ \protect \BOthers {.}}{%
{\protect \APACyear {2010}}%
}]{%
jackson2010}
\APACinsertmetastar {%
jackson2010}%
\begin{APACrefauthors}%
{Jackson}, I.%
, {Faul}, U\BPBI H.%
, {Suetsugu}, D.%
, {Bina}, C.%
, {Inoue}, T.%
\BCBL {}\ \BBA {} {Jellinek}, M.%
\end{APACrefauthors}%
\unskip\
\newblock
\APACrefYearMonthDay{2010}{{\APACmonth{11}}}{}.
\newblock
{\BBOQ}\APACrefatitle {{Grainsize-sensitive viscoelastic relaxation in olivine:
  Towards a robust laboratory-based model for seismological application}}
  {{Grainsize-sensitive viscoelastic relaxation in olivine: Towards a robust
  laboratory-based model for seismological application}}.{\BBCQ}
\newblock
\APACjournalVolNumPages{Physics of the Earth and Planetary
  Interiors}{183}{1-2}{151-163}.
\newblock
\begin{APACrefDOI} \doi{10.1016/j.pepi.2010.09.005} \end{APACrefDOI}
\PrintBackRefs{\CurrentBib}

\bibitem [\protect \citeauthoryear {%
{Jackson}%
, {Fitz Gerald}%
, {Faul}%
\BCBL {}\ \BBA {} {Tan}%
}{%
{Jackson}%
\ \protect \BOthers {.}}{%
{\protect \APACyear {2002}}%
}]{%
jackson2002}
\APACinsertmetastar {%
jackson2002}%
\begin{APACrefauthors}%
{Jackson}, I.%
, {Fitz Gerald}, J\BPBI D.%
, {Faul}, U\BPBI H.%
\BCBL {}\ \BBA {} {Tan}, B\BPBI H.%
\end{APACrefauthors}%
\unskip\
\newblock
\APACrefYearMonthDay{2002}{{\APACmonth{12}}}{}.
\newblock
{\BBOQ}\APACrefatitle {{Grain-size-sensitive seismic wave attenuation in
  polycrystalline olivine}} {{Grain-size-sensitive seismic wave attenuation in
  polycrystalline olivine}}.{\BBCQ}
\newblock
\APACjournalVolNumPages{Journal of Geophysical Research (Solid
  Earth)}{107}{B12}{2360}.
\newblock
\begin{APACrefDOI} \doi{10.1029/2001JB001225} \end{APACrefDOI}
\PrintBackRefs{\CurrentBib}

\bibitem [\protect \citeauthoryear {%
{Jacobs}%
\ \protect \BOthers {.}}{%
{Jacobs}%
\ \protect \BOthers {.}}{%
{\protect \APACyear {2022}}%
}]{%
jacobs2022}
\APACinsertmetastar {%
jacobs2022}%
\begin{APACrefauthors}%
{Jacobs}, M\BPBI H\BPBI G.%
, {van den Berg}, A\BPBI P.%
, {Schmid-Fetzer}, R.%
, {de Vries}, J.%
, {van Westrenen}, W.%
\BCBL {}\ \BBA {} {Zhao}, Y.%
\end{APACrefauthors}%
\unskip\
\newblock
\APACrefYearMonthDay{2022}{{\APACmonth{07}}}{}.
\newblock
{\BBOQ}\APACrefatitle {{Thermodynamic properties of geikielite (MgTiO$_{3}$)
  and ilmenite (FeTiO$_{3}$) derived from vibrational methods combined with
  Raman and infrared spectroscopic data}} {{Thermodynamic properties of
  geikielite (MgTiO$_{3}$) and ilmenite (FeTiO$_{3}$) derived from vibrational
  methods combined with Raman and infrared spectroscopic data}}.{\BBCQ}
\newblock
\APACjournalVolNumPages{Physics and Chemistry of Minerals}{49}{7}{23}.
\newblock
\begin{APACrefDOI} \doi{10.1007/s00269-022-01195-5} \end{APACrefDOI}
\PrintBackRefs{\CurrentBib}

\bibitem [\protect \citeauthoryear {%
{Jacobsen}%
\ \protect \BOthers {.}}{%
{Jacobsen}%
\ \protect \BOthers {.}}{%
{\protect \APACyear {2008}}%
}]{%
jacobsen2008}
\APACinsertmetastar {%
jacobsen2008}%
\begin{APACrefauthors}%
{Jacobsen}, S\BPBI D.%
, {Jiang}, F.%
, {Mao}, Z.%
, {Duffy}, T\BPBI S.%
, {Smyth}, J\BPBI R.%
, {Holl}, C\BPBI M.%
\BCBL {}\ \BBA {} {Frost}, D\BPBI J.%
\end{APACrefauthors}%
\unskip\
\newblock
\APACrefYearMonthDay{2008}{{\APACmonth{07}}}{}.
\newblock
{\BBOQ}\APACrefatitle {{Effects of hydration on the elastic properties of
  olivine}} {{Effects of hydration on the elastic properties of
  olivine}}.{\BBCQ}
\newblock
\APACjournalVolNumPages{Geophysical Research Letters}{35}{14}{L14303}.
\newblock
\begin{APACrefDOI} \doi{10.1029/2008GL034398} \end{APACrefDOI}
\PrintBackRefs{\CurrentBib}

\bibitem [\protect \citeauthoryear {%
{Jones}%
\ \protect \BOthers {.}}{%
{Jones}%
\ \protect \BOthers {.}}{%
{\protect \APACyear {2022}}%
}]{%
jones2022}
\APACinsertmetastar {%
jones2022}%
\begin{APACrefauthors}%
{Jones}, M\BPBI J.%
, {Evans}, A\BPBI J.%
, {Johnson}, B\BPBI C.%
, {Weller}, M\BPBI B.%
, {Andrews-Hanna}, J\BPBI C.%
, {Tikoo}, S\BPBI M.%
\BCBL {}\ \BBA {} {Keane}, J\BPBI T.%
\end{APACrefauthors}%
\unskip\
\newblock
\APACrefYearMonthDay{2022}{{\APACmonth{04}}}{}.
\newblock
{\BBOQ}\APACrefatitle {{A South Pole{\textendash}Aitken impact origin of the
  lunar compositional asymmetry}} {{A South Pole{\textendash}Aitken impact
  origin of the lunar compositional asymmetry}}.{\BBCQ}
\newblock
\APACjournalVolNumPages{Science Advances}{8}{14}{eabm8475}.
\newblock
\begin{APACrefDOI} \doi{10.1126/sciadv.abm8475} \end{APACrefDOI}
\PrintBackRefs{\CurrentBib}

\bibitem [\protect \citeauthoryear {%
{Karato}%
}{%
{Karato}%
}{%
{\protect \APACyear {2013}}%
}]{%
karato2013}
\APACinsertmetastar {%
karato2013}%
\begin{APACrefauthors}%
{Karato}, S\BHBI i.%
\end{APACrefauthors}%
\unskip\
\newblock
\APACrefYearMonthDay{2013}{{\APACmonth{12}}}{}.
\newblock
{\BBOQ}\APACrefatitle {{Geophysical constraints on the water content of the
  lunar mantle and its implications for the origin of the Moon}} {{Geophysical
  constraints on the water content of the lunar mantle and its implications for
  the origin of the Moon}}.{\BBCQ}
\newblock
\APACjournalVolNumPages{Earth and Planetary Science Letters}{384}{}{144-153}.
\newblock
\begin{APACrefDOI} \doi{10.1016/j.epsl.2013.10.001} \end{APACrefDOI}
\PrintBackRefs{\CurrentBib}

\bibitem [\protect \citeauthoryear {%
Katz%
, Spiegelman%
\BCBL {}\ \BBA {} Langmuir%
}{%
Katz%
\ \protect \BOthers {.}}{%
{\protect \APACyear {2003}}%
}]{%
katz2003}
\APACinsertmetastar {%
katz2003}%
\begin{APACrefauthors}%
Katz, R\BPBI F.%
, Spiegelman, M.%
\BCBL {}\ \BBA {} Langmuir, C\BPBI H.%
\end{APACrefauthors}%
\unskip\
\newblock
\APACrefYearMonthDay{2003}{}{}.
\newblock
{\BBOQ}\APACrefatitle {A new parameterization of hydrous mantle melting} {A new
  parameterization of hydrous mantle melting}.{\BBCQ}
\newblock
\APACjournalVolNumPages{Geochemistry, Geophysics, Geosystems}{4}{9}{}.
\newblock
\begin{APACrefURL}
  \url{https://agupubs.onlinelibrary.wiley.com/doi/abs/10.1029/2002GC000433}
  \end{APACrefURL}
\newblock
\begin{APACrefDOI} \doi{https://doi.org/10.1029/2002GC000433} \end{APACrefDOI}
\PrintBackRefs{\CurrentBib}

\bibitem [\protect \citeauthoryear {%
{Kawamura}%
, {Lognonn{\'e}}%
, {Nishikawa}%
\BCBL {}\ \BBA {} {Tanaka}%
}{%
{Kawamura}%
\ \protect \BOthers {.}}{%
{\protect \APACyear {2017}}%
}]{%
kawamura2017}
\APACinsertmetastar {%
kawamura2017}%
\begin{APACrefauthors}%
{Kawamura}, T.%
, {Lognonn{\'e}}, P.%
, {Nishikawa}, Y.%
\BCBL {}\ \BBA {} {Tanaka}, S.%
\end{APACrefauthors}%
\unskip\
\newblock
\APACrefYearMonthDay{2017}{{\APACmonth{07}}}{}.
\newblock
{\BBOQ}\APACrefatitle {{Evaluation of deep moonquake source parameters:
  Implication for fault characteristics and thermal state}} {{Evaluation of
  deep moonquake source parameters: Implication for fault characteristics and
  thermal state}}.{\BBCQ}
\newblock
\APACjournalVolNumPages{Journal of Geophysical Research
  (Planets)}{122}{7}{1487-1504}.
\newblock
\begin{APACrefDOI} \doi{10.1002/2016JE005147} \end{APACrefDOI}
\PrintBackRefs{\CurrentBib}

\bibitem [\protect \citeauthoryear {%
{K{\^e}}%
}{%
{K{\^e}}%
}{%
{\protect \APACyear {1947}}%
}]{%
ke1947}
\APACinsertmetastar {%
ke1947}%
\begin{APACrefauthors}%
{K{\^e}}, T\BHBI S.%
\end{APACrefauthors}%
\unskip\
\newblock
\APACrefYearMonthDay{1947}{{\APACmonth{04}}}{}.
\newblock
{\BBOQ}\APACrefatitle {{Experimental Evidence of the Viscous Behavior of Grain
  Boundaries in Metals}} {{Experimental Evidence of the Viscous Behavior of
  Grain Boundaries in Metals}}.{\BBCQ}
\newblock
\APACjournalVolNumPages{Physical Review}{71}{8}{533-546}.
\newblock
\begin{APACrefDOI} \doi{10.1103/PhysRev.71.533} \end{APACrefDOI}
\PrintBackRefs{\CurrentBib}

\bibitem [\protect \citeauthoryear {%
{Kennedy}%
}{%
{Kennedy}%
}{%
{\protect \APACyear {1953}}%
}]{%
kennedy1953}
\APACinsertmetastar {%
kennedy1953}%
\begin{APACrefauthors}%
{Kennedy}, A\BPBI J.%
\end{APACrefauthors}%
\unskip\
\newblock
\APACrefYearMonthDay{1953}{{\APACmonth{04}}}{}.
\newblock
{\BBOQ}\APACrefatitle {{On the generality of the cubic creep function}} {{On
  the generality of the cubic creep function}}.{\BBCQ}
\newblock
\APACjournalVolNumPages{Journal of Mechanics Physics of Solids}{1}{3}{172-181}.
\newblock
\begin{APACrefDOI} \doi{10.1016/0022-5096(53)90035-0} \end{APACrefDOI}
\PrintBackRefs{\CurrentBib}

\bibitem [\protect \citeauthoryear {%
{Kervazo}%
, {Tobie}%
, {Choblet}%
, {Dumoulin}%
\BCBL {}\ \BBA {} {B{\v{e}}hounkov{\'a}}%
}{%
{Kervazo}%
\ \protect \BOthers {.}}{%
{\protect \APACyear {2021}}%
}]{%
kervazo2021}
\APACinsertmetastar {%
kervazo2021}%
\begin{APACrefauthors}%
{Kervazo}, M.%
, {Tobie}, G.%
, {Choblet}, G.%
, {Dumoulin}, C.%
\BCBL {}\ \BBA {} {B{\v{e}}hounkov{\'a}}, M.%
\end{APACrefauthors}%
\unskip\
\newblock
\APACrefYearMonthDay{2021}{{\APACmonth{06}}}{}.
\newblock
{\BBOQ}\APACrefatitle {{Solid tides in Io's partially molten interior.
  Contribution of bulk dissipation}} {{Solid tides in Io's partially molten
  interior. Contribution of bulk dissipation}}.{\BBCQ}
\newblock
\APACjournalVolNumPages{Astronomy \& Astrophysics}{650}{}{A72}.
\newblock
\begin{APACrefDOI} \doi{10.1051/0004-6361/202039433} \end{APACrefDOI}
\PrintBackRefs{\CurrentBib}

\bibitem [\protect \citeauthoryear {%
{Khan}%
, {Connolly}%
, {Pommier}%
\BCBL {}\ \BBA {} {Noir}%
}{%
{Khan}%
\ \protect \BOthers {.}}{%
{\protect \APACyear {2014}}%
}]{%
khan2014}
\APACinsertmetastar {%
khan2014}%
\begin{APACrefauthors}%
{Khan}, A.%
, {Connolly}, J\BPBI A\BPBI D.%
, {Pommier}, A.%
\BCBL {}\ \BBA {} {Noir}, J.%
\end{APACrefauthors}%
\unskip\
\newblock
\APACrefYearMonthDay{2014}{{\APACmonth{10}}}{}.
\newblock
{\BBOQ}\APACrefatitle {{Geophysical evidence for melt in the deep lunar
  interior and implications for lunar evolution}} {{Geophysical evidence for
  melt in the deep lunar interior and implications for lunar
  evolution}}.{\BBCQ}
\newblock
\APACjournalVolNumPages{Journal of Geophysical Research
  (Planets)}{119}{10}{2197-2221}.
\newblock
\begin{APACrefDOI} \doi{10.1002/2014JE004661} \end{APACrefDOI}
\PrintBackRefs{\CurrentBib}

\bibitem [\protect \citeauthoryear {%
Khan%
, Pommier%
, Neumann%
\BCBL {}\ \BBA {} Mosegaard%
}{%
Khan%
\ \protect \BOthers {.}}{%
{\protect \APACyear {2013}}%
}]{%
khan2013}
\APACinsertmetastar {%
khan2013}%
\begin{APACrefauthors}%
Khan, A.%
, Pommier, A.%
, Neumann, G.%
\BCBL {}\ \BBA {} Mosegaard, K.%
\end{APACrefauthors}%
\unskip\
\newblock
\APACrefYearMonthDay{2013}{}{}.
\newblock
{\BBOQ}\APACrefatitle {The lunar moho and the internal structure of the Moon: A
  geophysical perspective} {The lunar moho and the internal structure of the
  moon: A geophysical perspective}.{\BBCQ}
\newblock
\APACjournalVolNumPages{Tectonophysics}{609}{}{331-352}.
\newblock
\begin{APACrefURL}
  \url{https://www.sciencedirect.com/science/article/pii/S0040195113001236}
  \end{APACrefURL}
\newblock
\APACrefnote{Moho: 100 years after Andrija Mohorovicic}
\newblock
\begin{APACrefDOI} \doi{https://doi.org/10.1016/j.tecto.2013.02.024}
  \end{APACrefDOI}
\PrintBackRefs{\CurrentBib}

\bibitem [\protect \citeauthoryear {%
{Konopliv}%
\ \protect \BOthers {.}}{%
{Konopliv}%
\ \protect \BOthers {.}}{%
{\protect \APACyear {2013}}%
}]{%
konopliv2013}
\APACinsertmetastar {%
konopliv2013}%
\begin{APACrefauthors}%
{Konopliv}, A\BPBI S.%
, {Park}, R\BPBI S.%
, {Yuan}, D\BHBI N.%
, {Asmar}, S\BPBI W.%
, {Watkins}, M\BPBI M.%
, {Williams}, J\BPBI G.%
\BDBL {}{Zuber}, M\BPBI T.%
\end{APACrefauthors}%
\unskip\
\newblock
\APACrefYearMonthDay{2013}{{\APACmonth{07}}}{}.
\newblock
{\BBOQ}\APACrefatitle {{The JPL lunar gravity field to spherical harmonic
  degree 660 from the GRAIL Primary Mission}} {{The JPL lunar gravity field to
  spherical harmonic degree 660 from the GRAIL Primary Mission}}.{\BBCQ}
\newblock
\APACjournalVolNumPages{Journal of Geophysical Research
  (Planets)}{118}{7}{1415-1434}.
\newblock
\begin{APACrefDOI} \doi{10.1002/jgre.20097} \end{APACrefDOI}
\PrintBackRefs{\CurrentBib}

\bibitem [\protect \citeauthoryear {%
Kraettli%
, Schmidt%
\BCBL {}\ \BBA {} Liebske%
}{%
Kraettli%
\ \protect \BOthers {.}}{%
{\protect \APACyear {2022}}%
}]{%
kraettli2022}
\APACinsertmetastar {%
kraettli2022}%
\begin{APACrefauthors}%
Kraettli, G.%
, Schmidt, M\BPBI W.%
\BCBL {}\ \BBA {} Liebske, C.%
\end{APACrefauthors}%
\unskip\
\newblock
\APACrefYearMonthDay{2022}{}{}.
\newblock
{\BBOQ}\APACrefatitle {Fractional crystallization of a basal lunar magma ocean:
  A dense melt-bearing garnetite layer above the core?} {Fractional
  crystallization of a basal lunar magma ocean: A dense melt-bearing garnetite
  layer above the core?}{\BBCQ}
\newblock
\APACjournalVolNumPages{Icarus}{371}{}{114699}.
\newblock
\begin{APACrefURL}
  \url{https://www.sciencedirect.com/science/article/pii/S0019103521003547}
  \end{APACrefURL}
\newblock
\begin{APACrefDOI} \doi{https://doi.org/10.1016/j.icarus.2021.114699}
  \end{APACrefDOI}
\PrintBackRefs{\CurrentBib}

\bibitem [\protect \citeauthoryear {%
{Kronrod}%
\ \protect \BOthers {.}}{%
{Kronrod}%
\ \protect \BOthers {.}}{%
{\protect \APACyear {2022}}%
}]{%
kronrod2022}
\APACinsertmetastar {%
kronrod2022}%
\begin{APACrefauthors}%
{Kronrod}, E.%
, {Matsumoto}, K.%
, {Kuskov}, O\BPBI L.%
, {Kronrod}, V.%
, {Yamada}, R.%
\BCBL {}\ \BBA {} {Kamata}, S.%
\end{APACrefauthors}%
\unskip\
\newblock
\APACrefYearMonthDay{2022}{{\APACmonth{04}}}{}.
\newblock
{\BBOQ}\APACrefatitle {{Towards geochemical alternatives to geophysical models
  of the internal structure of the lunar mantle and core}} {{Towards
  geochemical alternatives to geophysical models of the internal structure of
  the lunar mantle and core}}.{\BBCQ}
\newblock
\APACjournalVolNumPages{Advances in Space Research}{69}{7}{2798-2824}.
\newblock
\begin{APACrefDOI} \doi{10.1016/j.asr.2022.01.012} \end{APACrefDOI}
\PrintBackRefs{\CurrentBib}

\bibitem [\protect \citeauthoryear {%
{Laneuville}%
, {Wieczorek}%
, {Breuer}%
\BCBL {}\ \BBA {} {Tosi}%
}{%
{Laneuville}%
\ \protect \BOthers {.}}{%
{\protect \APACyear {2013}}%
}]{%
laneuville2013}
\APACinsertmetastar {%
laneuville2013}%
\begin{APACrefauthors}%
{Laneuville}, M.%
, {Wieczorek}, M\BPBI A.%
, {Breuer}, D.%
\BCBL {}\ \BBA {} {Tosi}, N.%
\end{APACrefauthors}%
\unskip\
\newblock
\APACrefYearMonthDay{2013}{{\APACmonth{07}}}{}.
\newblock
{\BBOQ}\APACrefatitle {{Asymmetric thermal evolution of the Moon}} {{Asymmetric
  thermal evolution of the Moon}}.{\BBCQ}
\newblock
\APACjournalVolNumPages{Journal of Geophysical Research
  (Planets)}{118}{7}{1435-1452}.
\newblock
\begin{APACrefDOI} \doi{10.1002/jgre.20103} \end{APACrefDOI}
\PrintBackRefs{\CurrentBib}

\bibitem [\protect \citeauthoryear {%
{Lee}%
\ \BBA {} {Morris}%
}{%
{Lee}%
\ \BBA {} {Morris}%
}{%
{\protect \APACyear {2010}}%
}]{%
lee2010}
\APACinsertmetastar {%
lee2010}%
\begin{APACrefauthors}%
{Lee}, L\BPBI C.%
\BCBT {}\ \BBA {} {Morris}, S\BPBI J\BPBI S.%
\end{APACrefauthors}%
\unskip\
\newblock
\APACrefYearMonthDay{2010}{{\APACmonth{03}}}{}.
\newblock
{\BBOQ}\APACrefatitle {{Anelasticity and grain boundary sliding}}
  {{Anelasticity and grain boundary sliding}}.{\BBCQ}
\newblock
\APACjournalVolNumPages{Proceedings of the Royal Society of London Series
  A}{466}{2121}{2651-2671}.
\newblock
\begin{APACrefDOI} \doi{10.1098/rspa.2009.0624} \end{APACrefDOI}
\PrintBackRefs{\CurrentBib}

\bibitem [\protect \citeauthoryear {%
{Lee}%
, {Morris}%
\BCBL {}\ \BBA {} {Wilkening}%
}{%
{Lee}%
\ \protect \BOthers {.}}{%
{\protect \APACyear {2011}}%
}]{%
lee2011}
\APACinsertmetastar {%
lee2011}%
\begin{APACrefauthors}%
{Lee}, L\BPBI C.%
, {Morris}, S\BPBI J\BPBI S.%
\BCBL {}\ \BBA {} {Wilkening}, J.%
\end{APACrefauthors}%
\unskip\
\newblock
\APACrefYearMonthDay{2011}{{\APACmonth{06}}}{}.
\newblock
{\BBOQ}\APACrefatitle {{Stress concentrations, diffusionally accommodated grain
  boundary sliding and the viscoelasticity of polycrystals}} {{Stress
  concentrations, diffusionally accommodated grain boundary sliding and the
  viscoelasticity of polycrystals}}.{\BBCQ}
\newblock
\APACjournalVolNumPages{Proceedings of the Royal Society of London Series
  A}{467}{2130}{1624-1644}.
\newblock
\begin{APACrefDOI} \doi{10.1098/rspa.2010.0447} \end{APACrefDOI}
\PrintBackRefs{\CurrentBib}

\bibitem [\protect \citeauthoryear {%
{Lemoine}%
\ \protect \BOthers {.}}{%
{Lemoine}%
\ \protect \BOthers {.}}{%
{\protect \APACyear {2013}}%
}]{%
lemoine2013}
\APACinsertmetastar {%
lemoine2013}%
\begin{APACrefauthors}%
{Lemoine}, F\BPBI G.%
, {Goossens}, S.%
, {Sabaka}, T\BPBI J.%
, {Nicholas}, J\BPBI B.%
, {Mazarico}, E.%
, {Rowlands}, D\BPBI D.%
\BDBL {}{Zuber}, M\BPBI T.%
\end{APACrefauthors}%
\unskip\
\newblock
\APACrefYearMonthDay{2013}{{\APACmonth{08}}}{}.
\newblock
{\BBOQ}\APACrefatitle {{High{\ensuremath{-}}degree gravity models from GRAIL
  primary mission data}} {{High{\ensuremath{-}}degree gravity models from GRAIL
  primary mission data}}.{\BBCQ}
\newblock
\APACjournalVolNumPages{Journal of Geophysical Research
  (Planets)}{118}{8}{1676-1698}.
\newblock
\begin{APACrefDOI} \doi{10.1002/jgre.20118} \end{APACrefDOI}
\PrintBackRefs{\CurrentBib}

\bibitem [\protect \citeauthoryear {%
Li%
\ \protect \BOthers {.}}{%
Li%
\ \protect \BOthers {.}}{%
{\protect \APACyear {2019}}%
}]{%
li2019}
\APACinsertmetastar {%
li2019}%
\begin{APACrefauthors}%
Li, H.%
, Zhang, N.%
, Liang, Y.%
, Wu, B.%
, Dygert, N\BPBI J.%
, Huang, J.%
\BCBL {}\ \BBA {} Parmentier, E\BPBI M.%
\end{APACrefauthors}%
\unskip\
\newblock
\APACrefYearMonthDay{2019}{}{}.
\newblock
{\BBOQ}\APACrefatitle {Lunar Cumulate Mantle Overturn: A Model Constrained by
  Ilmenite Rheology} {Lunar cumulate mantle overturn: A model constrained by
  ilmenite rheology}.{\BBCQ}
\newblock
\APACjournalVolNumPages{Journal of Geophysical Research:
  Planets}{124}{5}{1357-1378}.
\newblock
\begin{APACrefURL}
  \url{https://agupubs.onlinelibrary.wiley.com/doi/abs/10.1029/2018JE005905}
  \end{APACrefURL}
\newblock
\begin{APACrefDOI} \doi{https://doi.org/10.1029/2018JE005905} \end{APACrefDOI}
\PrintBackRefs{\CurrentBib}

\bibitem [\protect \citeauthoryear {%
{Mao}%
\ \protect \BOthers {.}}{%
{Mao}%
\ \protect \BOthers {.}}{%
{\protect \APACyear {2015}}%
}]{%
mao2015}
\APACinsertmetastar {%
mao2015}%
\begin{APACrefauthors}%
{Mao}, Z.%
, {Fan}, D.%
, {Lin}, J\BHBI F.%
, {Yang}, J.%
, {Tkachev}, S\BPBI N.%
, {Zhuravlev}, K.%
\BCBL {}\ \BBA {} {Prakapenka}, V\BPBI B.%
\end{APACrefauthors}%
\unskip\
\newblock
\APACrefYearMonthDay{2015}{{\APACmonth{09}}}{}.
\newblock
{\BBOQ}\APACrefatitle {{Elasticity of single-crystal olivine at high pressures
  and temperatures}} {{Elasticity of single-crystal olivine at high pressures
  and temperatures}}.{\BBCQ}
\newblock
\APACjournalVolNumPages{Earth and Planetary Science Letters}{426}{}{204-215}.
\newblock
\begin{APACrefDOI} \doi{10.1016/j.epsl.2015.06.045} \end{APACrefDOI}
\PrintBackRefs{\CurrentBib}

\bibitem [\protect \citeauthoryear {%
{Marquardt}%
\ \BBA {} {Faul}%
}{%
{Marquardt}%
\ \BBA {} {Faul}%
}{%
{\protect \APACyear {2018}}%
}]{%
marquardt2018}
\APACinsertmetastar {%
marquardt2018}%
\begin{APACrefauthors}%
{Marquardt}, K.%
\BCBT {}\ \BBA {} {Faul}, U\BPBI H.%
\end{APACrefauthors}%
\unskip\
\newblock
\APACrefYearMonthDay{2018}{{\APACmonth{02}}}{}.
\newblock
{\BBOQ}\APACrefatitle {{The structure and composition of olivine grain
  boundaries: 40 years of studies, status and current developments}} {{The
  structure and composition of olivine grain boundaries: 40 years of studies,
  status and current developments}}.{\BBCQ}
\newblock
\APACjournalVolNumPages{Physics and Chemistry of Minerals}{45}{2}{139-172}.
\newblock
\begin{APACrefDOI} \doi{10.1007/s00269-017-0935-9} \end{APACrefDOI}
\PrintBackRefs{\CurrentBib}

\bibitem [\protect \citeauthoryear {%
{Matsumoto}%
\ \protect \BOthers {.}}{%
{Matsumoto}%
\ \protect \BOthers {.}}{%
{\protect \APACyear {2015}}%
}]{%
matsumoto2015}
\APACinsertmetastar {%
matsumoto2015}%
\begin{APACrefauthors}%
{Matsumoto}, K.%
, {Yamada}, R.%
, {Kikuchi}, F.%
, {Kamata}, S.%
, {Ishihara}, Y.%
, {Iwata}, T.%
\BDBL {}{Sasaki}, S.%
\end{APACrefauthors}%
\unskip\
\newblock
\APACrefYearMonthDay{2015}{{\APACmonth{09}}}{}.
\newblock
{\BBOQ}\APACrefatitle {{Internal structure of the Moon inferred from Apollo
  seismic data and selenodetic data from GRAIL and LLR}} {{Internal structure
  of the Moon inferred from Apollo seismic data and selenodetic data from GRAIL
  and LLR}}.{\BBCQ}
\newblock
\APACjournalVolNumPages{Geophysical Research Letters}{42}{18}{7351-7358}.
\newblock
\begin{APACrefDOI} \doi{10.1002/2015GL065335} \end{APACrefDOI}
\PrintBackRefs{\CurrentBib}

\bibitem [\protect \citeauthoryear {%
{Matsuyama}%
\ \protect \BOthers {.}}{%
{Matsuyama}%
\ \protect \BOthers {.}}{%
{\protect \APACyear {2016}}%
}]{%
matsuyama2016}
\APACinsertmetastar {%
matsuyama2016}%
\begin{APACrefauthors}%
{Matsuyama}, I.%
, {Nimmo}, F.%
, {Keane}, J\BPBI T.%
, {Chan}, N\BPBI H.%
, {Taylor}, G\BPBI J.%
, {Wieczorek}, M\BPBI A.%
\BDBL {}{Williams}, J\BPBI G.%
\end{APACrefauthors}%
\unskip\
\newblock
\APACrefYearMonthDay{2016}{{\APACmonth{08}}}{}.
\newblock
{\BBOQ}\APACrefatitle {{GRAIL, LLR, and LOLA constraints on the interior
  structure of the Moon}} {{GRAIL, LLR, and LOLA constraints on the interior
  structure of the Moon}}.{\BBCQ}
\newblock
\APACjournalVolNumPages{Geophysical Research Letters}{43}{16}{8365-8375}.
\newblock
\begin{APACrefDOI} \doi{10.1002/2016GL069952} \end{APACrefDOI}
\PrintBackRefs{\CurrentBib}

\bibitem [\protect \citeauthoryear {%
{Mazarico}%
, {Barker}%
, {Neumann}%
, {Zuber}%
\BCBL {}\ \BBA {} {Smith}%
}{%
{Mazarico}%
\ \protect \BOthers {.}}{%
{\protect \APACyear {2014}}%
}]{%
mazarico2014}
\APACinsertmetastar {%
mazarico2014}%
\begin{APACrefauthors}%
{Mazarico}, E.%
, {Barker}, M\BPBI K.%
, {Neumann}, G\BPBI A.%
, {Zuber}, M\BPBI T.%
\BCBL {}\ \BBA {} {Smith}, D\BPBI E.%
\end{APACrefauthors}%
\unskip\
\newblock
\APACrefYearMonthDay{2014}{{\APACmonth{04}}}{}.
\newblock
{\BBOQ}\APACrefatitle {{Detection of the lunar body tide by the Lunar Orbiter
  Laser Altimeter}} {{Detection of the lunar body tide by the Lunar Orbiter
  Laser Altimeter}}.{\BBCQ}
\newblock
\APACjournalVolNumPages{Geophysical Research Letters}{41}{7}{2282-2288}.
\newblock
\begin{APACrefDOI} \doi{10.1002/2013GL059085} \end{APACrefDOI}
\PrintBackRefs{\CurrentBib}

\bibitem [\protect \citeauthoryear {%
{Morris}%
\ \BBA {} {Jackson}%
}{%
{Morris}%
\ \BBA {} {Jackson}%
}{%
{\protect \APACyear {2009}}%
}]{%
morris2009}
\APACinsertmetastar {%
morris2009}%
\begin{APACrefauthors}%
{Morris}, S\BPBI J\BPBI S.%
\BCBT {}\ \BBA {} {Jackson}, I.%
\end{APACrefauthors}%
\unskip\
\newblock
\APACrefYearMonthDay{2009}{{\APACmonth{04}}}{}.
\newblock
{\BBOQ}\APACrefatitle {{Diffusionally assisted grain-boundary sliding and
  viscoelasticity of polycrystals}} {{Diffusionally assisted grain-boundary
  sliding and viscoelasticity of polycrystals}}.{\BBCQ}
\newblock
\APACjournalVolNumPages{Journal of Mechanics and Physics of
  Solids}{57}{4}{744-761}.
\newblock
\begin{APACrefDOI} \doi{10.1016/j.jmps.2008.12.006} \end{APACrefDOI}
\PrintBackRefs{\CurrentBib}

\bibitem [\protect \citeauthoryear {%
{Mosegaard}%
\ \BBA {} {Tarantola}%
}{%
{Mosegaard}%
\ \BBA {} {Tarantola}%
}{%
{\protect \APACyear {1995}}%
}]{%
mosegaard1995}
\APACinsertmetastar {%
mosegaard1995}%
\begin{APACrefauthors}%
{Mosegaard}, K.%
\BCBT {}\ \BBA {} {Tarantola}, A.%
\end{APACrefauthors}%
\unskip\
\newblock
\APACrefYearMonthDay{1995}{{\APACmonth{07}}}{}.
\newblock
{\BBOQ}\APACrefatitle {{Monte Carlo sampling of solutions to inverse problems}}
  {{Monte Carlo sampling of solutions to inverse problems}}.{\BBCQ}
\newblock
\APACjournalVolNumPages{Journal of Geophysical
  Research}{100}{B7}{12,431-12,447}.
\newblock
\begin{APACrefDOI} \doi{10.1029/94JB03097} \end{APACrefDOI}
\PrintBackRefs{\CurrentBib}

\bibitem [\protect \citeauthoryear {%
{Murase}%
\ \BBA {} {McBirney}%
}{%
{Murase}%
\ \BBA {} {McBirney}%
}{%
{\protect \APACyear {1973}}%
}]{%
murase1973}
\APACinsertmetastar {%
murase1973}%
\begin{APACrefauthors}%
{Murase}, T.%
\BCBT {}\ \BBA {} {McBirney}, A\BPBI R.%
\end{APACrefauthors}%
\unskip\
\newblock
\APACrefYearMonthDay{1973}{{\APACmonth{01}}}{}.
\newblock
{\BBOQ}\APACrefatitle {{Properties of Some Common Igneous Rocks and Their Melts
  at High Temperatures}} {{Properties of Some Common Igneous Rocks and Their
  Melts at High Temperatures}}.{\BBCQ}
\newblock
\APACjournalVolNumPages{Geological Society of America Bulletin}{84}{11}{3563}.
\newblock
\begin{APACrefDOI} \doi{10.1130/0016-7606(1973)84<3563:POSCIR>2.0.CO;2}
  \end{APACrefDOI}
\PrintBackRefs{\CurrentBib}

\bibitem [\protect \citeauthoryear {%
{Nakamura}%
}{%
{Nakamura}%
}{%
{\protect \APACyear {2005}}%
}]{%
Nakamura2005}
\APACinsertmetastar {%
Nakamura2005}%
\begin{APACrefauthors}%
{Nakamura}, Y.%
\end{APACrefauthors}%
\unskip\
\newblock
\APACrefYearMonthDay{2005}{{\APACmonth{01}}}{}.
\newblock
{\BBOQ}\APACrefatitle {{Farside deep moonquakes and deep interior of the Moon}}
  {{Farside deep moonquakes and deep interior of the Moon}}.{\BBCQ}
\newblock
\APACjournalVolNumPages{Journal of Geophysical Research
  (Planets)}{110}{E1}{E01001}.
\newblock
\begin{APACrefDOI} \doi{10.1029/2004JE002332} \end{APACrefDOI}
\PrintBackRefs{\CurrentBib}

\bibitem [\protect \citeauthoryear {%
{Nakamura}%
\ \BBA {} {Koyama}%
}{%
{Nakamura}%
\ \BBA {} {Koyama}%
}{%
{\protect \APACyear {1982}}%
}]{%
nakamura1982}
\APACinsertmetastar {%
nakamura1982}%
\begin{APACrefauthors}%
{Nakamura}, Y.%
\BCBT {}\ \BBA {} {Koyama}, J.%
\end{APACrefauthors}%
\unskip\
\newblock
\APACrefYearMonthDay{1982}{{\APACmonth{06}}}{}.
\newblock
{\BBOQ}\APACrefatitle {{Seismic Q of the lunar upper mantle.}} {{Seismic Q of
  the lunar upper mantle.}}{\BBCQ}
\newblock
\APACjournalVolNumPages{\jgr}{87}{}{4855-4861}.
\newblock
\begin{APACrefDOI} \doi{10.1029/JB087iB06p04855} \end{APACrefDOI}
\PrintBackRefs{\CurrentBib}

\bibitem [\protect \citeauthoryear {%
{Nakamura}%
\ \protect \BOthers {.}}{%
{Nakamura}%
\ \protect \BOthers {.}}{%
{\protect \APACyear {1973}}%
}]{%
nakamura1973}
\APACinsertmetastar {%
nakamura1973}%
\begin{APACrefauthors}%
{Nakamura}, Y.%
, {Lammlein}, D.%
, {Latham}, G.%
, {Ewing}, M.%
, {Dorman}, J.%
, {Press}, F.%
\BCBL {}\ \BBA {} {Toksoz}, N.%
\end{APACrefauthors}%
\unskip\
\newblock
\APACrefYearMonthDay{1973}{{\APACmonth{07}}}{}.
\newblock
{\BBOQ}\APACrefatitle {{New Seismic Data on the State of the Deep Lunar
  Interior}} {{New Seismic Data on the State of the Deep Lunar
  Interior}}.{\BBCQ}
\newblock
\APACjournalVolNumPages{Science}{181}{4094}{49-51}.
\newblock
\begin{APACrefDOI} \doi{10.1126/science.181.4094.49} \end{APACrefDOI}
\PrintBackRefs{\CurrentBib}

\bibitem [\protect \citeauthoryear {%
{Nakamura}%
\ \protect \BOthers {.}}{%
{Nakamura}%
\ \protect \BOthers {.}}{%
{\protect \APACyear {1974}}%
}]{%
Nakamura}
\APACinsertmetastar {%
Nakamura}%
\begin{APACrefauthors}%
{Nakamura}, Y.%
, {Latham}, G.%
, {Lammlein}, D.%
, {Ewing}, M.%
, {Duennebier}, F.%
\BCBL {}\ \BBA {} {Dorman}, J.%
\end{APACrefauthors}%
\unskip\
\newblock
\APACrefYearMonthDay{1974}{{\APACmonth{01}}}{}.
\newblock
{\BBOQ}\APACrefatitle {{Deep lunar interior inferred from recent seismic data}}
  {{Deep lunar interior inferred from recent seismic data}}.{\BBCQ}
\newblock
\APACjournalVolNumPages{Geophysical Research Letters}{1}{3}{137-140}.
\newblock
\begin{APACrefDOI} \doi{10.1029/GL001i003p00137} \end{APACrefDOI}
\PrintBackRefs{\CurrentBib}

\bibitem [\protect \citeauthoryear {%
{Nimmo}%
, {Faul}%
\BCBL {}\ \BBA {} {Garnero}%
}{%
{Nimmo}%
\ \protect \BOthers {.}}{%
{\protect \APACyear {2012}}%
}]{%
nimmo2012}
\APACinsertmetastar {%
nimmo2012}%
\begin{APACrefauthors}%
{Nimmo}, F.%
, {Faul}, U\BPBI H.%
\BCBL {}\ \BBA {} {Garnero}, E\BPBI J.%
\end{APACrefauthors}%
\unskip\
\newblock
\APACrefYearMonthDay{2012}{{\APACmonth{09}}}{}.
\newblock
{\BBOQ}\APACrefatitle {{Dissipation at tidal and seismic frequencies in a
  melt-free Moon}} {{Dissipation at tidal and seismic frequencies in a
  melt-free Moon}}.{\BBCQ}
\newblock
\APACjournalVolNumPages{Journal of Geophysical Research
  (Planets)}{117}{E9}{E09005}.
\newblock
\begin{APACrefDOI} \doi{10.1029/2012JE004160} \end{APACrefDOI}
\PrintBackRefs{\CurrentBib}

\bibitem [\protect \citeauthoryear {%
{Noyelles}%
, {Frouard}%
, {Makarov}%
\BCBL {}\ \BBA {} {Efroimsky}%
}{%
{Noyelles}%
\ \protect \BOthers {.}}{%
{\protect \APACyear {2014}}%
}]{%
2014Icar..241...26N}
\APACinsertmetastar {%
2014Icar..241...26N}%
\begin{APACrefauthors}%
{Noyelles}, B.%
, {Frouard}, J.%
, {Makarov}, V\BPBI V.%
\BCBL {}\ \BBA {} {Efroimsky}, M.%
\end{APACrefauthors}%
\unskip\
\newblock
\APACrefYearMonthDay{2014}{{\APACmonth{10}}}{}.
\newblock
{\BBOQ}\APACrefatitle {{Spin-orbit evolution of Mercury revisited}}
  {{Spin-orbit evolution of Mercury revisited}}.{\BBCQ}
\newblock
\APACjournalVolNumPages{Icarus}{241}{}{26-44}.
\newblock
\begin{APACrefDOI} \doi{10.1016/j.icarus.2014.05.045} \end{APACrefDOI}
\PrintBackRefs{\CurrentBib}

\bibitem [\protect \citeauthoryear {%
{Nunn}%
\ \protect \BOthers {.}}{%
{Nunn}%
\ \protect \BOthers {.}}{%
{\protect \APACyear {2020}}%
}]{%
nunn2020}
\APACinsertmetastar {%
nunn2020}%
\begin{APACrefauthors}%
{Nunn}, C.%
, {Garcia}, R\BPBI F.%
, {Nakamura}, Y.%
, {Marusiak}, A\BPBI G.%
, {Kawamura}, T.%
, {Sun}, D.%
\BDBL {}{Zhu}, P.%
\end{APACrefauthors}%
\unskip\
\newblock
\APACrefYearMonthDay{2020}{{\APACmonth{07}}}{}.
\newblock
{\BBOQ}\APACrefatitle {{Lunar Seismology: A Data and Instrumentation Review}}
  {{Lunar Seismology: A Data and Instrumentation Review}}.{\BBCQ}
\newblock
\APACjournalVolNumPages{Space Science Reviews}{216}{5}{89}.
\newblock
\begin{APACrefDOI} \doi{10.1007/s11214-020-00709-3} \end{APACrefDOI}
\PrintBackRefs{\CurrentBib}

\bibitem [\protect \citeauthoryear {%
{Panning}%
\ \protect \BOthers {.}}{%
{Panning}%
\ \protect \BOthers {.}}{%
{\protect \APACyear {2021}}%
}]{%
Panning}
\APACinsertmetastar {%
Panning}%
\begin{APACrefauthors}%
{Panning}, M.%
, {Kedar}, S.%
, {Bowles}, N.%
, {Calcutt}, S.%
, {Cutler}, J.%
, {Elliott}, J.%
\BDBL {}{Yana}, C.%
\end{APACrefauthors}%
\unskip\
\newblock
\APACrefYearMonthDay{2021}{{\APACmonth{12}}}{}.
\newblock
{\BBOQ}\APACrefatitle {{Farside Seismic Suite (FSS): First seismic data from
  the farside of the Moon delivered by a commercial lander}} {{Farside Seismic
  Suite (FSS): First seismic data from the farside of the Moon delivered by a
  commercial lander}}.{\BBCQ}
\newblock
\BIn{} \APACrefbtitle {AGU Fall Meeting Abstracts} {Agu fall meeting
  abstracts}\ (\BVOL\ 2021, \BPG~P54C-01).
\newblock
\begin{APACrefURL}
  \url{https://www.hou.usra.edu/meetings/lpsc2022/pdf/1576.pdf}
  \end{APACrefURL}
\PrintBackRefs{\CurrentBib}

\bibitem [\protect \citeauthoryear {%
{Pavlov}%
, {Williams}%
\BCBL {}\ \BBA {} {Suvorkin}%
}{%
{Pavlov}%
\ \protect \BOthers {.}}{%
{\protect \APACyear {2016}}%
}]{%
pavlov2016}
\APACinsertmetastar {%
pavlov2016}%
\begin{APACrefauthors}%
{Pavlov}, D\BPBI A.%
, {Williams}, J\BPBI G.%
\BCBL {}\ \BBA {} {Suvorkin}, V\BPBI V.%
\end{APACrefauthors}%
\unskip\
\newblock
\APACrefYearMonthDay{2016}{{\APACmonth{11}}}{}.
\newblock
{\BBOQ}\APACrefatitle {{Determining parameters of Moon's orbital and rotational
  motion from LLR observations using GRAIL and IERS-recommended models}}
  {{Determining parameters of Moon's orbital and rotational motion from LLR
  observations using GRAIL and IERS-recommended models}}.{\BBCQ}
\newblock
\APACjournalVolNumPages{Celestial Mechanics and Dynamical
  Astronomy}{126}{1-3}{61-88}.
\newblock
\begin{APACrefDOI} \doi{10.1007/s10569-016-9712-1} \end{APACrefDOI}
\PrintBackRefs{\CurrentBib}

\bibitem [\protect \citeauthoryear {%
{Petit}%
\ \BBA {} {Luzum}%
}{%
{Petit}%
\ \BBA {} {Luzum}%
}{%
{\protect \APACyear {2010}}%
}]{%
petit2010}
\APACinsertmetastar {%
petit2010}%
\begin{APACrefauthors}%
{Petit}, G.%
\BCBT {}\ \BBA {} {Luzum}, B.%
\end{APACrefauthors}%
\unskip\
\newblock
\APACrefYearMonthDay{2010}{{\APACmonth{05}}}{}.
\newblock
{\BBOQ}\APACrefatitle {{The new reference edition of the IERS Conventions}}
  {{The new reference edition of the IERS Conventions}}.{\BBCQ}
\newblock
\BIn{} \APACrefbtitle {EGU General Assembly Conference Abstracts} {Egu general
  assembly conference abstracts}\ (\BPG~2919).
\PrintBackRefs{\CurrentBib}

\bibitem [\protect \citeauthoryear {%
{Qin}%
, {Muirhead}%
\BCBL {}\ \BBA {} {Zhong}%
}{%
{Qin}%
\ \protect \BOthers {.}}{%
{\protect \APACyear {2012}}%
}]{%
qin2012}
\APACinsertmetastar {%
qin2012}%
\begin{APACrefauthors}%
{Qin}, C.%
, {Muirhead}, A\BPBI C.%
\BCBL {}\ \BBA {} {Zhong}, S.%
\end{APACrefauthors}%
\unskip\
\newblock
\APACrefYearMonthDay{2012}{{\APACmonth{07}}}{}.
\newblock
{\BBOQ}\APACrefatitle {{Correlation of deep moonquakes and mare basalts:
  Implications for lunar mantle structure and evolution}} {{Correlation of deep
  moonquakes and mare basalts: Implications for lunar mantle structure and
  evolution}}.{\BBCQ}
\newblock
\APACjournalVolNumPages{\icarus}{220}{1}{100-105}.
\newblock
\begin{APACrefDOI} \doi{10.1016/j.icarus.2012.04.023} \end{APACrefDOI}
\PrintBackRefs{\CurrentBib}

\bibitem [\protect \citeauthoryear {%
{Qu}%
, {Jackson}%
\BCBL {}\ \BBA {} {Faul}%
}{%
{Qu}%
\ \protect \BOthers {.}}{%
{\protect \APACyear {2021}}%
}]{%
qu2021}
\APACinsertmetastar {%
qu2021}%
\begin{APACrefauthors}%
{Qu}, T.%
, {Jackson}, I.%
\BCBL {}\ \BBA {} {Faul}, U\BPBI H.%
\end{APACrefauthors}%
\unskip\
\newblock
\APACrefYearMonthDay{2021}{{\APACmonth{10}}}{}.
\newblock
{\BBOQ}\APACrefatitle {{Low-Frequency Seismic Properties of
  Olivine-Orthopyroxene Mixtures}} {{Low-Frequency Seismic Properties of
  Olivine-Orthopyroxene Mixtures}}.{\BBCQ}
\newblock
\APACjournalVolNumPages{Journal of Geophysical Research (Solid
  Earth)}{126}{10}{e2021JB022504}.
\newblock
\begin{APACrefDOI} \doi{10.1029/2021JB022504} \end{APACrefDOI}
\PrintBackRefs{\CurrentBib}

\bibitem [\protect \citeauthoryear {%
{Raevskiy}%
, {Gudkova}%
, {Kuskov}%
\BCBL {}\ \BBA {} {Kronrod}%
}{%
{Raevskiy}%
\ \protect \BOthers {.}}{%
{\protect \APACyear {2015}}%
}]{%
raevskiy2015}
\APACinsertmetastar {%
raevskiy2015}%
\begin{APACrefauthors}%
{Raevskiy}, S\BPBI N.%
, {Gudkova}, T\BPBI V.%
, {Kuskov}, O\BPBI L.%
\BCBL {}\ \BBA {} {Kronrod}, V\BPBI A.%
\end{APACrefauthors}%
\unskip\
\newblock
\APACrefYearMonthDay{2015}{{\APACmonth{01}}}{}.
\newblock
{\BBOQ}\APACrefatitle {{On reconciling the models of the interior structure of
  the moon with gravity data}} {{On reconciling the models of the interior
  structure of the moon with gravity data}}.{\BBCQ}
\newblock
\APACjournalVolNumPages{Izvestiya, Physics of the Solid Earth}{51}{1}{134-142}.
\newblock
\begin{APACrefDOI} \doi{10.1134/S1069351315010127} \end{APACrefDOI}
\PrintBackRefs{\CurrentBib}

\bibitem [\protect \citeauthoryear {%
{Raj}%
\ \BBA {} {Ashby}%
}{%
{Raj}%
\ \BBA {} {Ashby}%
}{%
{\protect \APACyear {1971}}%
}]{%
raj1971}
\APACinsertmetastar {%
raj1971}%
\begin{APACrefauthors}%
{Raj}, R.%
\BCBT {}\ \BBA {} {Ashby}, M\BPBI F.%
\end{APACrefauthors}%
\unskip\
\newblock
\APACrefYearMonthDay{1971}{{\APACmonth{01}}}{}.
\newblock
{\BBOQ}\APACrefatitle {{On grain boundary sliding and diffusional creep}} {{On
  grain boundary sliding and diffusional creep}}.{\BBCQ}
\newblock
\APACjournalVolNumPages{Metallurgical Transactions}{2}{}{1113-1127}.
\newblock
\begin{APACrefDOI} \doi{10.1007/BF02664244} \end{APACrefDOI}
\PrintBackRefs{\CurrentBib}

\bibitem [\protect \citeauthoryear {%
{Renaud}%
\ \BBA {} {Henning}%
}{%
{Renaud}%
\ \BBA {} {Henning}%
}{%
{\protect \APACyear {2018}}%
}]{%
RenaudHenning2018}
\APACinsertmetastar {%
RenaudHenning2018}%
\begin{APACrefauthors}%
{Renaud}, J\BPBI P.%
\BCBT {}\ \BBA {} {Henning}, W\BPBI G.%
\end{APACrefauthors}%
\unskip\
\newblock
\APACrefYearMonthDay{2018}{{\APACmonth{04}}}{}.
\newblock
{\BBOQ}\APACrefatitle {{Increased Tidal Dissipation Using Advanced Rheological
  Models: Implications for Io and Tidally Active Exoplanets}} {{Increased Tidal
  Dissipation Using Advanced Rheological Models: Implications for Io and
  Tidally Active Exoplanets}}.{\BBCQ}
\newblock
\APACjournalVolNumPages{The Astrophysical Journal}{857}{2}{98}.
\newblock
\begin{APACrefDOI} \doi{10.3847/1538-4357/aab784} \end{APACrefDOI}
\PrintBackRefs{\CurrentBib}

\bibitem [\protect \citeauthoryear {%
Renner%
, Evans%
\BCBL {}\ \BBA {} Hirth%
}{%
Renner%
\ \protect \BOthers {.}}{%
{\protect \APACyear {2000}}%
}]{%
Renner2000}
\APACinsertmetastar {%
Renner2000}%
\begin{APACrefauthors}%
Renner, J.%
, Evans, B.%
\BCBL {}\ \BBA {} Hirth, G.%
\end{APACrefauthors}%
\unskip\
\newblock
\APACrefYearMonthDay{2000}{}{}.
\newblock
{\BBOQ}\APACrefatitle {On the rheologically critical melt fraction} {On the
  rheologically critical melt fraction}.{\BBCQ}
\newblock
\APACjournalVolNumPages{Earth and Planetary Science Letters}{181}{4}{585-594}.
\newblock
\begin{APACrefURL}
  \url{https://www.sciencedirect.com/science/article/pii/S0012821X00002223}
  \end{APACrefURL}
\newblock
\begin{APACrefDOI} \doi{https://doi.org/10.1016/S0012-821X(00)00222-3}
  \end{APACrefDOI}
\PrintBackRefs{\CurrentBib}

\bibitem [\protect \citeauthoryear {%
{Sabadini}%
\ \BBA {} {Vermeersen}%
}{%
{Sabadini}%
\ \BBA {} {Vermeersen}%
}{%
{\protect \APACyear {2004}}%
}]{%
sabadini2004}
\APACinsertmetastar {%
sabadini2004}%
\begin{APACrefauthors}%
{Sabadini}, R.%
\BCBT {}\ \BBA {} {Vermeersen}, B.%
\end{APACrefauthors}%
\unskip\
\newblock
\APACrefYear{2004}.
\newblock
\APACrefbtitle {{Global Dynamics of the Earth: Applications of Normal Mode
  Relaxation Theory to Solid-Earth Geophysics}} {{Global Dynamics of the Earth:
  Applications of Normal Mode Relaxation Theory to Solid-Earth Geophysics}}.
\newblock
\APACaddressPublisher{Dodrech, the Netherlands}{Kluwer Academic Publishers}.
\PrintBackRefs{\CurrentBib}

\bibitem [\protect \citeauthoryear {%
{Samuel}%
\ \protect \BOthers {.}}{%
{Samuel}%
\ \protect \BOthers {.}}{%
{\protect \APACyear {2021}}%
}]{%
samuel2021}
\APACinsertmetastar {%
samuel2021}%
\begin{APACrefauthors}%
{Samuel}, H.%
, {Ballmer}, M\BPBI D.%
, {Padovan}, S.%
, {Tosi}, N.%
, {Rivoldini}, A.%
\BCBL {}\ \BBA {} {Plesa}, A\BHBI C.%
\end{APACrefauthors}%
\unskip\
\newblock
\APACrefYearMonthDay{2021}{{\APACmonth{04}}}{}.
\newblock
{\BBOQ}\APACrefatitle {{The Thermo Chemical Evolution of Mars With a Strongly
  Stratified Mantle}} {{The Thermo Chemical Evolution of Mars With a Strongly
  Stratified Mantle}}.{\BBCQ}
\newblock
\APACjournalVolNumPages{Journal of Geophysical Research
  (Planets)}{126}{4}{e06613}.
\newblock
\begin{APACrefDOI} \doi{10.1029/2020JE006613} \end{APACrefDOI}
\PrintBackRefs{\CurrentBib}

\bibitem [\protect \citeauthoryear {%
Segatz%
, Spohn%
, Ross%
\BCBL {}\ \BBA {} Schubert%
}{%
Segatz%
\ \protect \BOthers {.}}{%
{\protect \APACyear {1988}}%
}]{%
segatz88}
\APACinsertmetastar {%
segatz88}%
\begin{APACrefauthors}%
Segatz, M.%
, Spohn, T.%
, Ross, M\BPBI N.%
\BCBL {}\ \BBA {} Schubert, G.%
\end{APACrefauthors}%
\unskip\
\newblock
\APACrefYearMonthDay{1988}{}{}.
\newblock
{\BBOQ}\APACrefatitle {{T}idal {D}issipation, {S}urface {H}eat {F}low, and
  {F}igure of {V}iscoelastic {M}odels of {Io}} {{T}idal {D}issipation,
  {S}urface {H}eat {F}low, and {F}igure of {V}iscoelastic {M}odels of
  {Io}}.{\BBCQ}
\newblock
\APACjournalVolNumPages{Icarus}{75}{2}{187-206}.
\newblock
\begin{APACrefDOI} \doi{10.1016/0019-1035(88)90001-2} \end{APACrefDOI}
\PrintBackRefs{\CurrentBib}

\bibitem [\protect \citeauthoryear {%
{Siegler}%
\ \BBA {} {Smrekar}%
}{%
{Siegler}%
\ \BBA {} {Smrekar}%
}{%
{\protect \APACyear {2014}}%
}]{%
siegler2014}
\APACinsertmetastar {%
siegler2014}%
\begin{APACrefauthors}%
{Siegler}, M\BPBI A.%
\BCBT {}\ \BBA {} {Smrekar}, S\BPBI E.%
\end{APACrefauthors}%
\unskip\
\newblock
\APACrefYearMonthDay{2014}{{\APACmonth{01}}}{}.
\newblock
{\BBOQ}\APACrefatitle {{Lunar heat flow: Regional prospective of the Apollo
  landing sites}} {{Lunar heat flow: Regional prospective of the Apollo landing
  sites}}.{\BBCQ}
\newblock
\APACjournalVolNumPages{Journal of Geophysical Research
  (Planets)}{119}{1}{47-63}.
\newblock
\begin{APACrefDOI} \doi{10.1002/2013JE004453} \end{APACrefDOI}
\PrintBackRefs{\CurrentBib}

\bibitem [\protect \citeauthoryear {%
{St{\"a}hler}%
\ \protect \BOthers {.}}{%
{St{\"a}hler}%
\ \protect \BOthers {.}}{%
{\protect \APACyear {2021}}%
}]{%
staehler2021}
\APACinsertmetastar {%
staehler2021}%
\begin{APACrefauthors}%
{St{\"a}hler}, S\BPBI C.%
, {Khan}, A.%
, {Banerdt}, W\BPBI B.%
, {Lognonn{\'e}}, P.%
, {Giardini}, D.%
, {Ceylan}, S.%
\BDBL {}{Smrekar}, S\BPBI E.%
\end{APACrefauthors}%
\unskip\
\newblock
\APACrefYearMonthDay{2021}{{\APACmonth{07}}}{}.
\newblock
{\BBOQ}\APACrefatitle {{Seismic detection of the martian core}} {{Seismic
  detection of the martian core}}.{\BBCQ}
\newblock
\APACjournalVolNumPages{Science}{373}{6553}{443-448}.
\newblock
\begin{APACrefDOI} \doi{10.1126/science.abi7730} \end{APACrefDOI}
\PrintBackRefs{\CurrentBib}

\bibitem [\protect \citeauthoryear {%
{Stark}%
\ \protect \BOthers {.}}{%
{Stark}%
\ \protect \BOthers {.}}{%
{\protect \APACyear {2022}}%
}]{%
stark2022}
\APACinsertmetastar {%
stark2022}%
\begin{APACrefauthors}%
{Stark}, A.%
, {Xiao}, H.%
, {Hu}, X.%
, {Fienga}, A.%
, {Hussmann}, H.%
, {Oberst}, J.%
\BDBL {}{Saliby}, C.%
\end{APACrefauthors}%
\unskip\
\newblock
\APACrefYearMonthDay{2022}{{\APACmonth{05}}}{}.
\newblock
{\BBOQ}\APACrefatitle {{Measurement of tidal deformation through
  self-registration of laser profiles: Application to Earth's Moon}}
  {{Measurement of tidal deformation through self-registration of laser
  profiles: Application to Earth's Moon}}.{\BBCQ}
\newblock
\BIn{} \APACrefbtitle {EGU General Assembly Conference Abstracts} {Egu general
  assembly conference abstracts}\ (\BPG~EGU22-10626).
\newblock
\begin{APACrefDOI} \doi{10.5194/egusphere-egu22-10626} \end{APACrefDOI}
\PrintBackRefs{\CurrentBib}

\bibitem [\protect \citeauthoryear {%
{Steinbr{\"u}gge}%
\ \protect \BOthers {.}}{%
{Steinbr{\"u}gge}%
\ \protect \BOthers {.}}{%
{\protect \APACyear {2021}}%
}]{%
steinbruegge2021}
\APACinsertmetastar {%
steinbruegge2021}%
\begin{APACrefauthors}%
{Steinbr{\"u}gge}, G.%
, {Dumberry}, M.%
, {Rivoldini}, A.%
, {Schubert}, G.%
, {Cao}, H.%
, {Schroeder}, D\BPBI M.%
\BCBL {}\ \BBA {} {Soderlund}, K\BPBI M.%
\end{APACrefauthors}%
\unskip\
\newblock
\APACrefYearMonthDay{2021}{{\APACmonth{02}}}{}.
\newblock
{\BBOQ}\APACrefatitle {{Challenges on Mercury's Interior Structure Posed by the
  New Measurements of its Obliquity and Tides}} {{Challenges on Mercury's
  Interior Structure Posed by the New Measurements of its Obliquity and
  Tides}}.{\BBCQ}
\newblock
\APACjournalVolNumPages{\grl}{48}{3}{e89895}.
\newblock
\begin{APACrefDOI} \doi{10.1029/2020GL089895} \end{APACrefDOI}
\PrintBackRefs{\CurrentBib}

\bibitem [\protect \citeauthoryear {%
Sundberg%
\ \BBA {} Cooper%
}{%
Sundberg%
\ \BBA {} Cooper%
}{%
{\protect \APACyear {2010}}%
}]{%
Sundberg}
\APACinsertmetastar {%
Sundberg}%
\begin{APACrefauthors}%
Sundberg, M.%
\BCBT {}\ \BBA {} Cooper, R\BPBI F.%
\end{APACrefauthors}%
\unskip\
\newblock
\APACrefYearMonthDay{2010}{}{}.
\newblock
{\BBOQ}\APACrefatitle {A composite viscoelastic model for incorporating grain
  boundary sliding and transient diffusion creep; correlating creep and
  attenuation responses for materials with a fine grain size} {A composite
  viscoelastic model for incorporating grain boundary sliding and transient
  diffusion creep; correlating creep and attenuation responses for materials
  with a fine grain size}.{\BBCQ}
\newblock
\APACjournalVolNumPages{Philosophical Magazine}{90}{20}{2817-2840}.
\newblock
\begin{APACrefURL} \url{https://doi.org/10.1080/14786431003746656}
  \end{APACrefURL}
\newblock
\begin{APACrefDOI} \doi{10.1080/14786431003746656} \end{APACrefDOI}
\PrintBackRefs{\CurrentBib}

\bibitem [\protect \citeauthoryear {%
{Takeuchi}%
\ \BBA {} {Saito}%
}{%
{Takeuchi}%
\ \BBA {} {Saito}%
}{%
{\protect \APACyear {1972}}%
}]{%
takeuchi1972}
\APACinsertmetastar {%
takeuchi1972}%
\begin{APACrefauthors}%
{Takeuchi}, H.%
\BCBT {}\ \BBA {} {Saito}, M.%
\end{APACrefauthors}%
\unskip\
\newblock
\APACrefYearMonthDay{1972}{}{}.
\newblock
{\BBOQ}\APACrefatitle {{Seismic Surface Waves}} {{Seismic Surface
  Waves}}.{\BBCQ}
\newblock
\APACjournalVolNumPages{Methods in Computational Physics: Advances in Research
  and Applications}{11}{}{217-295}.
\newblock
\begin{APACrefDOI} \doi{10.1016/B978-0-12-460811-5.50010-6} \end{APACrefDOI}
\PrintBackRefs{\CurrentBib}

\bibitem [\protect \citeauthoryear {%
B\BPBI H.~{Tan}%
, {Jackson}%
\BCBL {}\ \BBA {} {Fitz Gerald}%
}{%
B\BPBI H.~{Tan}%
\ \protect \BOthers {.}}{%
{\protect \APACyear {2001}}%
}]{%
tan2001}
\APACinsertmetastar {%
tan2001}%
\begin{APACrefauthors}%
{Tan}, B\BPBI H.%
, {Jackson}, I.%
\BCBL {}\ \BBA {} {Fitz Gerald}, J\BPBI D.%
\end{APACrefauthors}%
\unskip\
\newblock
\APACrefYearMonthDay{2001}{{\APACmonth{01}}}{}.
\newblock
{\BBOQ}\APACrefatitle {{High-temperature viscoelasticity of fine-grained
  polycrystalline olivine}} {{High-temperature viscoelasticity of fine-grained
  polycrystalline olivine}}.{\BBCQ}
\newblock
\APACjournalVolNumPages{Physics and Chemistry of Minerals}{28}{9}{641-664}.
\newblock
\begin{APACrefDOI} \doi{10.1007/s002690100189} \end{APACrefDOI}
\PrintBackRefs{\CurrentBib}

\bibitem [\protect \citeauthoryear {%
Y.~{Tan}%
\ \BBA {} {Harada}%
}{%
Y.~{Tan}%
\ \BBA {} {Harada}%
}{%
{\protect \APACyear {2021}}%
}]{%
tan2021}
\APACinsertmetastar {%
tan2021}%
\begin{APACrefauthors}%
{Tan}, Y.%
\BCBT {}\ \BBA {} {Harada}, Y.%
\end{APACrefauthors}%
\unskip\
\newblock
\APACrefYearMonthDay{2021}{{\APACmonth{09}}}{}.
\newblock
{\BBOQ}\APACrefatitle {{Tidal constraints on the low-viscosity zone of the
  Moon}} {{Tidal constraints on the low-viscosity zone of the Moon}}.{\BBCQ}
\newblock
\APACjournalVolNumPages{Icarus}{365}{}{114361}.
\newblock
\begin{APACrefDOI} \doi{10.1016/j.icarus.2021.114361} \end{APACrefDOI}
\PrintBackRefs{\CurrentBib}

\bibitem [\protect \citeauthoryear {%
{Thor}%
\ \protect \BOthers {.}}{%
{Thor}%
\ \protect \BOthers {.}}{%
{\protect \APACyear {2021}}%
}]{%
thor2021}
\APACinsertmetastar {%
thor2021}%
\begin{APACrefauthors}%
{Thor}, R\BPBI N.%
, {Kallenbach}, R.%
, {Christensen}, U\BPBI R.%
, {Gl{\"a}ser}, P.%
, {Stark}, A.%
, {Steinbr{\"u}gge}, G.%
\BCBL {}\ \BBA {} {Oberst}, J.%
\end{APACrefauthors}%
\unskip\
\newblock
\APACrefYearMonthDay{2021}{{\APACmonth{01}}}{}.
\newblock
{\BBOQ}\APACrefatitle {{Determination of the lunar body tide from global laser
  altimetry data}} {{Determination of the lunar body tide from global laser
  altimetry data}}.{\BBCQ}
\newblock
\APACjournalVolNumPages{Journal of Geodesy}{95}{1}{4}.
\newblock
\begin{APACrefURL}
  \url{https://www.hou.usra.edu/meetings/lpsc2022/pdf/1576.pdf}
  \end{APACrefURL}
\newblock
\begin{APACrefDOI} \doi{10.1007/s00190-020-01455-8} \end{APACrefDOI}
\PrintBackRefs{\CurrentBib}

\bibitem [\protect \citeauthoryear {%
Tobie%
, Grasset%
, Lunine%
, Mocquet%
\BCBL {}\ \BBA {} Sotin%
}{%
Tobie%
\ \protect \BOthers {.}}{%
{\protect \APACyear {2005}}%
}]{%
tobie05b}
\APACinsertmetastar {%
tobie05b}%
\begin{APACrefauthors}%
Tobie, G.%
, Grasset, O.%
, Lunine, J\BPBI I.%
, Mocquet, A.%
\BCBL {}\ \BBA {} Sotin, S.%
\end{APACrefauthors}%
\unskip\
\newblock
\APACrefYearMonthDay{2005}{}{}.
\newblock
{\BBOQ}\APACrefatitle {{Tidal dissipation within large icy satellites:
  Applications to Europa and Titan}} {{Tidal dissipation within large icy
  satellites: Applications to Europa and Titan}}.{\BBCQ}
\newblock
\APACjournalVolNumPages{Icarus}{175}{2}{496-502}.
\newblock
\begin{APACrefDOI} \doi{10.1016/j.icarus.2004.12.007} \end{APACrefDOI}
\PrintBackRefs{\CurrentBib}

\bibitem [\protect \citeauthoryear {%
{van Kan Parker}%
\ \protect \BOthers {.}}{%
{van Kan Parker}%
\ \protect \BOthers {.}}{%
{\protect \APACyear {2012}}%
}]{%
vankanparker2012}
\APACinsertmetastar {%
vankanparker2012}%
\begin{APACrefauthors}%
{van Kan Parker}, M.%
, {Sanloup}, C.%
, {Sator}, N.%
, {Guillot}, B.%
, {Tronche}, E\BPBI J.%
, {Perrillat}, J\BHBI P.%
\BDBL {}{van Westrenen}, W.%
\end{APACrefauthors}%
\unskip\
\newblock
\APACrefYearMonthDay{2012}{{\APACmonth{03}}}{}.
\newblock
{\BBOQ}\APACrefatitle {{Neutral buoyancy of titanium-rich melts in the deep
  lunar interior}} {{Neutral buoyancy of titanium-rich melts in the deep lunar
  interior}}.{\BBCQ}
\newblock
\APACjournalVolNumPages{Nature Geoscience}{5}{3}{186-189}.
\newblock
\begin{APACrefDOI} \doi{10.1038/ngeo1402} \end{APACrefDOI}
\PrintBackRefs{\CurrentBib}

\bibitem [\protect \citeauthoryear {%
{Viswanathan}%
\ \protect \BOthers {.}}{%
{Viswanathan}%
\ \protect \BOthers {.}}{%
{\protect \APACyear {2018}}%
}]{%
viswanathan2018}
\APACinsertmetastar {%
viswanathan2018}%
\begin{APACrefauthors}%
{Viswanathan}, V.%
, {Fienga}, A.%
, {Minazzoli}, O.%
, {Bernus}, L.%
, {Laskar}, J.%
\BCBL {}\ \BBA {} {Gastineau}, M.%
\end{APACrefauthors}%
\unskip\
\newblock
\APACrefYearMonthDay{2018}{{\APACmonth{05}}}{}.
\newblock
{\BBOQ}\APACrefatitle {{The new lunar ephemeris INPOP17a and its application to
  fundamental physics}} {{The new lunar ephemeris INPOP17a and its application
  to fundamental physics}}.{\BBCQ}
\newblock
\APACjournalVolNumPages{Monthly Notices of the Royal Astronomical
  Society}{476}{2}{1877-1888}.
\newblock
\begin{APACrefDOI} \doi{10.1093/mnras/sty096} \end{APACrefDOI}
\PrintBackRefs{\CurrentBib}

\bibitem [\protect \citeauthoryear {%
{Viswanathan}%
, {Rambaux}%
, {Fienga}%
, {Laskar}%
\BCBL {}\ \BBA {} {Gastineau}%
}{%
{Viswanathan}%
\ \protect \BOthers {.}}{%
{\protect \APACyear {2019}}%
}]{%
viswanathan2019}
\APACinsertmetastar {%
viswanathan2019}%
\begin{APACrefauthors}%
{Viswanathan}, V.%
, {Rambaux}, N.%
, {Fienga}, A.%
, {Laskar}, J.%
\BCBL {}\ \BBA {} {Gastineau}, M.%
\end{APACrefauthors}%
\unskip\
\newblock
\APACrefYearMonthDay{2019}{{\APACmonth{07}}}{}.
\newblock
{\BBOQ}\APACrefatitle {{Observational Constraint on the Radius and Oblateness
  of the Lunar Core-Mantle Boundary}} {{Observational Constraint on the Radius
  and Oblateness of the Lunar Core-Mantle Boundary}}.{\BBCQ}
\newblock
\APACjournalVolNumPages{Geophysical Research Letters}{46}{13}{7295-7303}.
\newblock
\begin{APACrefDOI} \doi{10.1029/2019GL082677} \end{APACrefDOI}
\PrintBackRefs{\CurrentBib}

\bibitem [\protect \citeauthoryear {%
{Walterov\'{a}}%
, {B\v{e}hounkov\'{a}}%
\BCBL {}\ \BBA {} {Efroimsky}%
}{%
{Walterov\'{a}}%
\ \protect \BOthers {.}}{%
{\protect \APACyear {2023}}%
}]{%
walterova2023data}
\APACinsertmetastar {%
walterova2023data}%
\begin{APACrefauthors}%
{Walterov\'{a}}, M.%
, {B\v{e}hounkov\'{a}}, M.%
\BCBL {}\ \BBA {} {Efroimsky}, M.%
\end{APACrefauthors}%
\unskip\
\newblock
\APACrefYearMonthDay{2023}{{\APACmonth{03}}}{}.
\newblock
\APACrefbtitle {{Is there a semi-molten layer at the base of the lunar mantle?
  [software]}.} {{Is there a semi-molten layer at the base of the lunar mantle?
  [software]}.}
\newblock
\APACaddressPublisher{}{Zenodo}.
\newblock
\begin{APACrefURL} \url{https://doi.org/10.5281/zenodo.7788121}
  \end{APACrefURL}
\newblock
\begin{APACrefDOI} \doi{10.5281/zenodo.7788121} \end{APACrefDOI}
\PrintBackRefs{\CurrentBib}

\bibitem [\protect \citeauthoryear {%
{Weber}%
, {Lin}%
, {Garnero}%
, {Williams}%
\BCBL {}\ \BBA {} {Lognonn{\'e}}%
}{%
{Weber}%
\ \protect \BOthers {.}}{%
{\protect \APACyear {2011}}%
}]{%
Weber}
\APACinsertmetastar {%
Weber}%
\begin{APACrefauthors}%
{Weber}, R\BPBI C.%
, {Lin}, P\BHBI Y.%
, {Garnero}, E\BPBI J.%
, {Williams}, Q.%
\BCBL {}\ \BBA {} {Lognonn{\'e}}, P.%
\end{APACrefauthors}%
\unskip\
\newblock
\APACrefYearMonthDay{2011}{{\APACmonth{01}}}{}.
\newblock
{\BBOQ}\APACrefatitle {{Seismic Detection of the Lunar Core}} {{Seismic
  Detection of the Lunar Core}}.{\BBCQ}
\newblock
\APACjournalVolNumPages{Science}{331}{6015}{309}.
\newblock
\begin{APACrefDOI} \doi{10.1126/science.1199375} \end{APACrefDOI}
\PrintBackRefs{\CurrentBib}

\bibitem [\protect \citeauthoryear {%
{Wieczorek}%
\ \protect \BOthers {.}}{%
{Wieczorek}%
\ \protect \BOthers {.}}{%
{\protect \APACyear {2013}}%
}]{%
wieczorek2013}
\APACinsertmetastar {%
wieczorek2013}%
\begin{APACrefauthors}%
{Wieczorek}, M\BPBI A.%
, {Neumann}, G\BPBI A.%
, {Nimmo}, F.%
, {Kiefer}, W\BPBI S.%
, {Taylor}, G\BPBI J.%
, {Melosh}, H\BPBI J.%
\BDBL {}{Zuber}, M\BPBI T.%
\end{APACrefauthors}%
\unskip\
\newblock
\APACrefYearMonthDay{2013}{{\APACmonth{02}}}{}.
\newblock
{\BBOQ}\APACrefatitle {{The Crust of the Moon as Seen by GRAIL}} {{The Crust of
  the Moon as Seen by GRAIL}}.{\BBCQ}
\newblock
\APACjournalVolNumPages{Science}{339}{6120}{671-675}.
\newblock
\begin{APACrefDOI} \doi{10.1126/science.1231530} \end{APACrefDOI}
\PrintBackRefs{\CurrentBib}

\bibitem [\protect \citeauthoryear {%
{Williams}%
\ \BBA {} {Boggs}%
}{%
{Williams}%
\ \BBA {} {Boggs}%
}{%
{\protect \APACyear {2009}}%
}]{%
WilliamsBoggs2009}
\APACinsertmetastar {%
WilliamsBoggs2009}%
\begin{APACrefauthors}%
{Williams}, J\BPBI G.%
\BCBT {}\ \BBA {} {Boggs}, D\BPBI H.%
\end{APACrefauthors}%
\unskip\
\newblock
\APACrefYearMonthDay{2009}{}{}.
\newblock
{\BBOQ}\APACrefatitle {{Lunar Core and Mantle. What Does LLR See?}} {{Lunar
  Core and Mantle. What Does LLR See?}}{\BBCQ}
\newblock
\BIn{} \APACrefbtitle {Proceedings of the 16th International Workshop on Laser
  Ranging held on 13-17 October 2008 in Pozna\'n, Poland. Edited by S.
  Schilliak. Published by: Space Research Centre, Polish Academy of Sciences,
  Warsaw} {Proceedings of the 16th international workshop on laser ranging held
  on 13-17 october 2008 in pozna\'n, poland. edited by s. schilliak. published
  by: Space research centre, polish academy of sciences, warsaw}\
  (\BPG~101-120).
\newblock
\begin{APACrefURL}
  \url{http://cddis.gsfc.nasa.gov/lw16/docs/papers/proceedings\_vol2.pdf,
  http://cddis.gsfc.nasa.gov/lw16/docs/papers/sci\_1\_Williams\_p.pdf}
  \end{APACrefURL}
\PrintBackRefs{\CurrentBib}

\bibitem [\protect \citeauthoryear {%
{Williams}%
\ \BBA {} {Boggs}%
}{%
{Williams}%
\ \BBA {} {Boggs}%
}{%
{\protect \APACyear {2015}}%
}]{%
williams2015}
\APACinsertmetastar {%
williams2015}%
\begin{APACrefauthors}%
{Williams}, J\BPBI G.%
\BCBT {}\ \BBA {} {Boggs}, D\BPBI H.%
\end{APACrefauthors}%
\unskip\
\newblock
\APACrefYearMonthDay{2015}{{\APACmonth{04}}}{}.
\newblock
{\BBOQ}\APACrefatitle {{Tides on the Moon: Theory and determination of
  dissipation}} {{Tides on the Moon: Theory and determination of
  dissipation}}.{\BBCQ}
\newblock
\APACjournalVolNumPages{Journal of Geophysical Research
  (Planets)}{120}{4}{689-724}.
\newblock
\begin{APACrefDOI} \doi{10.1002/2014JE004755} \end{APACrefDOI}
\PrintBackRefs{\CurrentBib}

\bibitem [\protect \citeauthoryear {%
{Williams}%
, {Boggs}%
\BCBL {}\ \BBA {} {Ratcliff}%
}{%
{Williams}%
\ \protect \BOthers {.}}{%
{\protect \APACyear {2012}}%
}]{%
williams2012}
\APACinsertmetastar {%
williams2012}%
\begin{APACrefauthors}%
{Williams}, J\BPBI G.%
, {Boggs}, D\BPBI H.%
\BCBL {}\ \BBA {} {Ratcliff}, J\BPBI T.%
\end{APACrefauthors}%
\unskip\
\newblock
\APACrefYearMonthDay{2012}{{\APACmonth{03}}}{}.
\newblock
{\BBOQ}\APACrefatitle {{Lunar Moment of Inertia, Love Number, and Core}}
  {{Lunar Moment of Inertia, Love Number, and Core}}.{\BBCQ}
\newblock
\BIn{} \APACrefbtitle {43rd Annual Lunar and Planetary Science Conference}
  {43rd annual lunar and planetary science conference}\ (\BPG~2230).
\PrintBackRefs{\CurrentBib}

\bibitem [\protect \citeauthoryear {%
{Williams}%
, {Boggs}%
, {Yoder}%
, {Ratcliff}%
\BCBL {}\ \BBA {} {Dickey}%
}{%
{Williams}%
\ \protect \BOthers {.}}{%
{\protect \APACyear {2001}}%
}]{%
williams2001}
\APACinsertmetastar {%
williams2001}%
\begin{APACrefauthors}%
{Williams}, J\BPBI G.%
, {Boggs}, D\BPBI H.%
, {Yoder}, C\BPBI F.%
, {Ratcliff}, J\BPBI T.%
\BCBL {}\ \BBA {} {Dickey}, J\BPBI O.%
\end{APACrefauthors}%
\unskip\
\newblock
\APACrefYearMonthDay{2001}{{\APACmonth{11}}}{}.
\newblock
{\BBOQ}\APACrefatitle {{Lunar rotational dissipation in solid body and molten
  core}} {{Lunar rotational dissipation in solid body and molten core}}.{\BBCQ}
\newblock
\APACjournalVolNumPages{Journal of Geophysical
  Research}{106}{E11}{27933-27968}.
\newblock
\begin{APACrefDOI} \doi{10.1029/2000JE001396} \end{APACrefDOI}
\PrintBackRefs{\CurrentBib}

\bibitem [\protect \citeauthoryear {%
{Williams}%
\ \protect \BOthers {.}}{%
{Williams}%
\ \protect \BOthers {.}}{%
{\protect \APACyear {2014}}%
}]{%
williams2014}
\APACinsertmetastar {%
williams2014}%
\begin{APACrefauthors}%
{Williams}, J\BPBI G.%
, {Konopliv}, A\BPBI S.%
, {Boggs}, D\BPBI H.%
, {Park}, R\BPBI S.%
, {Yuan}, D\BHBI N.%
, {Lemoine}, F\BPBI G.%
\BDBL {}{Zuber}, M\BPBI T.%
\end{APACrefauthors}%
\unskip\
\newblock
\APACrefYearMonthDay{2014}{{\APACmonth{07}}}{}.
\newblock
{\BBOQ}\APACrefatitle {{Lunar interior properties from the GRAIL mission}}
  {{Lunar interior properties from the GRAIL mission}}.{\BBCQ}
\newblock
\APACjournalVolNumPages{Journal of Geophysical Research
  (Planets)}{119}{7}{1546-1578}.
\newblock
\begin{APACrefDOI} \doi{10.1002/2013JE004559} \end{APACrefDOI}
\PrintBackRefs{\CurrentBib}

\bibitem [\protect \citeauthoryear {%
{Williams}%
\ \protect \BOthers {.}}{%
{Williams}%
\ \protect \BOthers {.}}{%
{\protect \APACyear {2015}}%
}]{%
williams2015lpsc}
\APACinsertmetastar {%
williams2015lpsc}%
\begin{APACrefauthors}%
{Williams}, J\BPBI G.%
, {Konopliv}, A\BPBI S.%
, {Park}, R\BPBI S.%
, {Yuan}, D\BPBI N.%
, {Asmar}, S\BPBI W.%
, {Watkins}, M\BPBI M.%
\BDBL {}{Zuber}, M\BPBI T.%
\end{APACrefauthors}%
\unskip\
\newblock
\APACrefYearMonthDay{2015}{{\APACmonth{03}}}{}.
\newblock
{\BBOQ}\APACrefatitle {{The Deep Lunar Interior from GRAIL}} {{The Deep Lunar
  Interior from GRAIL}}.{\BBCQ}
\newblock
\BIn{} \APACrefbtitle {46th Annual Lunar and Planetary Science Conference}
  {46th annual lunar and planetary science conference}\ (\BPG~1380).
\PrintBackRefs{\CurrentBib}

\bibitem [\protect \citeauthoryear {%
Wu%
\ \BBA {} Peltier%
}{%
Wu%
\ \BBA {} Peltier%
}{%
{\protect \APACyear {1982}}%
}]{%
wu1982}
\APACinsertmetastar {%
wu1982}%
\begin{APACrefauthors}%
Wu, P.%
\BCBT {}\ \BBA {} Peltier, W\BPBI R.%
\end{APACrefauthors}%
\unskip\
\newblock
\APACrefYearMonthDay{1982}{}{}.
\newblock
{\BBOQ}\APACrefatitle {Viscous gravitational relaxation} {Viscous gravitational
  relaxation}.{\BBCQ}
\newblock
\APACjournalVolNumPages{Geophysical Journal International}{70}{2}{435-485}.
\newblock
\begin{APACrefDOI} \doi{10.1111/j.1365-246X.1982.tb04976.x} \end{APACrefDOI}
\PrintBackRefs{\CurrentBib}

\bibitem [\protect \citeauthoryear {%
{Wyatt}%
}{%
{Wyatt}%
}{%
{\protect \APACyear {1977}}%
}]{%
wyatt1977}
\APACinsertmetastar {%
wyatt1977}%
\begin{APACrefauthors}%
{Wyatt}, B\BPBI A.%
\end{APACrefauthors}%
\unskip\
\newblock
\APACrefYearMonthDay{1977}{{\APACmonth{01}}}{}.
\newblock
{\BBOQ}\APACrefatitle {{The melting and crystallisation behaviour of a natural
  clinopyroxene-ilmenite intergrowth}} {{The melting and crystallisation
  behaviour of a natural clinopyroxene-ilmenite intergrowth}}.{\BBCQ}
\newblock
\APACjournalVolNumPages{Contributions to Mineralogy and Petrology}{61}{1}{1-9}.
\newblock
\begin{APACrefDOI} \doi{10.1007/BF00375941} \end{APACrefDOI}
\PrintBackRefs{\CurrentBib}

\bibitem [\protect \citeauthoryear {%
{Xiao}%
\ \protect \BOthers {.}}{%
{Xiao}%
\ \protect \BOthers {.}}{%
{\protect \APACyear {2022}}%
}]{%
xiao2022}
\APACinsertmetastar {%
xiao2022}%
\begin{APACrefauthors}%
{Xiao}, C.%
, {Wu}, Y.%
, {Yan}, J.%
, {Harada}, Y.%
, {Zhang}, Y.%
\BCBL {}\ \BBA {} {Li}, F.%
\end{APACrefauthors}%
\unskip\
\newblock
\APACrefYearMonthDay{2022}{{\APACmonth{12}}}{}.
\newblock
{\BBOQ}\APACrefatitle {{Comparison of the Effects of Different Viscoelastic and
  Temperature Models on the Theoretical Tidal Response of the Moon}}
  {{Comparison of the Effects of Different Viscoelastic and Temperature Models
  on the Theoretical Tidal Response of the Moon}}.{\BBCQ}
\newblock
\APACjournalVolNumPages{Journal of Geophysical Research
  (Planets)}{127}{12}{e2022JE007215}.
\newblock
\begin{APACrefDOI} \doi{10.1029/2022JE007215} \end{APACrefDOI}
\PrintBackRefs{\CurrentBib}

\bibitem [\protect \citeauthoryear {%
{Yan}%
\ \protect \BOthers {.}}{%
{Yan}%
\ \protect \BOthers {.}}{%
{\protect \APACyear {2012}}%
}]{%
yan2012}
\APACinsertmetastar {%
yan2012}%
\begin{APACrefauthors}%
{Yan}, J.%
, {Goossens}, S.%
, {Matsumoto}, K.%
, {Ping}, J.%
, {Harada}, Y.%
, {Iwata}, T.%
\BDBL {}{Kawano}, N.%
\end{APACrefauthors}%
\unskip\
\newblock
\APACrefYearMonthDay{2012}{{\APACmonth{03}}}{}.
\newblock
{\BBOQ}\APACrefatitle {{CEGM02: An improved lunar gravity model using Chang'E-1
  orbital tracking data}} {{CEGM02: An improved lunar gravity model using
  Chang'E-1 orbital tracking data}}.{\BBCQ}
\newblock
\APACjournalVolNumPages{Planetary and Space Science}{62}{1}{1-9}.
\newblock
\begin{APACrefDOI} \doi{10.1016/j.pss.2011.11.010} \end{APACrefDOI}
\PrintBackRefs{\CurrentBib}

\bibitem [\protect \citeauthoryear {%
{Yan}%
\ \protect \BOthers {.}}{%
{Yan}%
\ \protect \BOthers {.}}{%
{\protect \APACyear {2020}}%
}]{%
yan2020}
\APACinsertmetastar {%
yan2020}%
\begin{APACrefauthors}%
{Yan}, J.%
, {Liu}, S.%
, {Xiao}, C.%
, {Ye}, M.%
, {Cao}, J.%
, {Harada}, Y.%
\BDBL {}{Barriot}, J\BHBI P.%
\end{APACrefauthors}%
\unskip\
\newblock
\APACrefYearMonthDay{2020}{{\APACmonth{04}}}{}.
\newblock
{\BBOQ}\APACrefatitle {{A degree-100 lunar gravity model from the Chang'e 5T1
  mission}} {{A degree-100 lunar gravity model from the Chang'e 5T1
  mission}}.{\BBCQ}
\newblock
\APACjournalVolNumPages{Astronomy \& Astrophysics}{636}{}{A45}.
\newblock
\begin{APACrefDOI} \doi{10.1051/0004-6361/201936802} \end{APACrefDOI}
\PrintBackRefs{\CurrentBib}

\bibitem [\protect \citeauthoryear {%
{Zhang}%
, {Parmentier}%
\BCBL {}\ \BBA {} {Liang}%
}{%
{Zhang}%
\ \protect \BOthers {.}}{%
{\protect \APACyear {2013}}%
}]{%
zhang2013}
\APACinsertmetastar {%
zhang2013}%
\begin{APACrefauthors}%
{Zhang}, N.%
, {Parmentier}, E\BPBI M.%
\BCBL {}\ \BBA {} {Liang}, Y.%
\end{APACrefauthors}%
\unskip\
\newblock
\APACrefYearMonthDay{2013}{{\APACmonth{09}}}{}.
\newblock
{\BBOQ}\APACrefatitle {{A 3-D numerical study of the thermal evolution of the
  Moon after cumulate mantle overturn: The importance of rheology and core
  solidification}} {{A 3-D numerical study of the thermal evolution of the Moon
  after cumulate mantle overturn: The importance of rheology and core
  solidification}}.{\BBCQ}
\newblock
\APACjournalVolNumPages{Journal of Geophysical Research.
  Planets}{118}{9}{1789-1804}.
\newblock
\begin{APACrefDOI} \doi{10.1002/jgre.20121} \end{APACrefDOI}
\PrintBackRefs{\CurrentBib}

\bibitem [\protect \citeauthoryear {%
{Zhao}%
, {de Vries}%
, {van den Berg}%
, {Jacobs}%
\BCBL {}\ \BBA {} {van Westrenen}%
}{%
{Zhao}%
\ \protect \BOthers {.}}{%
{\protect \APACyear {2019}}%
}]{%
zhao2019}
\APACinsertmetastar {%
zhao2019}%
\begin{APACrefauthors}%
{Zhao}, Y.%
, {de Vries}, J.%
, {van den Berg}, A\BPBI P.%
, {Jacobs}, M\BPBI H\BPBI G.%
\BCBL {}\ \BBA {} {van Westrenen}, W.%
\end{APACrefauthors}%
\unskip\
\newblock
\APACrefYearMonthDay{2019}{{\APACmonth{04}}}{}.
\newblock
{\BBOQ}\APACrefatitle {{The participation of ilmenite-bearing cumulates in
  lunar mantle overturn}} {{The participation of ilmenite-bearing cumulates in
  lunar mantle overturn}}.{\BBCQ}
\newblock
\APACjournalVolNumPages{Earth and Planetary Science Letters}{511}{}{1-11}.
\newblock
\begin{APACrefDOI} \doi{10.1016/j.epsl.2019.01.022} \end{APACrefDOI}
\PrintBackRefs{\CurrentBib}

\bibitem [\protect \citeauthoryear {%
{Zharkov}%
\ \BBA {} {Gudkova}%
}{%
{Zharkov}%
\ \BBA {} {Gudkova}%
}{%
{\protect \APACyear {2005}}%
}]{%
zharkov2005}
\APACinsertmetastar {%
zharkov2005}%
\begin{APACrefauthors}%
{Zharkov}, V\BPBI N.%
\BCBT {}\ \BBA {} {Gudkova}, T\BPBI V.%
\end{APACrefauthors}%
\unskip\
\newblock
\APACrefYearMonthDay{2005}{{\APACmonth{09}}}{}.
\newblock
{\BBOQ}\APACrefatitle {{Construction of Martian Interior Model}} {{Construction
  of Martian Interior Model}}.{\BBCQ}
\newblock
\APACjournalVolNumPages{Solar System Research}{39}{5}{343-373}.
\newblock
\begin{APACrefDOI} \doi{10.1007/s11208-005-0049-7} \end{APACrefDOI}
\PrintBackRefs{\CurrentBib}

\bibitem [\protect \citeauthoryear {%
{Zhu}%
, {W{\"u}nnemann}%
, {Potter}%
, {Kleine}%
\BCBL {}\ \BBA {} {Morbidelli}%
}{%
{Zhu}%
\ \protect \BOthers {.}}{%
{\protect \APACyear {2019}}%
}]{%
zhu2019}
\APACinsertmetastar {%
zhu2019}%
\begin{APACrefauthors}%
{Zhu}, M\BHBI H.%
, {W{\"u}nnemann}, K.%
, {Potter}, R\BPBI W\BPBI K.%
, {Kleine}, T.%
\BCBL {}\ \BBA {} {Morbidelli}, A.%
\end{APACrefauthors}%
\unskip\
\newblock
\APACrefYearMonthDay{2019}{{\APACmonth{08}}}{}.
\newblock
{\BBOQ}\APACrefatitle {{Are the Moon's Nearside-Farside Asymmetries the Result
  of a Giant Impact?}} {{Are the Moon's Nearside-Farside Asymmetries the Result
  of a Giant Impact?}}{\BBCQ}
\newblock
\APACjournalVolNumPages{Journal of Geophysical Research
  (Planets)}{124}{8}{2117-2140}.
\newblock
\begin{APACrefDOI} \doi{10.1029/2018JE005826} \end{APACrefDOI}
\PrintBackRefs{\CurrentBib}

\end{thebibliography}

%
%
%
%
%

\end{document}